\documentclass[useAMS,usenatbib]{mn2e}
\usepackage{graphicx}
\usepackage{Times}
\title[A numerical simulation of the evolution and fate of a FRI jet. The case of 3C~31]
{A numerical simulation of the evolution and fate of a FRI jet.
The case of 3C~31}
\author[M. Perucho \& J.-M.$^{\underline{a}}$ Mart\'{\i}]
  {M.~Perucho,$^1$
  J.-M.$^{\underline{a}}$~Mart\'{\i},$^2$ \\
  $^1$Max-Planck-Institut f\"ur Radioastronomie,
           Auf dem H\"ugel 69, 53121 Bonn, Germany\\
  $^2$Departament d'Astronomia i Astrof\'{\i}sica, Universitat de Val\`encia, Dr. Moliner 50, 46100 Burjassot (Val\`encia), Spain}
\date{Released 2007 Xxxxx XX}

\pagerange{\pageref{firstpage}--\pageref{lastpage}} \pubyear{2007}
\begin{document}

\label{firstpage}

\maketitle

\begin{abstract}
The evolution of FRI jets has been long studied in the framework
of the FRI-FRII dichotomy. The present paradigm consists of the
expansion of overpressured jets in the ambient medium and the
generation of standing recollimation shocks, followed by mass
entrainment from the external medium that decelerates the jets to
subsonic speeds. In this paper, we test the present theoretical
and observational models via a relativistic numerical simulation
of the jets in the radio galaxy 3C~31. We use the parameters
derived from the modelling presented by \cite{lb02a,lb02b} as
input parameters for the simulation of the evolution of the
source, thus assuming that they have not varied over the lifetime
of the source. We simulate about 10~\% of the total lifetime of
the jets in 3C~31. Realistic density and pressure gradients for
the atmosphere are used. The simulation includes an equation of
state for a two-component relativistic gas that allows a separate
treatment of leptonic and baryonic matter. We compare our results
with the modelling of the observational data of the source. Our
results show that the bow shock evolves self-similarly at a
quasi-constant speed, with slight deceleration by the end of the
simulation, in agreement with recent X-ray observations that show
the presence of bow shocks in FRI sources. The jet expands until
it becomes underpressured with respect to the ambient medium, and
then recollimates. Subsequent oscillations around pressure
equilibrium and generation of standing shocks lead to the mass
loading and disruption of the jet flow. We derive an estimate for
the minimum age of the source of $t>1.\,10^8\,\rm{yrs}$, which may
imply continuous activity of 3C~31 since the triggering of its
activity. The simulation shows that weak CSS sources may be the
young counterparts of FRIs. We conclude that the observed
properties of the jets in 3C~31 are basically recovered by the
standing shock scenario.
\end{abstract}


\section{Introduction}

The morphology and evolution of jets in low-power radiogalaxies,
i.e., FRI radiogalaxies in \cite{fr74} classification, have been
addressed by a number of authors, from theoretical
\citep[e.g.,][]{bic84,bic94,komi90a,komi90b,komi94,you93},
numerical \citep[e.g.,][]{you86,bow96} and observational
\citep[e.g.,][]{par02,lb02a,lb02b} points of view. FRI jets show
bright regions close to the core dimming into diffuse emission at
kiloparsec scales. The jets in FRI sources are relativistic at the
parsec scales \citep[see, e.g.,][]{la93,la96}, but decelerate
between the inner regions and the kiloparsec scales \citep[see,
e.g.,][]{la96}.

The theoretical paradigm for the evolution of jets in FRI sources
\citep{bic84,la93,la96} comprises the expansion of an
overpressured jet, followed by the generation of shocks due to
subsequent compressions and expansions around pressure
equilibrium; the jet is decelerated in these shocks and entraines
the external medium through turbulent mixing. The last stage of
the evolution, when the jet has already been decelerated to
subsonic speeds, is dominated by turbulence. Two main processes
for the entrainment of the ambient medium in the jets have been
invoked and studied in the literature: 1) entrainment through
mixing in a turbulent shear layer between the jet and the medium,
and 2) entrainment from stellar mass losses.
\cite{komi90a,komi90b} developed a theoretical model for the mass
entrainment at the jet boundary and made steady-state numerical
calculations, succeeding in explaining the main features of FRI
jets. A series of authors have concentrated on this topic.
\cite{you86} and \cite{bic94} studied the entrainment in jets.
\cite{you93} studied the importance of a sufficiently dense
environment in decelerating jets through entrainment.
\cite{komi94} studied the mass load of a leptonic jet by stellar
winds, concluding that this entrainment can be very important for
the process of deceleration. \cite{bow96} performed a series of
steady-state numerical simulations using an equation of state for
a two-component relativistic gas \citep[see also][]{bow94} and
including a term for the entrainment of the stellar mass loss in
the code according to the model of \cite{komi94}.

\cite{lb02a,lb02b} --LB02a,b from now on-- presented a model which
accurately describes the kinematics and dynamics of the jets in
the FRI radiogalaxy 3C~31, mapping the emission and magnetic
fields of the jets. In LB02a, the observed brightness and
polarization distributions were fitted by modelling the velocity,
synchrotron emissivity and ordering of the magnetic field. In
LB02b, a dynamical model was presented based on the results of
LB02a and estimates for external pressure and density profiles
from \emph{Chandra} \citep{hard02}, applying conservation of
particles, momentum and energy, and assuming that the jets are in
pressure equilibrium with the external medium at large distances
from the nucleus. \cite{lb04} studied the validity of so-called
adiabatic models in explaining the structure of the magnetic and
velocity fields, and the brightness and polarization distributions
in the jets of 3C~31. In these models, the radiating particles are
supposed to be accelerated before entering the region of interest
and then lose energy only by adiabatic processes. The authors
concluded that the adiabatic models give a good description of the
outer regions of the jet, whereas closer to the nucleus, the jet
shows a non-adiabatic behavior. Their fit to the data is inferior
to that of the free models of LB02a, but is obtained with fewer
free parameters.

In this paper, we present the results from a relativistic, purely
hydrodynamical simulation with input parameters taken from
LB02a,b. It is not clear if the magnetic field in the jets of
3C~31 is dynamically important, although it seems, from the
results obtained by LB02a,b, that it is close to equipartition.
However, we will assume in this work that the magnetic field is
not important for the dynamics of the problem. This assumption
remains to be checked by future relativistic magnetohydrodynamic
simulations. Our aim is to compare the results from the simulation
with those from observations and modelling. We also address the
problem of the jet evolution and young counterparts to FRI jets.
The nature of compact radio-sources and their relation with large
scale jets has been studied during the last years. Compact
Symmetric Objects (CSO), as a morphological subclass of Gigahertz
Peaked Spectrum sources (GPS) and Compact Steep Spectrum sources
(CSS), are understood as the young counterparts of FRII jets
\citep[e.g.,][]{fan95,pm02}. However, there is still some debate
on the compact radio sources that could be the young FRIs.
\cite{dra04} and \cite{kun05} have observed low-power CSS from the
IRAS sample and the FIRST survey, respectively. \cite{dra04}
proposed these weak CSS sources as the possible young counterparts
of FRI jets. We study this possibility in this work.

To our knowledge this is the first work to present long term
relativistic simulations of FRI jets. In contrast, the evolution
of FRII jets has been addressed by several authors \citep[see,
e.g.,][]{kofa98,sch02,km03,kra05}. Our work covers a parameter
space complementary to that used in \cite{sch02}, who discussed
the dependence of the morphology and dynamics of jets on their
composition, using the same equation of state as used here.

The plan of the paper is as follows. In Section~\ref{sec:setup},
we present the setup of our simulation. In
Section~\ref{sec:results} we summarize the main results of the
simulation, and we compare these results with new simulations
performed in order to test different evolutionary scenarios by
changing the initial conditions. In Section~\ref{sec:discus} we
discuss the main results of the work, and we present the
conclusions of the paper in Section~\ref{sec:concl}.

\section{Setup of the numerical simulation}\label{sec:setup}

\subsection{The model of Laing \& Bridle}

The numerical simulation presented in this paper is based on the
work by Laing \& Bridle (LB02a,b), as stated in the Introduction.
In this paragraph we review their main assumptions and
conclusions. In the modelling presented in LB02a the jet and
counter-jet are assumed to be identical, antiparallel,
axisymmetric and stationary, and the differences between them are
assumed to be due to relativistic aberration. The models are
designed to fit the observed brightness and polarization
distributions (assuming optically thin emission), taking into
account Doppler boosting effects and relying on simple
prescriptions for the variations of the flow velocity, synchrotron
emissivity an magnetic-field structure within the jet. The
rotation of the line of sight relative to the magnetic field
structure between the emitted and the observed frames is also
taken into account. The model focuses on the region enclosing the
inner 12 kpc of the jets, which is split in three parts: the inner
region (from 0 to 1.1 kpc, or from 0 to 2.5 arcsec), the flaring
region (from 1.1 to 3.5 kpc, or from 2.5 to 8.3 arcsec) and the
outer region (from 3.5 to 12 kpc, or from 8.3 to 28.3 arcsec). The
linear distances are calculated assuming a Hubble constant $H_0=
70 \,\rm{km\,s^{-1}\,Mpc^{-1}}$, taking the redshift of the parent
galaxy of 3C~31 (NGC~383, $z = 0.0169$) and a viewing angle of
$52\degr$. This modelling has been extended to other FRI sources
\citep[e.g., NGC~315 and 3C~296,][]{can05,la06}. In LB02b, the
authors present a dynamical model for the jet, based on: 1) the
results obtained in LB02a, 2) estimates of pressure and density
profiles from \emph{Chandra} and \emph{ROSAT}
\citep{hard02,kom99}, 3) the conservation of particles, momentum
and energy, 4) the assumption that the jets are in pressure
equilibrium with the external medium at large distances from the
nucleus, and 5) the momentum flux being $\Pi=\Phi/c$, where $\Phi$
is the energy flux (a good approximation for light, hot and/or
relativistic jets). The model is \emph{quasi-one-dimensional}, as,
although the widening of the jet is considered, only the axial
velocities are used in the analysis. The authors conclude that
after the expansion of the jet in the inner region, the jet flow
is overpressured and decelerated at the beginning of the flaring
region. In this region, they find local minima of pressure and
density and maxima in the Mach number and entrainment rate. At the
end of the flaring region, the jets are slightly underpressured
but close to pressure equilibrium with the ambient medium. Changes
in the outer region are smooth, with almost constant density and
monotonically increasing entrainment rate. The Mach number in the
outer region is $\sim 1-2$. Comparison of the jet density with the
number of radiating particles required by the observed synchrotron
emissivity lead the authors to conclude that the jet is probably
initially composed of a pair plasma ($e^- - e^+$), and mass-loaded
with baryons inside the galaxy. This mass load is attributed
dominantly to stellar wind material. The exact origin of the
majority of the gas entrained by the jets is, however, still
uncertain.

\subsection{Setup}
  The numerical simulation was performed using a finite-difference
code based on a high-resolution shock-capturing scheme which
solves the equations of relativistic hydrodynamics in two
dimensions written in conservation form \citep{mart97}. The code
is parallelized using OMP directives and it has been modified in
order to include the equation of state for relativistic Boltzmann
gases \citep[][, and Appendix A in this paper]{syn57,fal96} with
the routines used in the simulations of \cite{sch02}. The use of
this equation of state allows us to distinguish electrons,
positrons and protons in the simulation. Under the assumption of
charge neutrality, only one  conservation equation (for example,
that for the evolution of the leptonic density) must be added in
order to know the composition of the fluid at each computational
cell. The integration of this extra equation together with the use
of the Synge equation of state (involving the computation of
Bessel functions) increases the computational time per iteration
by $\sim 50\,\%$ with respect to the case of the one component,
ideal gas equation of state \citep{sch02}.

The equations of relativistic hydrodynamics in conservation form,
which are solved by the code, assuming axisymmetry and in
two-dimensional cylindrical coordinates $(R,z)$, are the following
(we use units in which the speed of light is set to 1):

\begin{equation}  \label{eq:claws}
  \frac{\partial \textbf{U}}{\partial t} + \frac{1}{R}\frac{\partial R
  \textbf{F}^R}{\partial R}  + \frac{\partial \textbf{F}^z}{\partial z} = \textbf{S}
  \quad ,
\end{equation}

\noindent with the vector of unknowns

\begin{equation}
  \textbf{U} = (D, D_l, S^R, S^z, \tau)^T \quad ,
\end{equation}

\noindent
fluxes

\begin{equation}
  \textbf{F}^R = (D v^R, D_l v^R, S^R v^R + p, S^z v^R, S^R - D v^R)^T \quad ,
\end{equation}

\begin{equation}
  \textbf{F}^z = (D v^z, D_l v^z, S^R v^z, S^z v^z + p, S^z - D v^z)^T \quad ,
\end{equation}

\noindent and source terms

\begin{equation} \label{eq:sources}
  \textbf{S} = (0, 0, p/R + g^R, g^z, v^R g^R + v^z g^z)^T.
\end{equation}

  The five unknowns $D, D_l, S^R, S^z$ and $\tau$ refer to the
densities of five conserved quantities, namely the total and
leptonic rest masses, the radial and axial components of the
momentum, and the energy (excluding rest mass energy). They are
all measured in the laboratory frame, and are related to the
quantities in the local rest frame of the fluid (primitive
variables) according to

\begin{equation} \label{eq:D}
  D = \rho\, W   \quad ,
\end{equation}

\begin{equation}
  D_l = \rho_l\, W   \quad ,
\end{equation}

\begin{equation}
  S^{R,z} = \rho \, h\, W^2 \, v^{R,z}   \quad ,
\end{equation}

\begin{equation} \label{eq:tau}
  \tau = \rho\, h\, W^2 \, - p\, -\, D   \quad ,
\end{equation}

\noindent where $\rho$ and $\rho_l$ are the total and the leptonic
rest mass density, respectively, $v^{R,z}$ are the components of
the velocity of the fluid, $W$ is the Lorentz factor ($W =
1/\sqrt{1-v^i v_i}$, where summation over repeated indices is
implied), and $h$ is the specific enthalpy defined as

\begin{equation}  \label{eq:h}
  h = 1 + \varepsilon + p/\rho \quad ,
\end{equation}

\noindent where $\varepsilon$ is the specific internal energy and
$p$ is the pressure. Finally, $g^{R}$, $g^{z}$ appearing in the
definition of the source vector, eq.~(\ref{eq:sources}), are the
components of an external force that keeps the atmosphere in
equilibrium (see below). The system (\ref{eq:claws}) is closed by
means of the Synge equation of state described in Appendix A.

%%%%%%%%%%%%%%%%%%%%%%%%%%%%%%%%%%%%%%%%%%%%%%%%%%%%%%%%%%%%%%%%%%%%%%%%%%
%
\begin{figure*}
\centerline{
\includegraphics[width=0.5\textwidth]{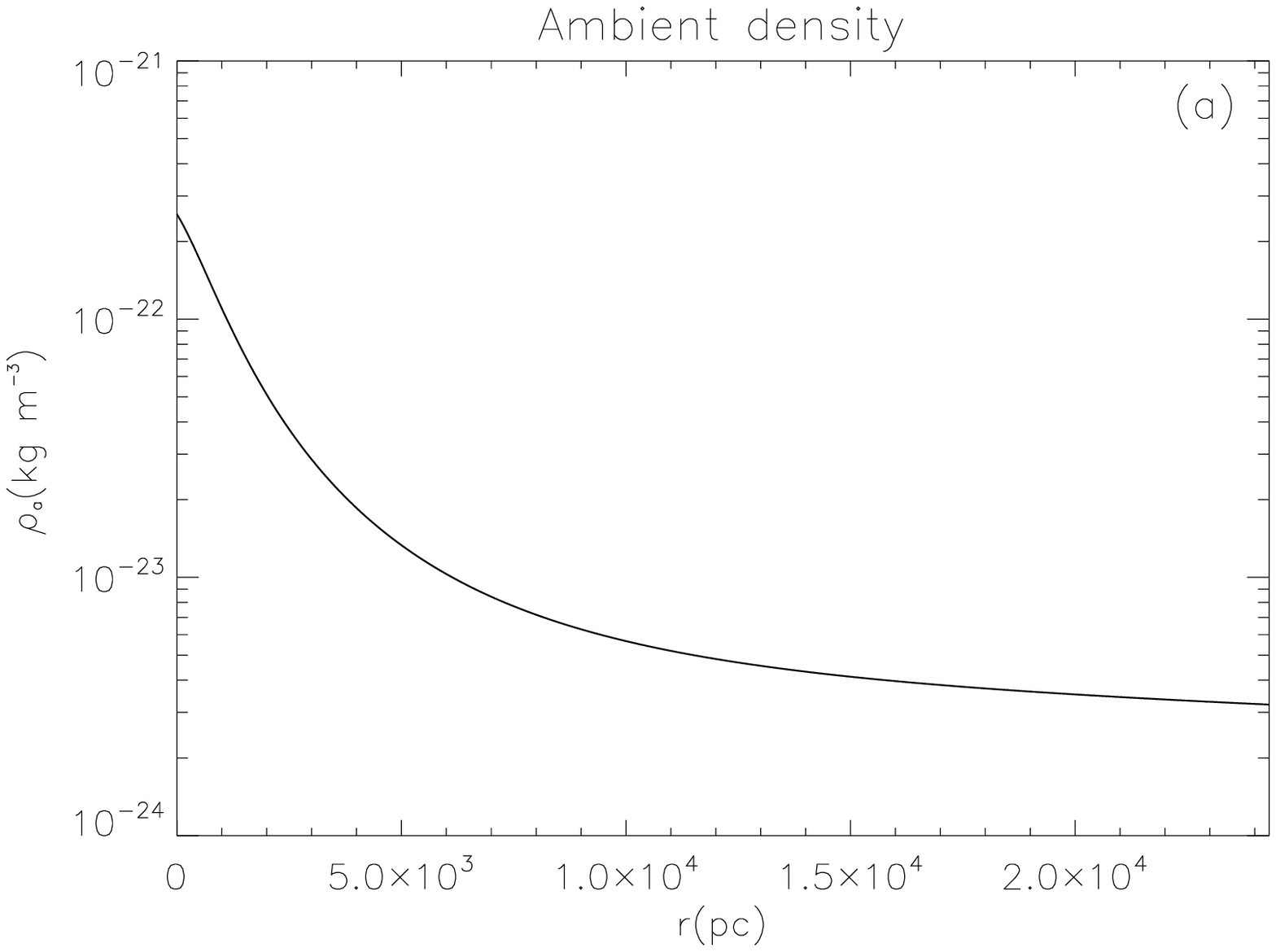}
\includegraphics[width=0.5\textwidth]{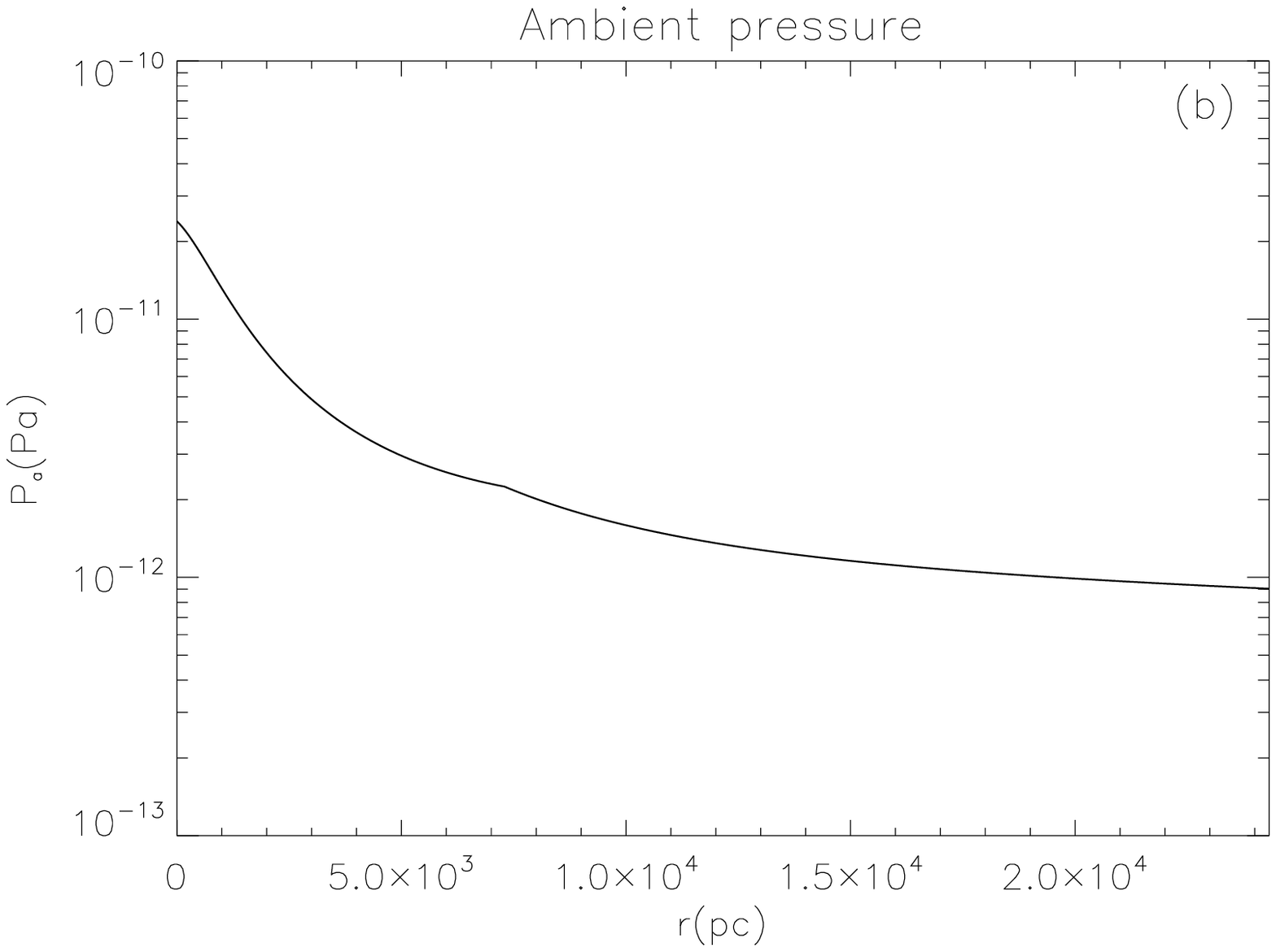}}
 \caption{Initial profiles of ambient rest mass density (a)
and pressure (b).}
 \label{fig:atm}
\end{figure*}
%
%%%%%%%%%%%%%%%%%%%%%%%%%%%%%%%%%%%%%%%%%%%%%%%%%%%%%%%%%%%%%%%%%%%%%%%%%%

%%%%%%%%%%%%%%%%%%%%%%%%%%%%%%%%%%%%%%%%%%%%%%%%%%%%%%%%%%%%%%%%%%%%%%%%%%
%

\begin{table*}
\begin{center}
\caption{Ambient medium parameters (Table~1 in LB02b). Subscripts
$c$ and $g$ refer to the galaxy core and to the surrounding group,
respectively.} \label{tab:LB02}
\begin{tabular}{@{}ccccc}
\hline
Component& Central density & Form factor & Core radius & Temperature\\
\hline
Galaxy&$n_c=1.8\,10^5 \rm{m^{-3}}$&$\beta_{atm,c}=0.73$&$r_c=1.2\,\rm{kpc}$&$T_c=4.9\,10^6 K$\\
Group&$n_g=1.9\,10^3 \rm{m^{-3}}$&$\beta_{atm,g}=0.38$&$r_g=52\,\rm{kpc}$&$T_g=1.7\,10^7 K$\\
\hline
\end{tabular}
\end{center}
\end{table*}
%
%%%%%%%%%%%%%%%%%%%%%%%%%%%%%%%%%%%%%%%%%%%%%%%%%%%%%%%%%%%%%%%%%%%%%%%%%%

  The code also integrates an equation for the jet mass fraction, $f$.
This quantity, set to 1 for the injected beam material and 0
otherwise, is used as a tracer of the jet material through the
grid and allows to study phenomena like the entrainment of ambient
material in the beam and the mixing in the cocoon.

  The medium in which the jet is injected consists of a decreasing
density atmosphere composed of Hydrogen\footnote{LB02b use the
standard composition with $74\%$ hydrogen, but this treatment
would require the inclusion of new populations of particles (in
order to account for the remaining $26\%$) in the code, involving
longer computational time, so that we discarded this option.}. The
dynamical equilibrium of the atmosphere is attained by introducing
an external force which compensates initial pressure gradients in
the radial and axial directions. The profile for the number
density of such a medium is \citep{hard02}:

\begin{equation}\label{profrho}
  n_{ext}(r) =
  n_c\left(1 + \frac{r^2}{r_c^2}\right)^{-3\beta_{atm,c}/2} + \nonumber {}\\
  n_g\left(1 + \frac{r^2}{r_g^2}\right)^{-3\beta_{atm,g}/2},
\end{equation}

\noindent
where $r$ is the spherical radial coordinate, $n_c$ and $n_g$ are
the core densities of the galaxy and the surrounding group, $r_c$
and $r_g$ are the radii of those cores, and $\beta_{atm,c}$ and
$\beta_{atm,g}$ are the exponents giving the profile for each
medium. The temperature profile is:

\begin{eqnarray}\label{temprof}
    T=T_c+(T_g-T_c)\frac{r}{r_m} \quad (r<r_m)\nonumber {}\\
    T=T_g   \qquad \qquad (r\geq r_m),
\end{eqnarray}

\noindent
with $r_m = 7.8\, \rm{kpc}$. The pressure is derived from the
following equation of state:

\begin{equation}\label{pres}
    P_{ext} = \frac{k_B T}{\mu X}n_{ext}(r),
\end{equation}

\noindent
where $\mu$ is the mass per particle in a.m.u. ($\mu = 0.5$ in our
case, cf. $0.6$ in LB02b), $X$ is the abundance of hydrogen by
mass ($X = 1$ here, cf. $0.74$ in LB02b). In Table~\ref{tab:LB02} we
reproduce Table~1 in LB02b, where the parameters for the equations
above are listed. The initial profiles for ambient medium pressure
and density along the axis of the jet are plotted in
Fig.~\ref{fig:atm}.

%%%%%%%%%%%%%%%%%%%%%%%%%%%%%%%%%%%%%%%%%%%%%%%%%%%%%%%%%%%%%%%%%%%%%%%%%%
%
\begin{table}
\begin{center}
\caption{Table of parameters used in the simulation. The
one-dimensional velocity estimation given in the Table stands for
the theoretical advance velocity of the jet, computed using the
equation derived in Mart\'{\i} et al. (1997) for a
pressure-matched jet propagating in one dimension (i.e., without
sideways expansion) through an homogeneous medium.}
\label{tab:sim3c31}
\begin{tabular}{@{}ll}
\hline
Velocity ($v_j$)&$0.87\,c$\\
Mach number ($M_j$)&$2.5$\\
Temperature ($T_j$, jet)&$4.1\,10^9 \rm{K}$\\
Temperature ($T_c$, ambient$^1$)&$5.7\,10^6 \rm{K}$\\
Temperature ($T_g$, ambient$^2$)&$1.7\,10^7 \rm{K}$\\
Density ($\rho_j$, jet)&$3\,10^{-27} \rm{kg/m^{3}}$\\
Density ($\rho_{a,c}$, ambient$^1$)&$3\,10^{-22} \rm{kg/m^{3}}$\\
Density ratio ($\eta$)&$10^{-5}$\\
Leptonic number ($X_l$, jet)&$1.0$\\
Specific int. energy ($\varepsilon_j$, jet)&$1.54\,\rm{c^2}$\\
Specific int. energy ($\varepsilon_{a,c}$, ambient$^1$)&$1.57\,10^{-6}\,\rm{c^2}$\\
Specific int. energy ($\varepsilon_{a,g}$, ambient$^2$)&$4.69\,10^{-6}\,\rm{c^2}$\\
Pressure ($P_j$, jet)&$6.91\,10^{-6}\,\rm{\rho_{a,c}\,c^2}$\\
Pressure ($P_{a,c}$, ambient$^1$)&$8.84\,10^{-7}\,\rm{\rho_{a,c}\,c^2}$\\
Pressure ($P_{a,g}$, ambient$^2$)&$3.07\,10^{-8}\,\rm{\rho_{a,c}\,c^2}$\\
Pressure ratio ($P_j/P_{a,c}$)&$7.8$\\
Adiabatic exponent ($\Gamma_j$, jet)&$1.38$\\
Adiabatic exponent ($\Gamma_a$, ambient)&$1.66$\\
1D velocity estimation ($v_h^{1d}$)&$9.9\,10^{-3}\,\rm{c}$\\
Time unit ($R_j/c$)& $60\,\rm{pc}/c \sim 195\, \rm{yrs}$ \\
\hline
\end{tabular}

\medskip
$^1$ and $^2$ stand for values in the ambient medium at the
injection and point furthest from injection in the grid,
respectively.
\end{center}
\end{table}
%
%%%%%%%%%%%%%%%%%%%%%%%%%%%%%%%%%%%%%%%%%%%%%%%%%%%%%%%%%%%%%%%%%%%%%

  The numerical code units are the jet radius ($R_j$), the speed of
light ($c$) and the density of the ambient medium at injection
($\rho_{a,c}$). Thus, the appropriate unit transformations are
performed in the code from physical to code units.

  The simulation was performed in cylindrical coordinates with axial
symmetry, i.e., only one half of the jet is computed. The grid
involved $2880 \times 1800$ cells, with a resolution of 8
$\rm{cells}/R_j$ in the axial direction up to 300 $R_j$, plus an
extended grid with geometrically increasing cell size (up to 450
$R_j$), and 16 $\rm{cells}/R_j$ in the radial direction up to 100
$R_j$, plus an extended grid (also with geometrically increasing
cell size) up to 200 $R_j$. Outflow boundary conditions were used
at the end of the grid in the axial direction and also far from
the jet axis in the radial direction. Reflecting boundary
conditions were taken for the jet axis in order to account for the
cylindrical symmetry.

  Injection of the jet in the atmosphere is done at $r=500\,
\rm{pc}$, the point where the modelling of the jet in LB02b
starts. The radius of the jet at that point is calculated from the
opening angle of $6.7\degr$ given for the jet in LB02a ($R_j=60\,
\rm{pc}$). The uniform grid is then $17.5\,\rm{kpc}\, \times\,
6\,\rm{kpc}$, and, with the extended grid, $26.25\,\rm{kpc}\,
\times\, 12\,\rm{kpc}$.

  The jet is injected with a speed $v_j = 0.87\,c$ (jet Lorentz factor,
$W_j \sim 2$), internal relativistic Mach number $M_j = 2.5$,
temperature $T_j = 4.1\,10^9 \rm{K}$, density ratio with the
external medium $\eta = 1.\,10^{-5}$, purely leptonic composition
($X_l = 1.0$), and overpressured by a factor $7.8$ with respect to
the ambient medium. In Table \ref{tab:sim3c31} we give the
complete list of parameters. The parameters given for the ambient
medium at injection are calculated for $r = 500\,\rm{pc}$ (this is
the reason for differences between Table~\ref{tab:LB02} and
Table~\ref{tab:sim3c31}). The numbers in this table give an energy
flux for the jet of $\Phi \sim 10^{37}\,\rm{W}$, which is very
close to the value given in LB02b ($\Phi = 1.1\,
10^{37}\,\rm{W}$), the difference being due to approximations in
variables considered. The quantity $L_{kin}$ (kinetic luminosity
of the jet) used for the simulations in \cite{sch02}, where the
authors make a study of the influence of jet composition on its
long term evolution, is equivalent to the jet energy flux defined
in LB02b. Comparing the value of $L_{kin}$ in our simulation to
that used in \cite{sch02} we use $L_{kin} \sim 10^{37}\,\rm{W}$,
on the upper end of the power distribution of FRI sources, whereas
in their work the jets have $L_{kin}\sim 10^{39}\,\rm{W}$,
appropriate for an FRII source.

\section{Results} \label{sec:results}
\subsection{Evolution} \label{sec:evol}

\subsubsection{Numerical Results}
  The simulation was run for $\sim 2000\, \rm{hours}\,(\sim 83\, \rm{days}$)
on eight processors in the SGI~Altix computer CERCA, at the
\emph{Universitat de Val\`encia}, corresponding to a source
lifetime of $7.26\,10^6\,\rm{yrs}$. At the moment when it was
stopped, the bow shock of the jet was located at
$\sim14.5\,\rm{kpc}$ from the injection point in the grid, i.e.,
at $\sim15\,\rm{kpc}$ from the source. The large amount of
computational time needed is a consequence of the small advance
speed of a FRI-like source and of the numerical effort invested in
the computation of the relativistic equation of state.

%%%%%%%%%%%%%%%%%%%%%%%%%%%%%%%%%%%%%%%%%%%%%%%%%%%%%%%%%%%%%%%%%%%%%%%%%
%
\begin{figure*}
\centerline{
\includegraphics[width=0.5\textwidth]{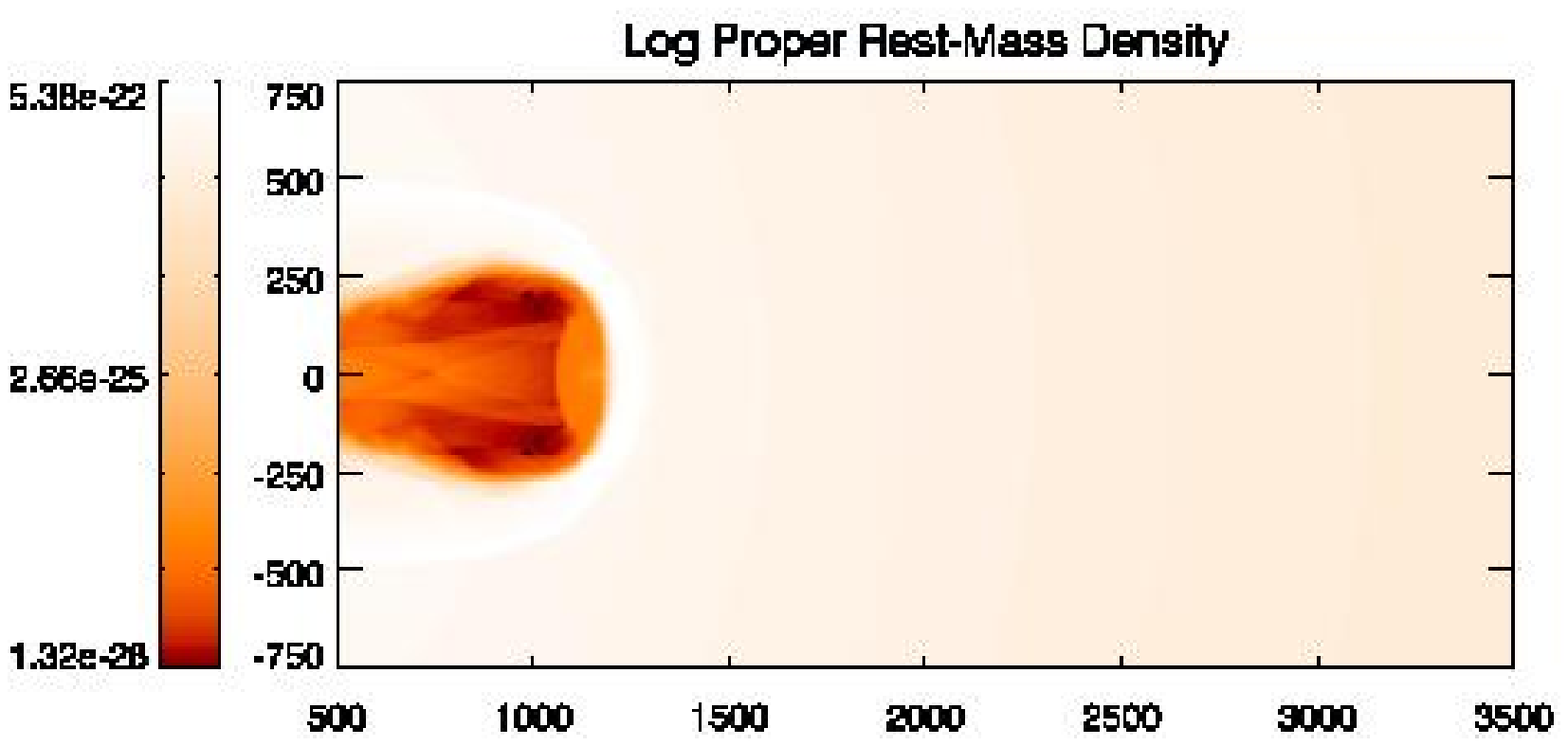}
\includegraphics[width=0.5\textwidth]{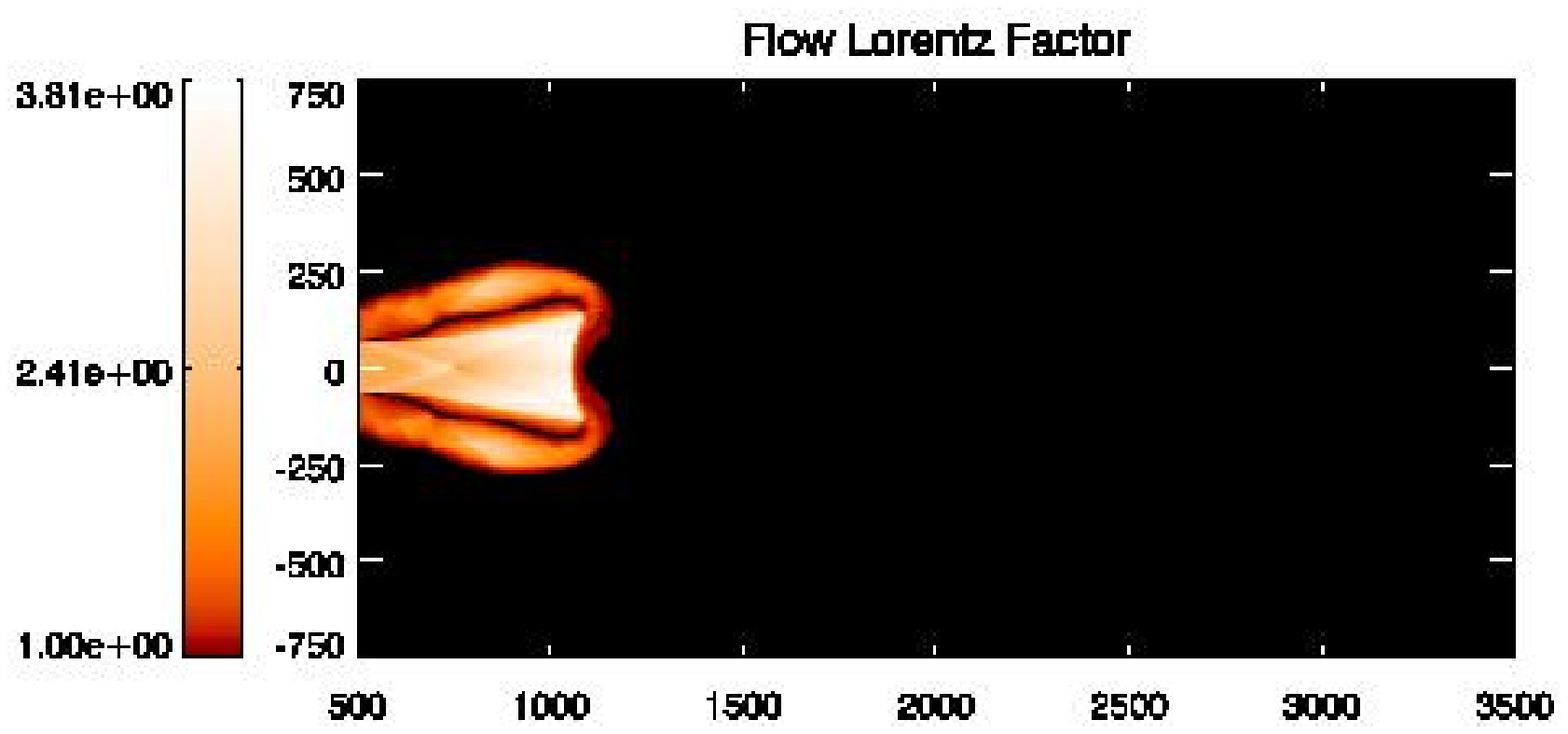}}
\centerline{
\includegraphics[width=0.5\textwidth]{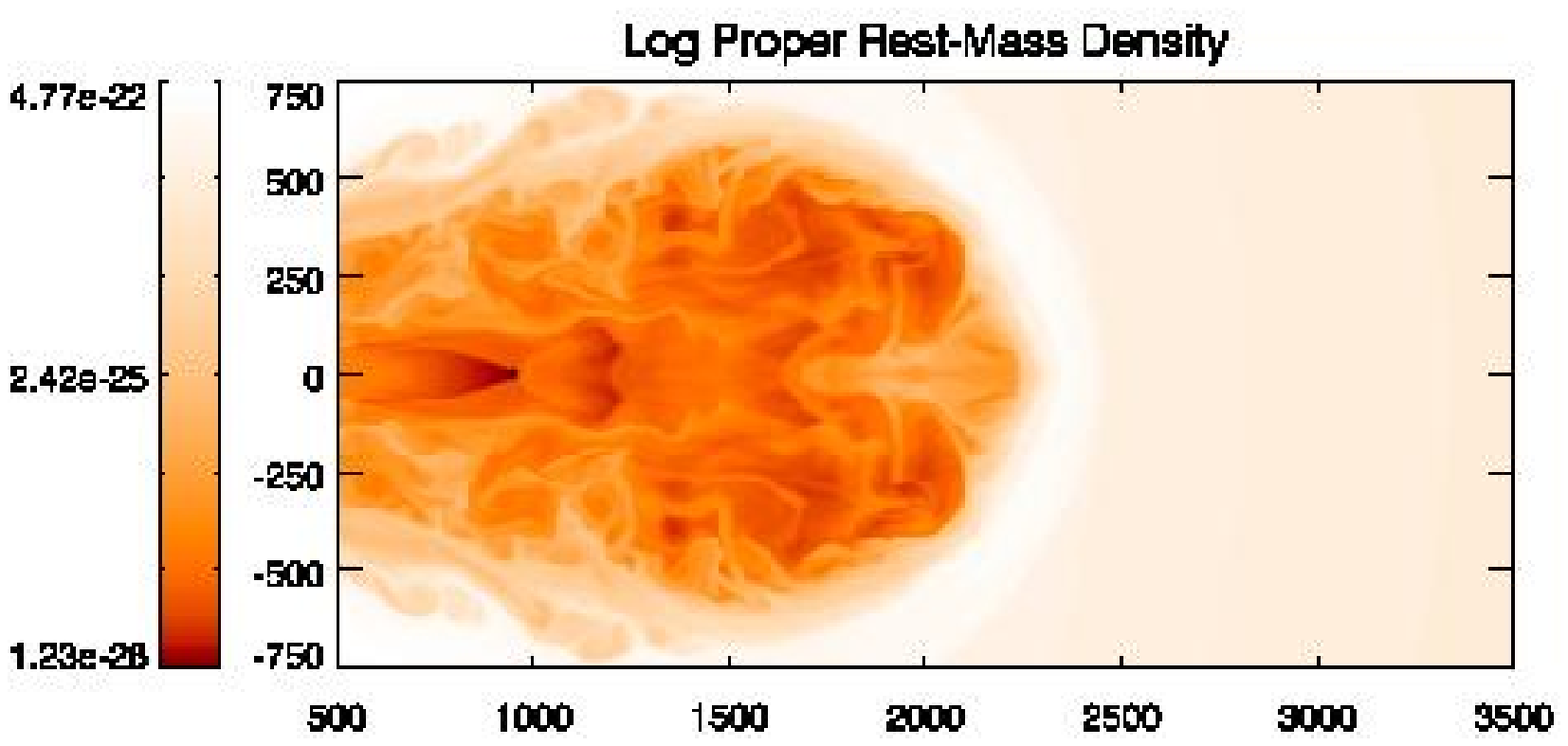}
\includegraphics[width=0.5\textwidth]{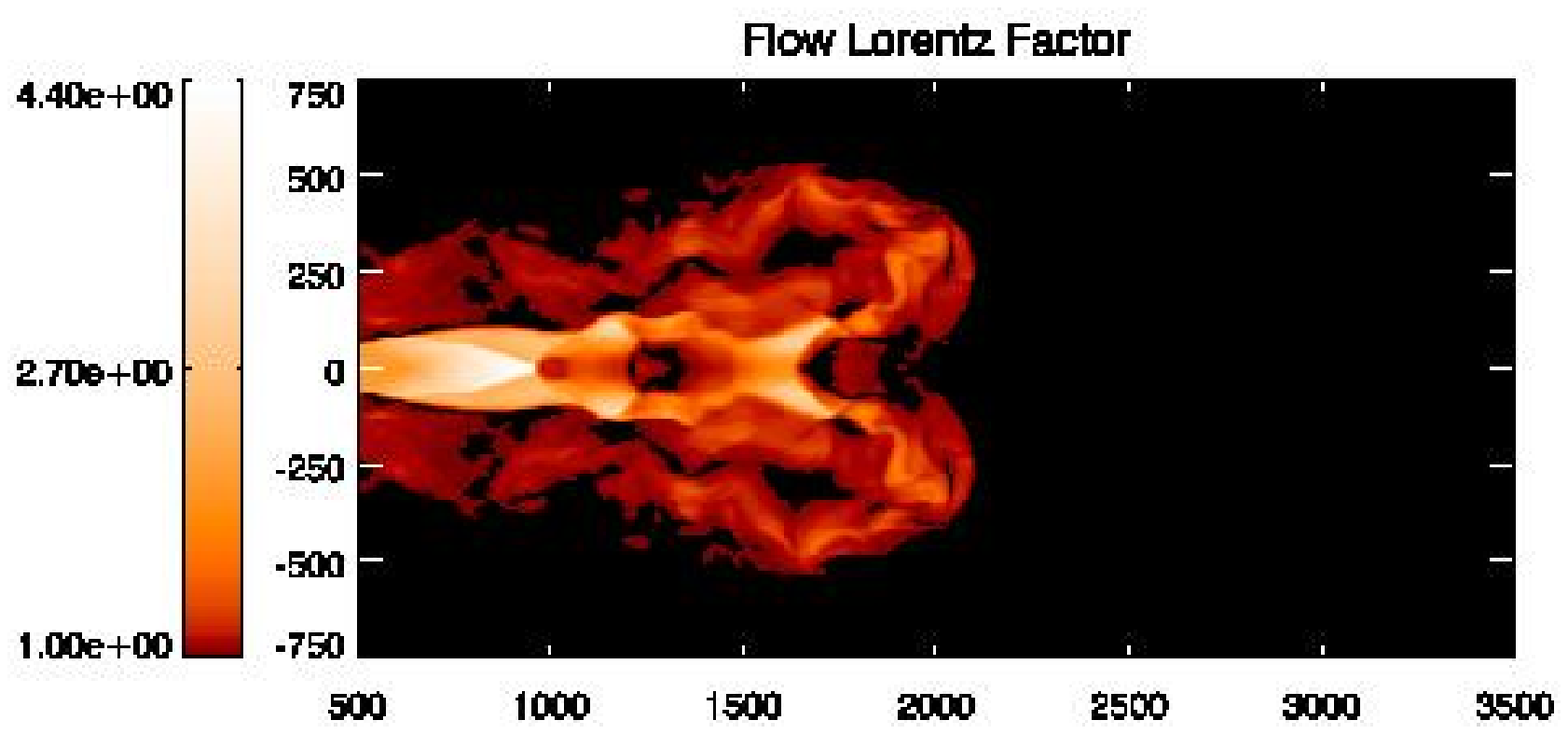}}
\centerline{
\includegraphics[width=0.5\textwidth]{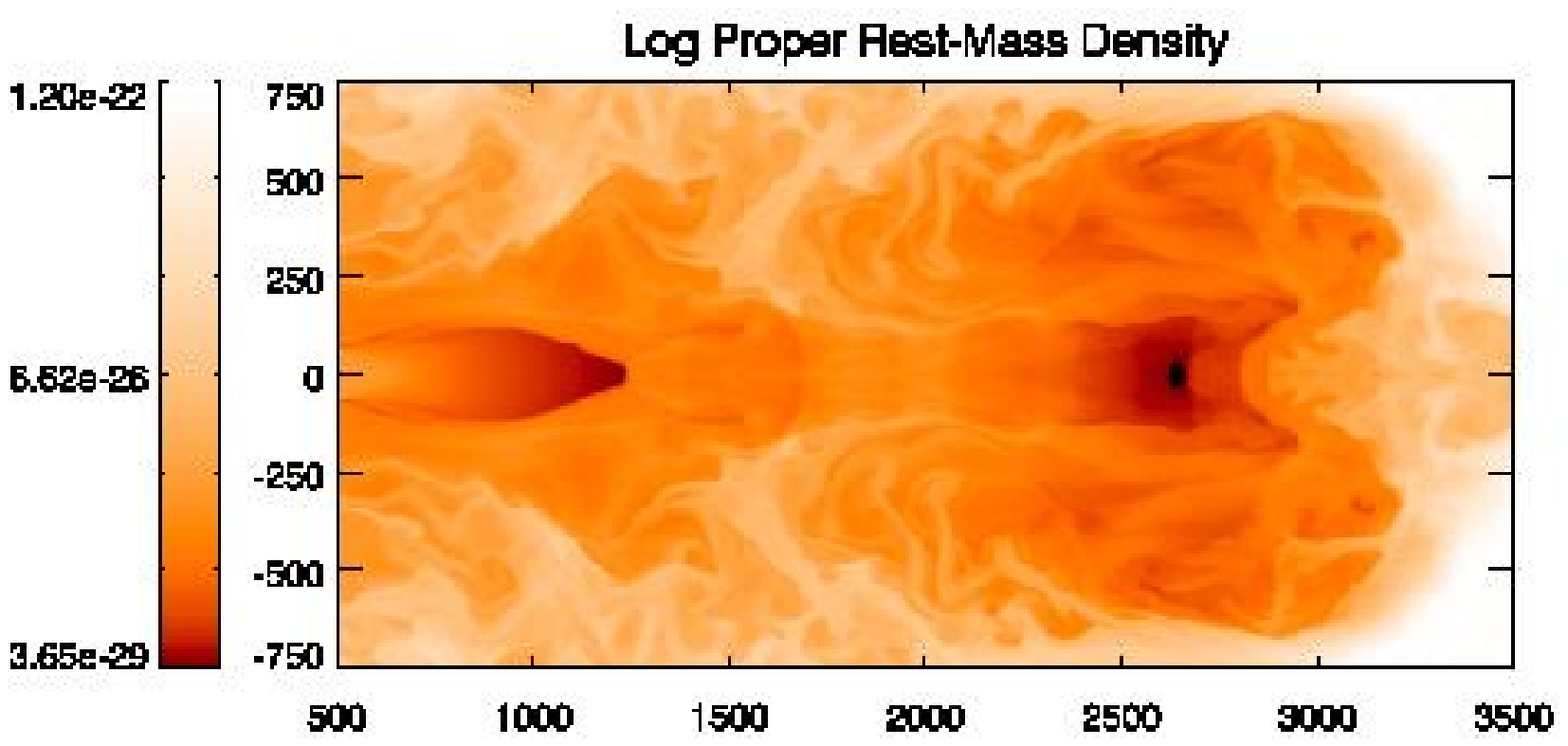}
\includegraphics[width=0.5\textwidth]{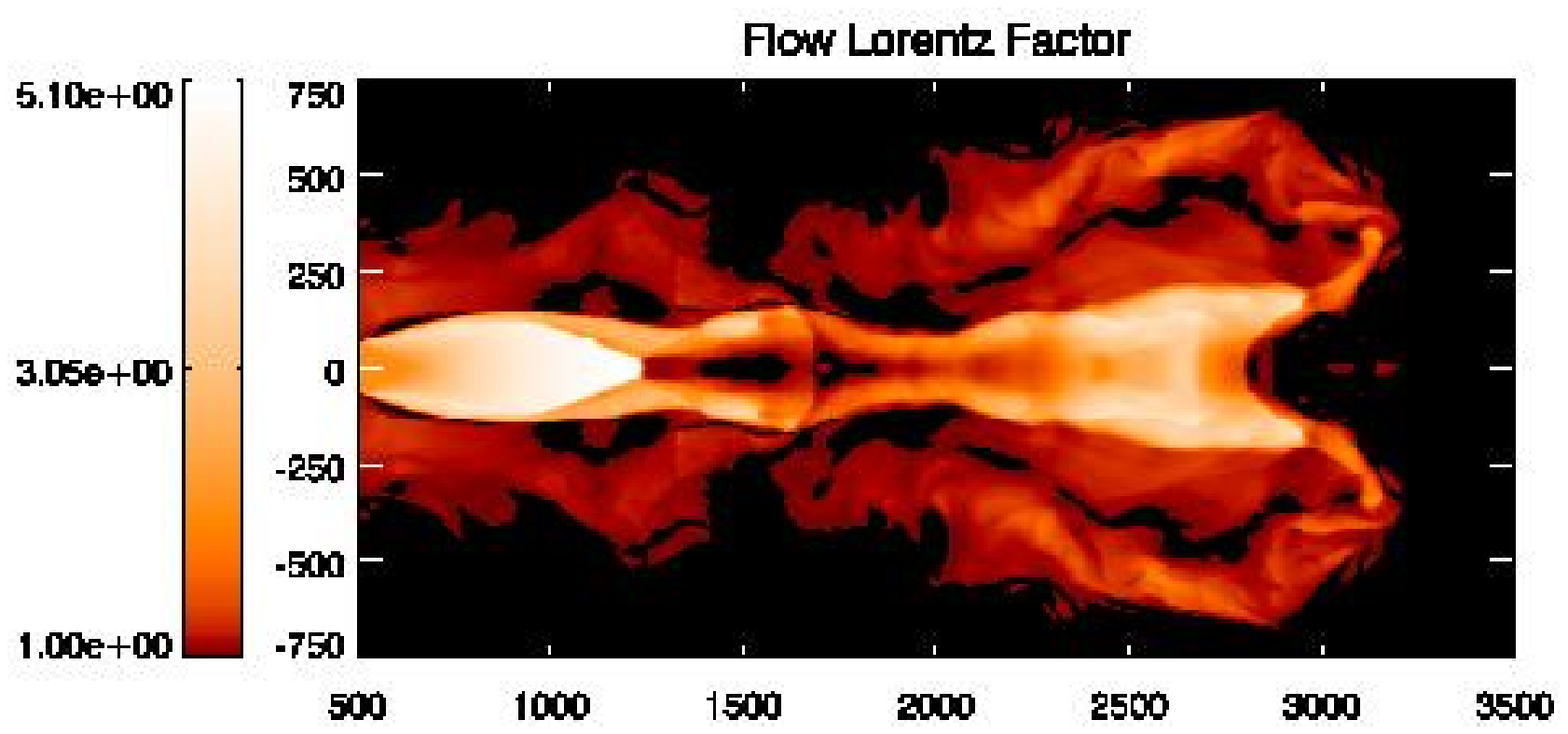}}
 \caption{Logarithm of rest mass density (left panels) in kg/m$^3$
and Lorentz factor (right panels) at three different times
($2.7\,10^5$, $8.2\,10^5$ and $1.4\,10^6\,\rm{yrs}$) during the
compact phase of the evolution. Coordinates are in parsecs.}
 \label{fig:compact}
\end{figure*}
%
%%%%%%%%%%%%%%%%%%%%%%%%%%%%%%%%%%%%%%%%%%%%%%%%%%%%%%%%%%%%%%%%%%%%%%%%

  Fig.~\ref{fig:compact} shows panels of the logarithm of rest mass
density and Lorentz factor at different times of the first stages
of evolution. The compact phase, defined as that in which the
source has a linear size smaller than $5\,\rm{kpc}$ can be divided
into two main epochs, based on these maps: 1) the \emph{CSO-like}
phase, when the source is smaller than $1\,\rm{kpc}$, and 2) the
\emph{weak CSS-like} phase, when the source is between $1$ and
$5\,\rm{kpc}$ long. During the \emph{CSO-like} phase (top panels
in Fig.~\ref{fig:compact}), the jet shows a large opening angle,
due to overpressure with respect to the ambient medium and a
strong Mach disk at its head. Hence the morphology of the source
is dominated by a short and featureless beam and a strong hot-spot
downstream of the terminal Mach shock. During the \emph{weak
CSS-like} phase (mid and bottom panels in Fig.~\ref{fig:compact}),
the Mach disk is weaker, as the jet is slowed down in
recollimation shocks. The whole structure of the jet appears to be
much more irregular during the latter phase. The transition
between the phases occurs in the course of the evolution, when the
terminal Mach shock disappears for the first time leading to a
conical shock, at $t = 4.3\,10^5\,\rm{yrs}$, when the linear size
of the jet is in the range $1-1.5\,\rm{kpc}$, i.e. in the
transition between CSO and CSS phases of young radio sources. The
flow dynamics behind the shock then changes affecting the global
evolution of the jet. This behaviour was already noted by
\cite{sch02} in the context of the early evolution of FRII jets.
It is important to remark that the axial symmetry imposed on the
simulation affects the internal structure of the jet (internal and
terminal shocks) and hence the jet dynamics and advance speed as
discussed, e.g., by \cite{alo99}. However, for the short
times/sizes involved in the previous discussion, the hypothetical
three dimensional effects are expected to be negligible.

\begin{figure*}
\centerline{
\includegraphics[width=0.5\textwidth]{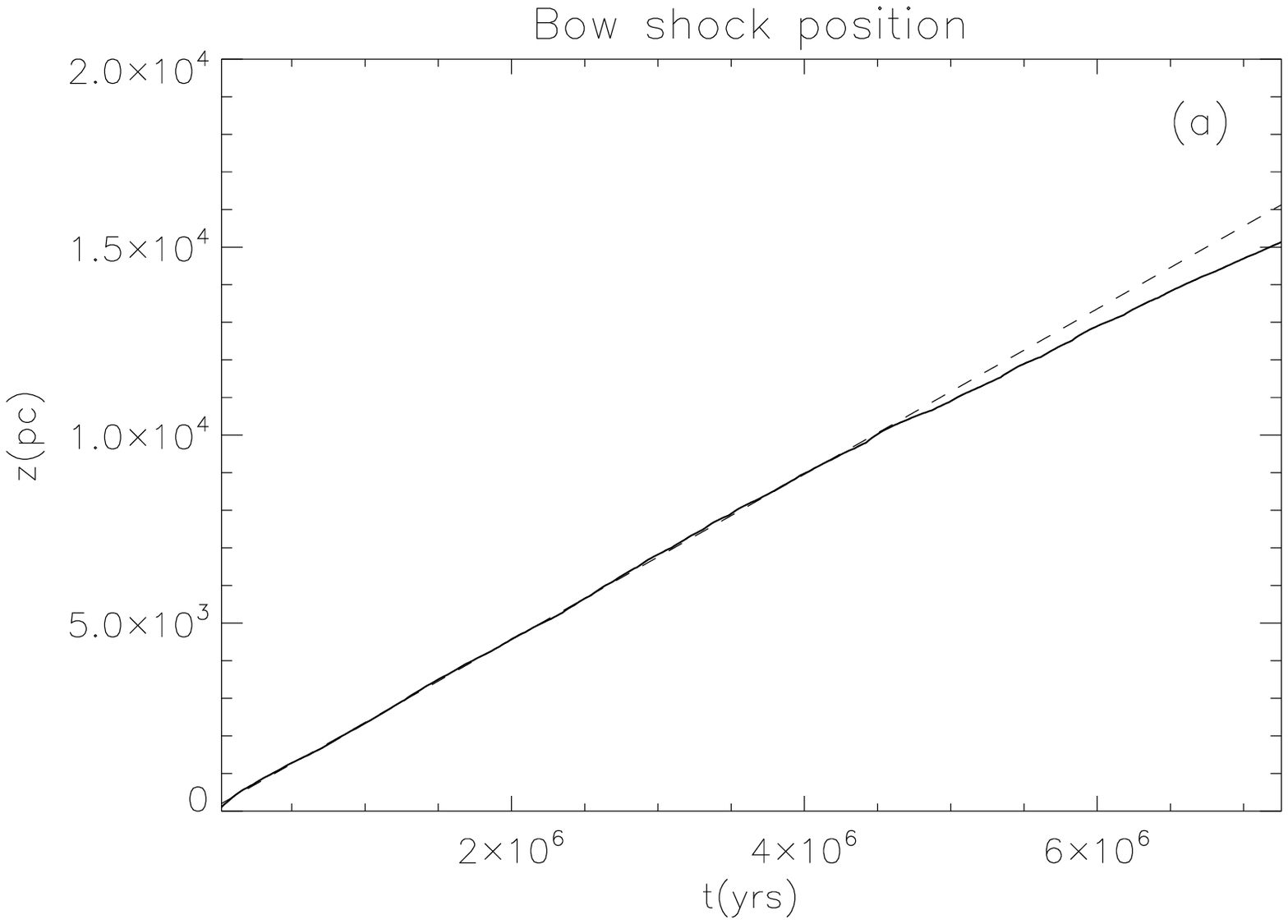}
\includegraphics[width=0.5\textwidth]{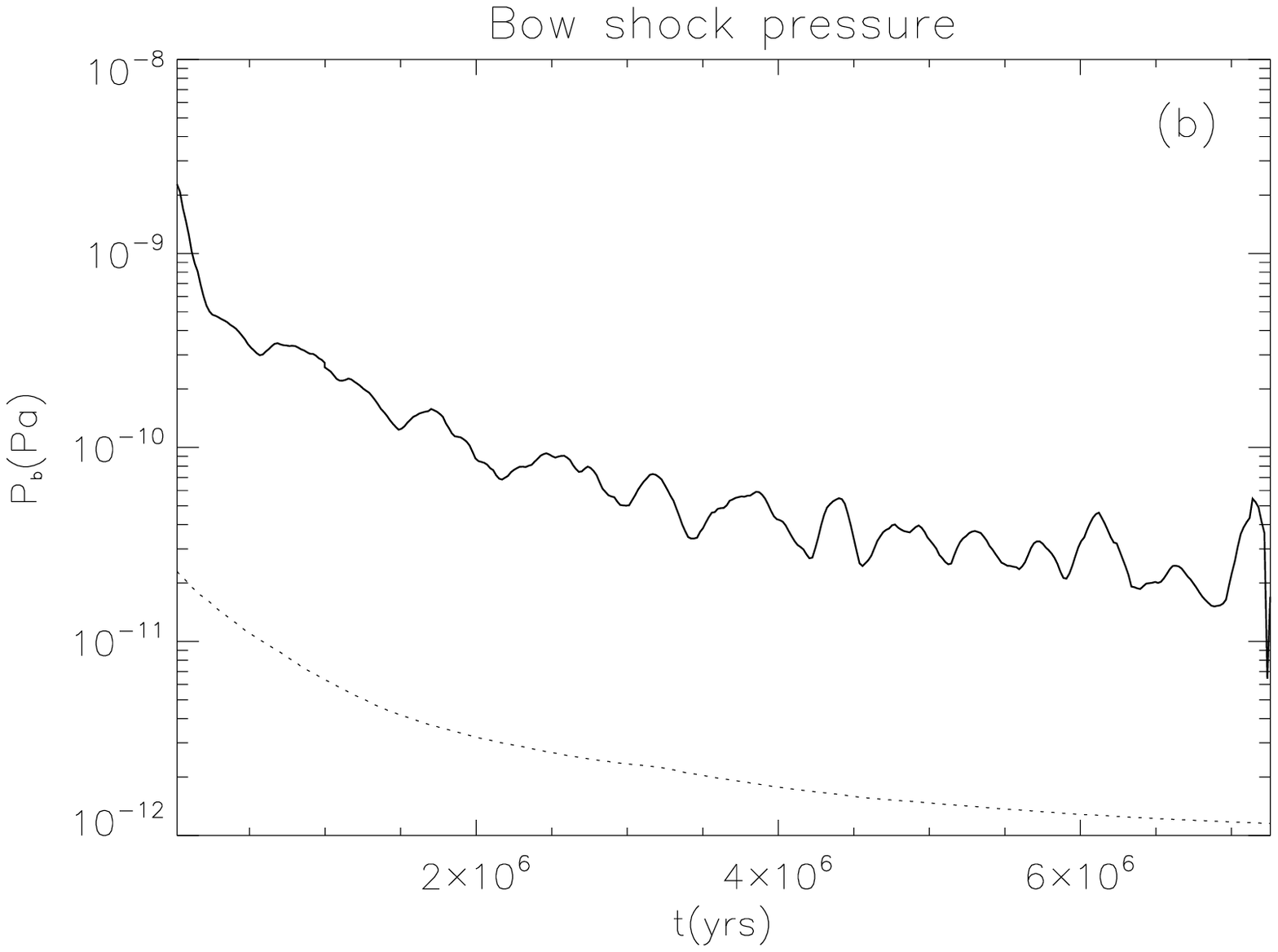}}
\centerline{
\includegraphics[width=0.5\textwidth]{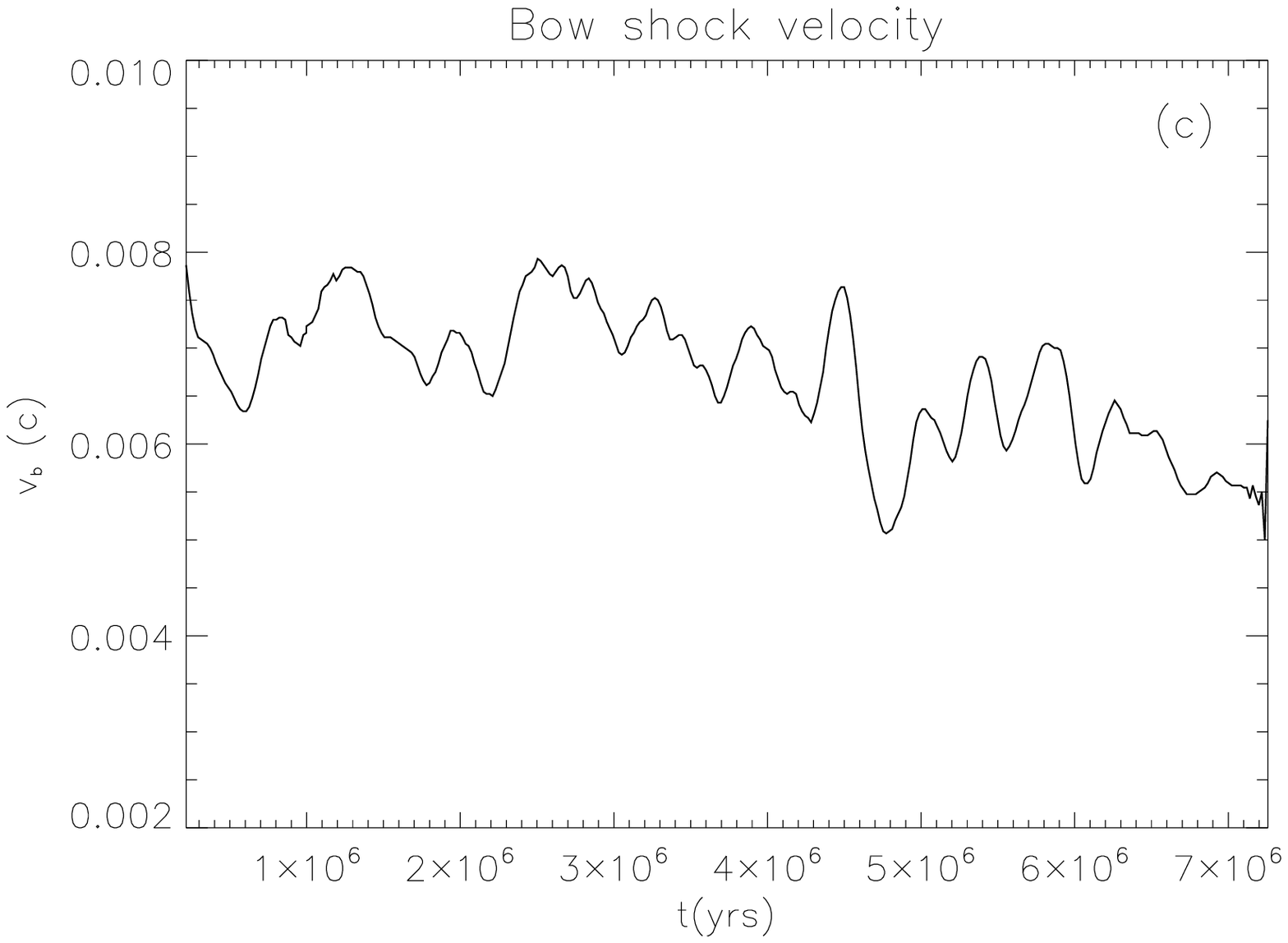}
\includegraphics[width=0.5\textwidth]{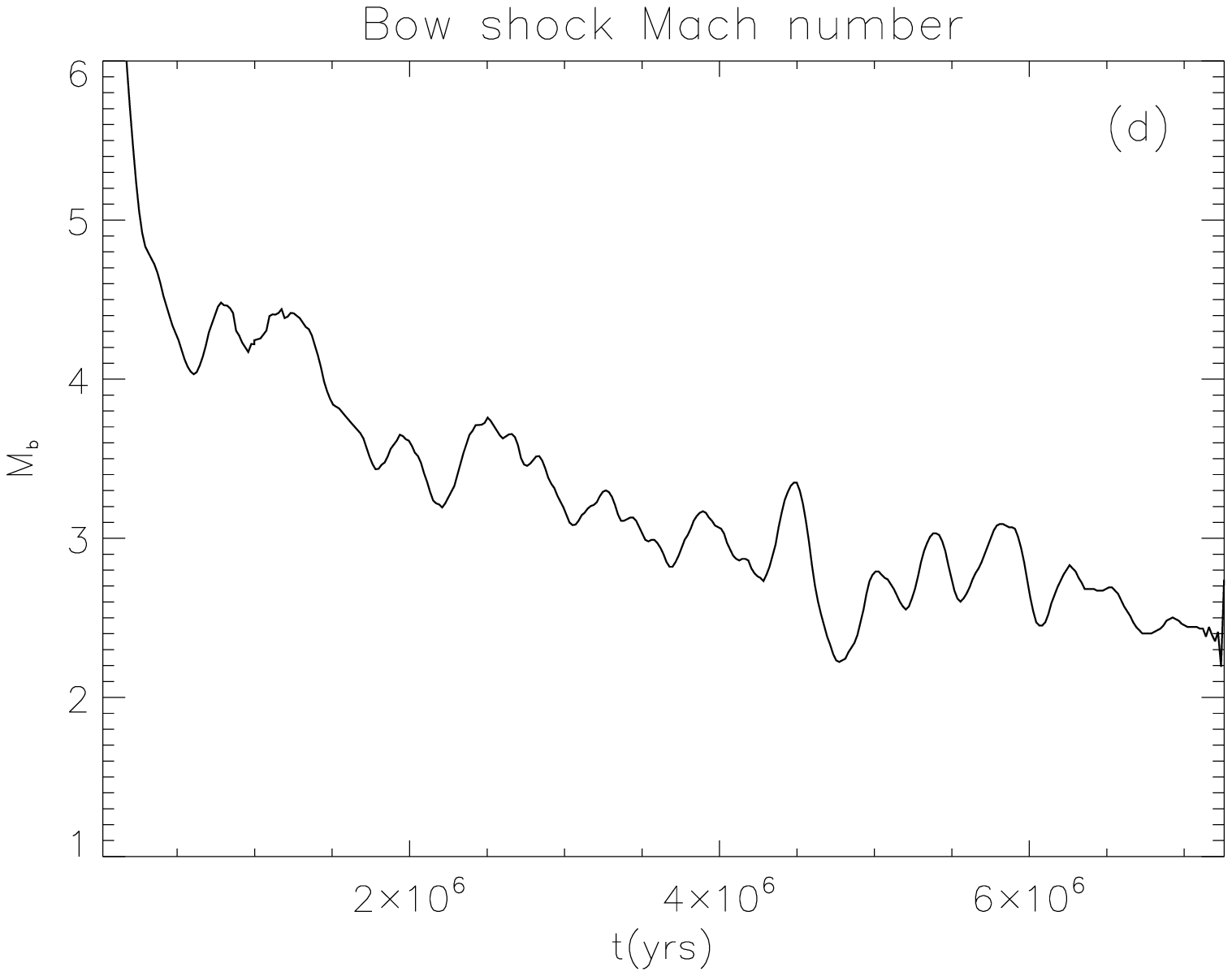}}

\caption{Evolution with time of the position (a), pressure (b),
instantaneous velocity (c) and Mach number (d) of the bow shock in
the axial direction. The bow shock position is selected as the
first point of the grid (starting from its end) where the speed is
larger than $10^{-4}\,c$. The dashed line in panel a indicates the
best fit to the first half of the evolution
($z\propto0.0022\,t[\rm{yrs}]\,\rm{pc}$); deceleration at the last
stages is clear. The bow shock pressure is taken as the value in
the first maximum, also searched from the end of the grid. The
dotted line in panel b shows the pressure of the ambient medium in
the region where the bow shock is located at the given time. The
instantaneous velocity of the bow shock is computed with the
discretized derivative of position with respect to time at each
instant, and Mach number of the bow shock propagating in the
ambient medium is computed with that advance speed and the mean
value of the sound speed in 16 cells ($2\,R_j\equiv120\,\rm{pc}$)
ahead of the bow shock position. Bow shock pressure, velocity and
Mach number curves have been smoothed using an IDL routine (with a
10 point smoothing) in order to avoid the noise coming from the
numerical derivative of the bow shock position.} \label{fig:evol0}
\end{figure*}

  Fig.~\ref{fig:evol0} shows the evolution of several
quantities versus time during the simulation.
Fig.~\ref{fig:evol0}a shows the position of the bow shock as a
function of time. The curve is consistent with a constant-velocity
expansion up to $t = 4.5\,10^6$ yrs (corresponding source size in
the axial direction, 9.5 kpc), and with a decelerating expansion
afterwards. Despite the variations, the advance velocity of the
bow shock, in Fig.~\ref{fig:evol0}c, shows this trend. Initially,
the advance speed is $v_{bs} \sim 7\,10^{-3}c$, close to the
one-dimensional estimate of the head of the jet for homogeneous
ambient medium, $v_h^{1d} = 9.9\,10^{-3}\,\rm{c}$
(Table~\ref{tab:sim3c31}), but by the end of the simulation, it
has decreased to $v_{bs} < 6\,10^{-3}c$. The duration of the phase
with constant advance speed is longer than in the case of
simulations with a uniform ambient medium \citep[e.g.,][]{sch02}
because the deceleration caused by the widening of the jet along
the evolution is now approximately balanced by the decrease of the
inertia of the ambient medium with distance. As stated in the
previous paragraph, three dimensional effects can modify the
change of jet advance speed with time. In particular, the {\it one
dimensional phase} can be shortened and phases of acceleration can
appear occasionally when the terminal shock changes from planar to
oblique. Hence the results discussed in this paragraph must be
treated with caution. The pressure behind the bow shock, in
Fig.~\ref{fig:evol0}b, drops and oscillates until the end of the
simulation following the behaviour of the ambient pressure
although with a larger instantaneous slope. The effect of
deceleration of the bow shock and the increase of temperature in
the ambient medium at larger distances to the source combine to
give a mild reduction of the Mach number (Fig.~\ref{fig:evol0}d)
along the evolution, although the shock is still supersonic
($M_{bs} \sim 2-3$) by the end of the simulation. Recent works by
\cite{kra03} and \cite{cro07} have detected the existence of shock
waves surrounding the radio lobes of the relatively young
($t\sim2\,10^6\,\rm{yrs}$) FRI jets of Centaurus A and NGC~3801,
respectively. The shock waves are revealed by shells of hot
interstellar gas emitting in the X-rays. The authors derive Mach
numbers between 3 and 8 for these shocks. The ages and Mach
numbers obtained in our simulation are in agreement with the
results reported in those papers. We will further discuss this
issue in the next Section.

%%%%%%%%%%%%%%%%%%%%%%%%%%%%%%%%%%%%%%%%%%%%%%%%%%%%%%%%%%%%%%%%%%%%%%
%
%
%%%%%%%%%%%%%%%%%%%%%%%%%%%%%%%%%%%%%%%%%%%%%%%%%%%%%%%%%%%%%%%%%%%%%%

  The propagation of the shock leaves behind a region of shocked
ambient material through which the jet propagates. In the case of
powerful jets, the shock is so strong and its expansion so fast
that the region encompassed by the bow shock is almost evacuated
with most of the matter concentrated in a thin shell of shocked
ambient material behind the shock. The evacuated region is
continuously fed by the jet to form an extended cocoon. In the
case of weak jets like the one considered here, the bow shocks are
correspondingly weaker and the dense region behind the shock,
correspondingly wider. Hence the cavity/shell division suitable
for the shocked regions surrounding powerful jets transforms into
a cocoon/shocked-ambient-medium division. This structure is easily
seen in the color maps of rest-mass density and tracer at the end
of the simulation, to be discussed in the next section.

%%%%%%%%%%%%%%%%%%%%%%%%%%%%%%%%%%%%%%%%%%%%%%%%%%%%%%%%%%%%%%%%%%
%
\begin{figure*}
\centerline{
\includegraphics[width=0.5\textwidth]{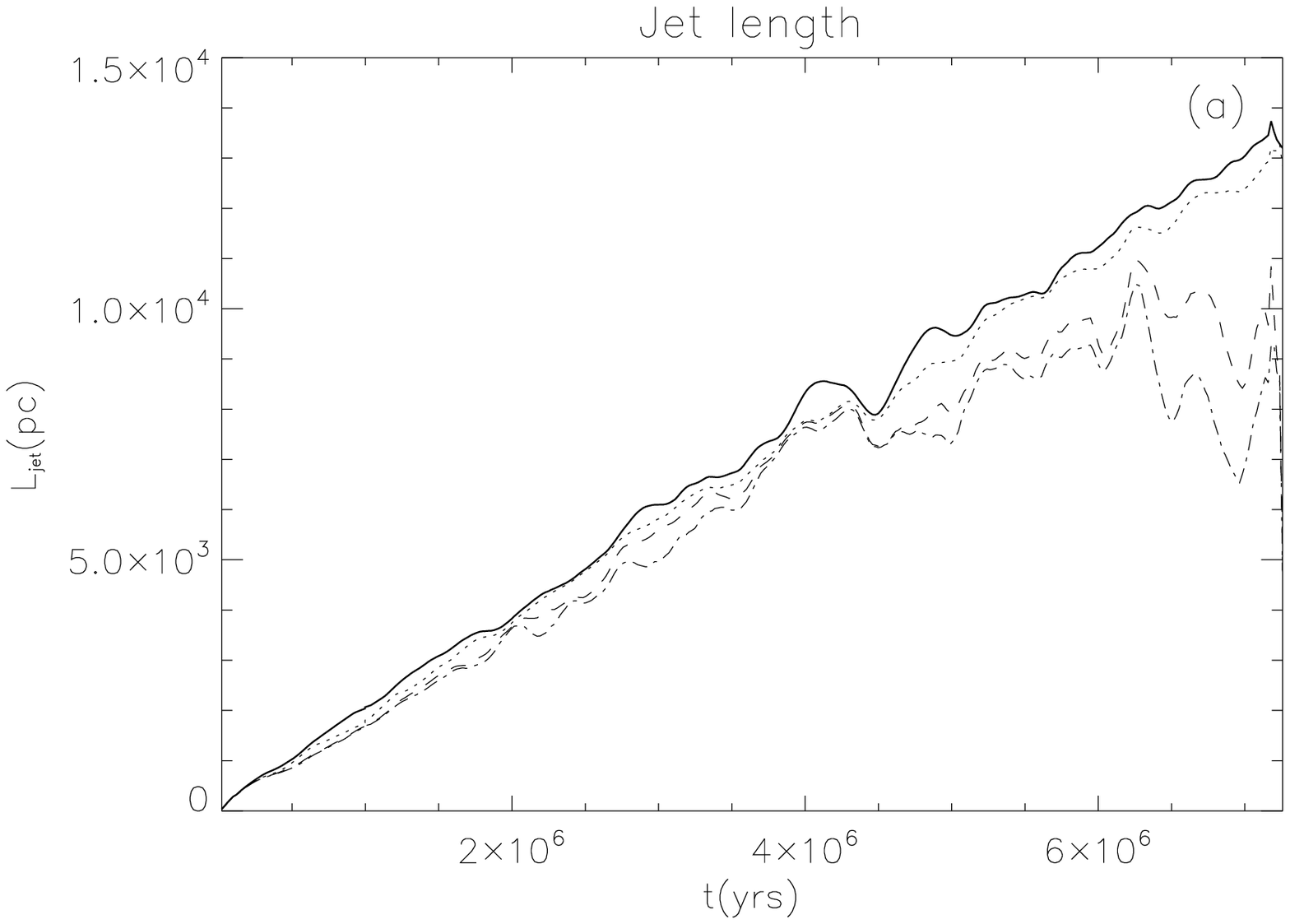}
\includegraphics[width=0.5\textwidth]{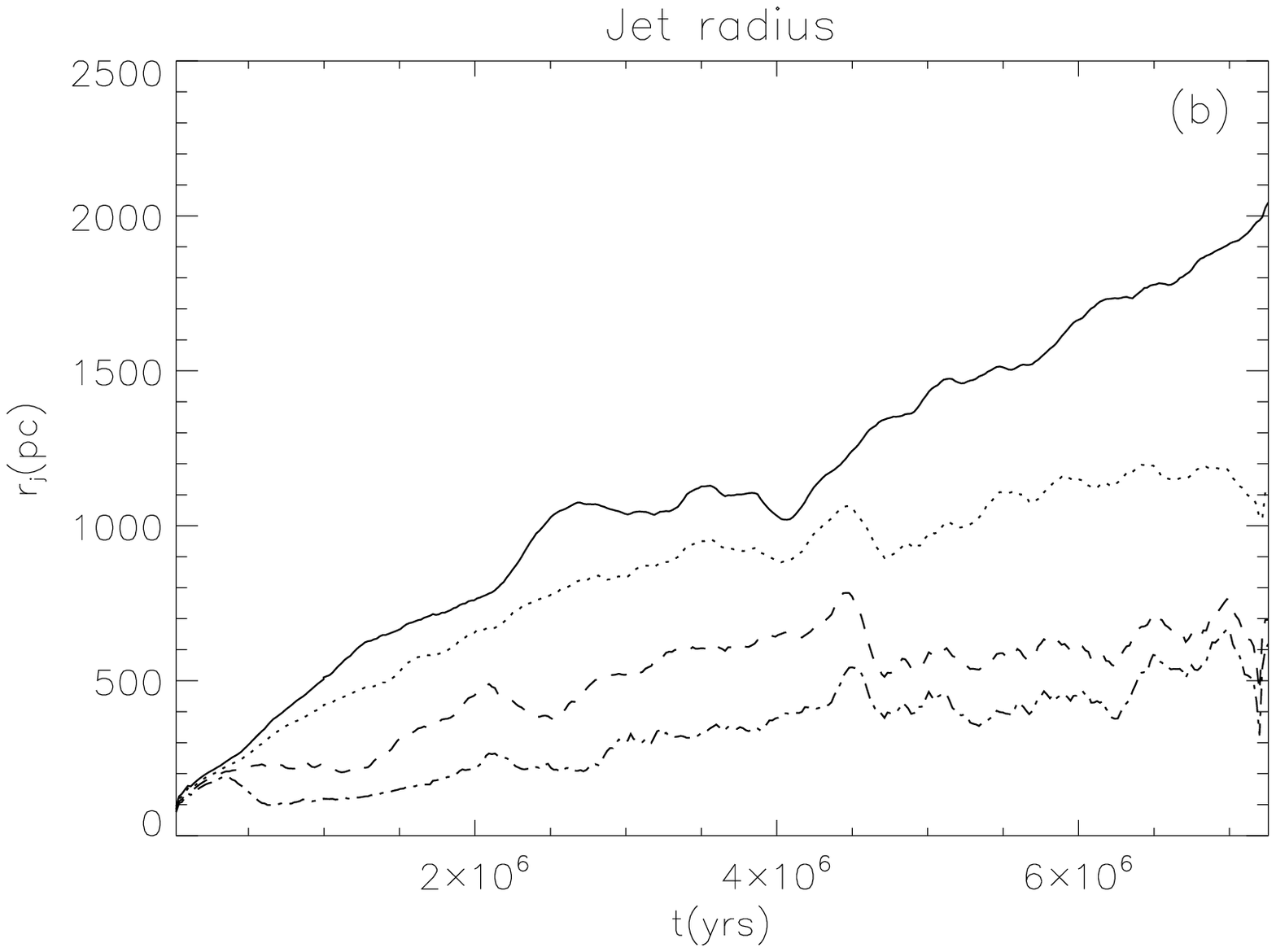}}
\caption{Jet length (a) and jet radius (b) versus time. The
different lines stand for the different criteria used to define
the jet/ambient transition: solid line is for jet mass fraction $f
= 0.01$, dotted line for $f = 0.1$, dashed line for $f = 0.5$ and
dash-dotted line for $f = 0.9$.} \label{fig:evol2}
\end{figure*}
%
%%%%%%%%%%%%%%%%%%%%%%%%%%%%%%%%%%%%%%%%%%%%%%%%%%%%%%%%%%%%%%%%%%

  Fig.~\ref{fig:evol2} shows the evolution of the jet length
(Fig.~\ref{fig:evol2}a) and the jet radius (Fig.~\ref{fig:evol2}b)
with time. Both quantities are calculated for different values of
the jet mass fraction. The spreading of the lines in the jet
length plot for $t > 4.5 \, 10^6$ yrs results from the entrainment
of ambient material in the head of the jet which causes the
deceleration in the jet advance mentioned in the preceding
paragraphs. The values of the jet radius for the different mass
fractions diverge right from the start of the simulation, implying
progressive jet stratification, shearing and mixing, as the jet
propagates outwards. The aspect ratio of the jet increases for
larger values of the jet mass fraction and is always larger than
the aspect ratio of the shocked region.

\subsubsection{Comparison with Analytical Models}
\label{ss:analmod}

\cite{bc89} developed a simple model for the evolution of the
cavities/cocoons surrounding powerful jets in homogeneous ambient
media under the assumptions that the jet propagation speed and the
power injected into the cavity are independent of time, and that
the pressure of the external medium is negligible. The model can
still be applied to describe the evolution of the shocked regions
surrounding weak jets. Then, the mean pressure in the shocked
region is given by

\begin{equation}
  P_s \propto \frac{L_s}{v_{bs} A_s},
\end{equation}

\noindent where $L_s$, $v_{bs}$ and $A_s$ are the power injected
into the shocked region from the jet through the terminal shock,
the bow shock speed in the direction of propagation of the jet,
and the transversal cross-section of the shocked region (i.e.,
$\pi R_s^2$, $R_s$ being the radius), respectively. The pressure
in the shocked region causes it to expand with velocity
$\dot{R}_s$ according to

\begin{equation} \label{eq:shock}
  P_s = \rho_a \dot{R}_s^2,
\end{equation}

\noindent
where $\rho_a$ is the ambient density. This implies $1/R_s \propto \dot{R}_s$, and hence

\begin{equation}
  R_s \propto t^{1/2}, \quad P_s \propto t^{-1}, \quad
  l_s/R_s \propto t^{1/2}
\end{equation}

\noindent
(where in this last expression, $l_s$ stands for the longitudinal size of the shocked region).

 In \cite{sch02}, and also in the context of powerful jets,
the authors developed a simple extension of Begelman \& Cioffis's
model to account for the secular deceleration of the jet advance
speed due to the expansion of the jet cross-section. According to
this model, $v_{bs} \propto t^\alpha$, and

\begin{equation}
  R_s \propto t^{1/2 - \alpha/4}, \quad
  P_s \propto t^{-1 - \alpha/2}, \quad l_s/R_s \propto t^{1/2 + 5\alpha/4}.
\end{equation}

\noindent
In the simulations presented in that paper, a value of $\alpha \sim -1/3$ was found and, accordingly,

\begin{equation}
  R_s \propto t^{7/12}, \quad P_s \propto t^{-5/6},
  \quad l_s/R_s \propto t^{1/12}.
\end{equation}

 In the present work, the model is generalized to consider the
expansion of the shocked region through an ambient medium with
decreasing density. A power law fit of the ambient density (see
Fig.~\ref{fig:atm}) with respect to the distance to the source
gives $\rho_a \propto R_s^{-1}$. With this and taking $\alpha \sim
-0.1$ (from the simulation), we finally have

\begin{equation}
  R_s \propto t^{0.7}, \quad P_s \propto t^{-1.3}, \quad l_s/R_s \propto t^{0.2}.
\end{equation}

\begin{figure*}
\centerline{
\includegraphics[width=0.5\textwidth]{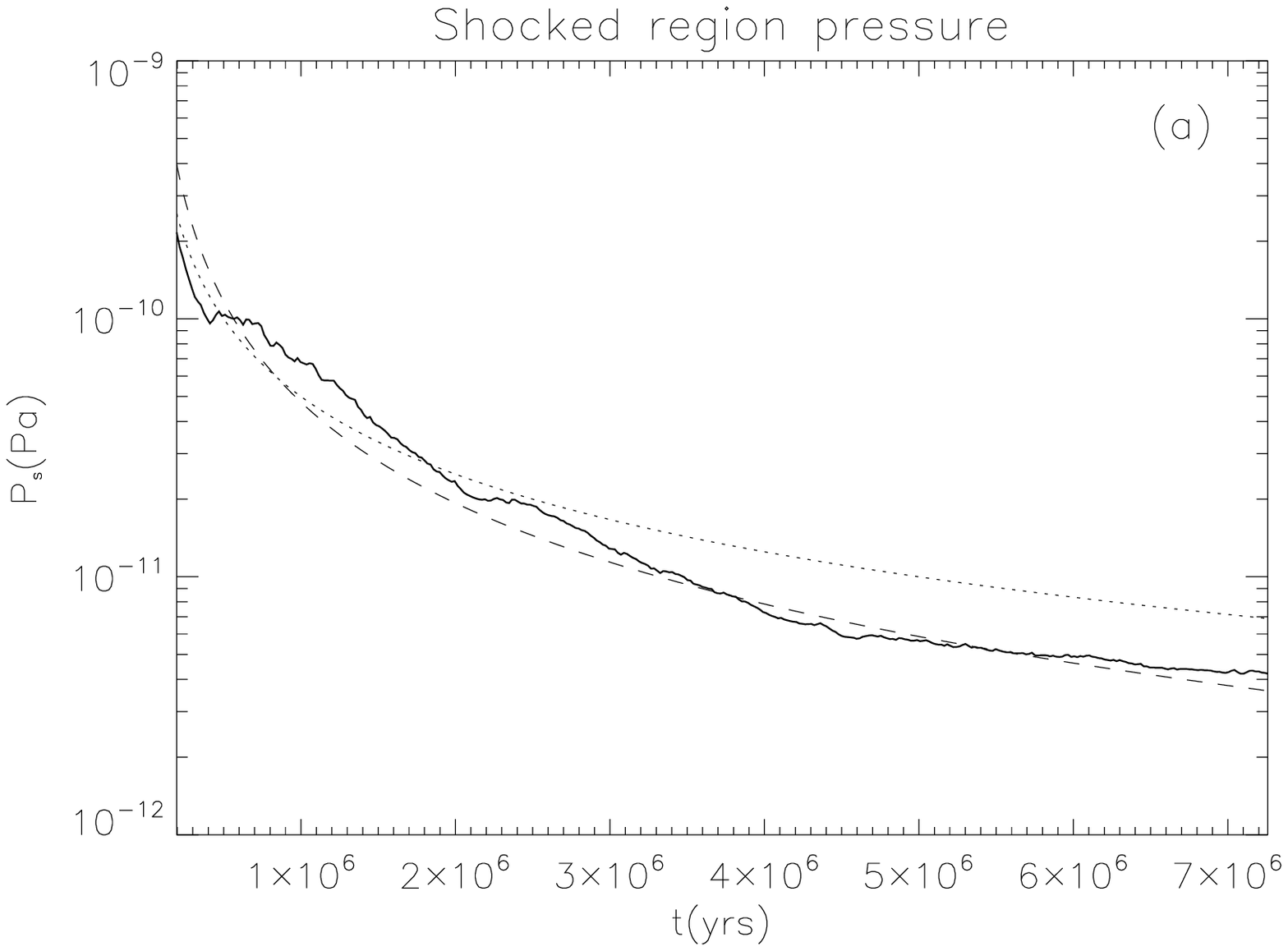}
\includegraphics[width=0.5\textwidth]{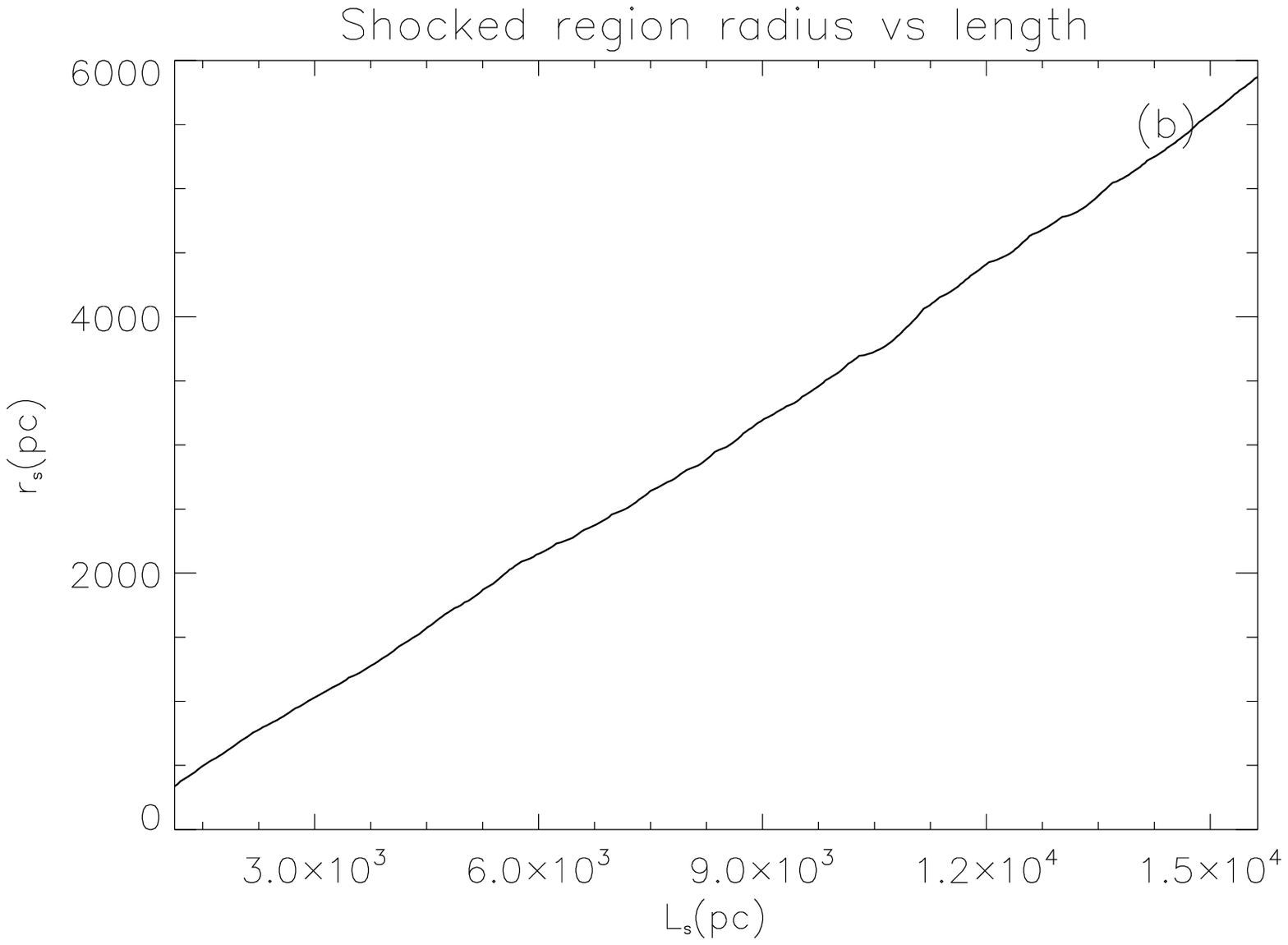}}
\caption{The panels show the pressure in the shocked region versus
time (a) and the radius of the shocked region versus its length
(b). The pressure is computed as the mean pressure in all the
cells with shocked ambient medium. The dotted line shows the
evolution as $\propto t^{-1}$ and the dashed line gives the best
fit of the curve $t^{-1.3}$. The radius is calculated as the mean
radius of the bow shock, with this position determined by a
lateral motion larger than $10^{-4}\,c$, see caption of
Fig.~\ref{fig:evol0}. The length of the shocked region used in
this plot as the abscissa coincides with the bow shock position
shown in the panel a of Fig.~\ref{fig:evol0}.} \label{fig:evol1}
\end{figure*}

\noindent These results are remarkably consistent with those
derived from the simulation ($R_s \propto t^{0.8}$, $P_s \propto
t^{-1.3}$, $l_s/R_s \propto t^{0.2}$). Fig.~\ref{fig:evol1}a
displays the evolution of the pressure in the shocked region
versus time, together with the $t^{-1}$ (as in Begelman \& Cioffi
and Scheck et al. models) and $t^{-1.3}$ fits. At this point, we
must remark that in the simulation we used an open boundary
condition in the symmetry plane at the jet basis that allows the
leakage of gas and, consequently, a faster pressure drop in the
shocked region. A crude {\it a posteriori} analysis does not allow
us to rule out the possibility that the gas internal energy that
left the shocked region through the open boundary is of the same
order (or $\sim 1/10$) than that remaining in it. However, the
agreement between the results from the model discussed above and
those derived from the simulation suggests that the flow through
the open boundary is effectively negligible. Despite the
continuous decrease of the pressure, the shocked region is still
overpressured by a factor of 4 with respect to its environment at
the end of the simulation. This fact might be alleviated with the
introduction of cooling processes. However, \cite{kra03} and
\cite{cro07} also derived strong overpressure (as much as two
orders of magnitude) of the X-ray emitting shells in Cen~A and
NGC~3801 with respect to the surrounding interstellar media. This
is in agreement with our results that give a pressure in the
shocked region at $t \sim 2 \, 10^6 \, \rm{yrs}$ (the approximate
age of the jets in Cen~A and NGC~3801) more than one order of
magnitude larger than that of the ambient medium. Finally, the
plot of shocked region radius versus linear size, in
Fig.~\ref{fig:evol1}b, shows a self-similar growth for the late
stages of the simulation (with an aspect ratio for the shocked
region of $\approx 2.7$). The deceleration of the jet advance for
$t > 4.5 \, 10^6$ yrs, produces a change in the evolution of the
aspect ratio of the shocked region that by the end of the
simulation has decreased to a value of 2.6.

%%%%%%%%%%%%%%%%%%%%%%%%%%%%%%%%%%%%%%%%%%%%%%%%%%%%%%%%%%%%%%%%%
%
%
%%%%%%%%%%%%%%%%%%%%%%%%%%%%%%%%%%%%%%%%%%%%%%%%%%%%%%%%%%%%%%%%%%

%%%%%%%%%%%%%%%%%%%%%%%%%%%%%%%%%%%%%%%%%%%%%%%%%%%%%%%%%%%%%%%%%%
%
\begin{figure*}
\centerline{
\includegraphics[width=0.5\textwidth]{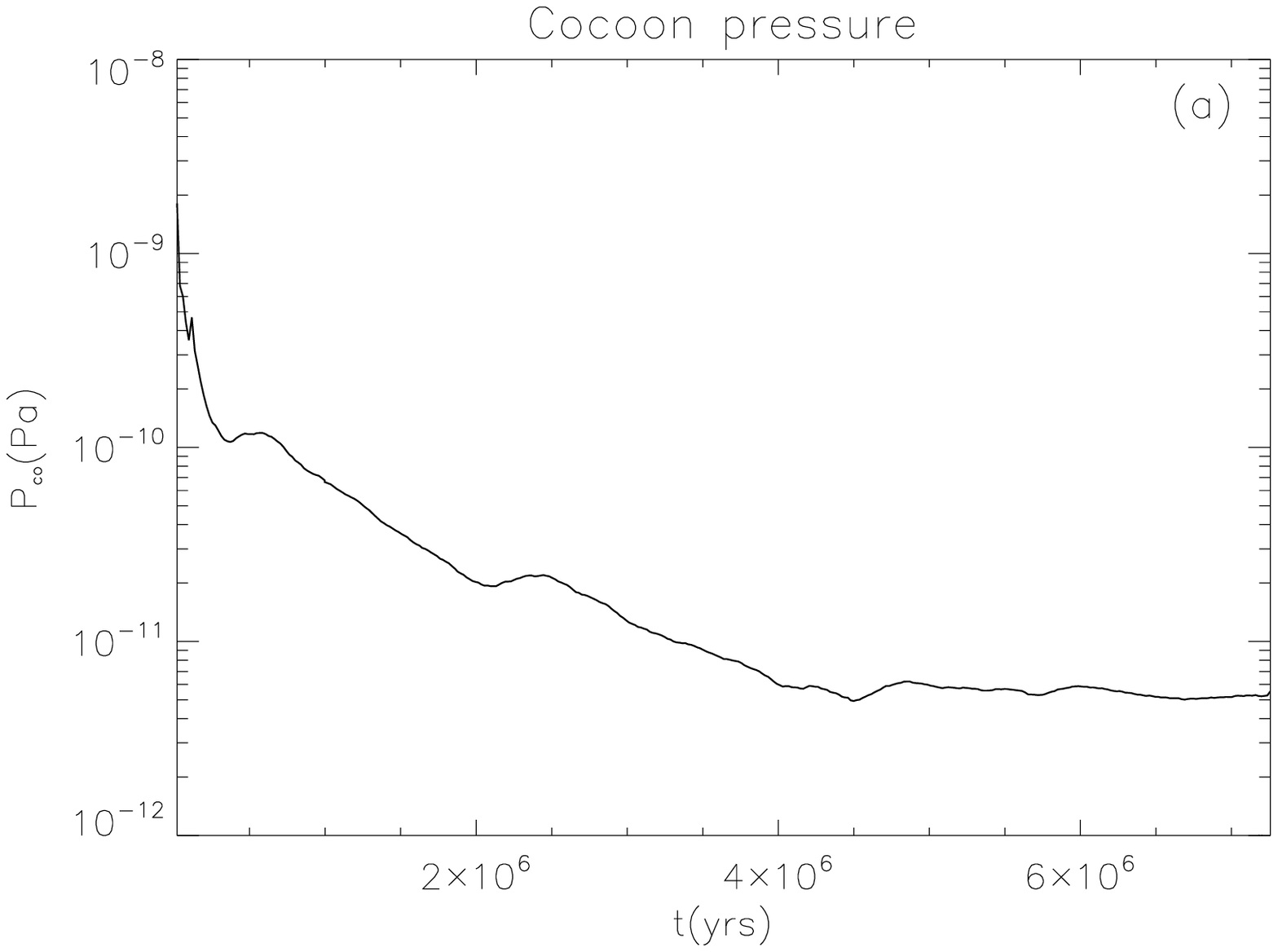}
\includegraphics[width=0.5\textwidth]{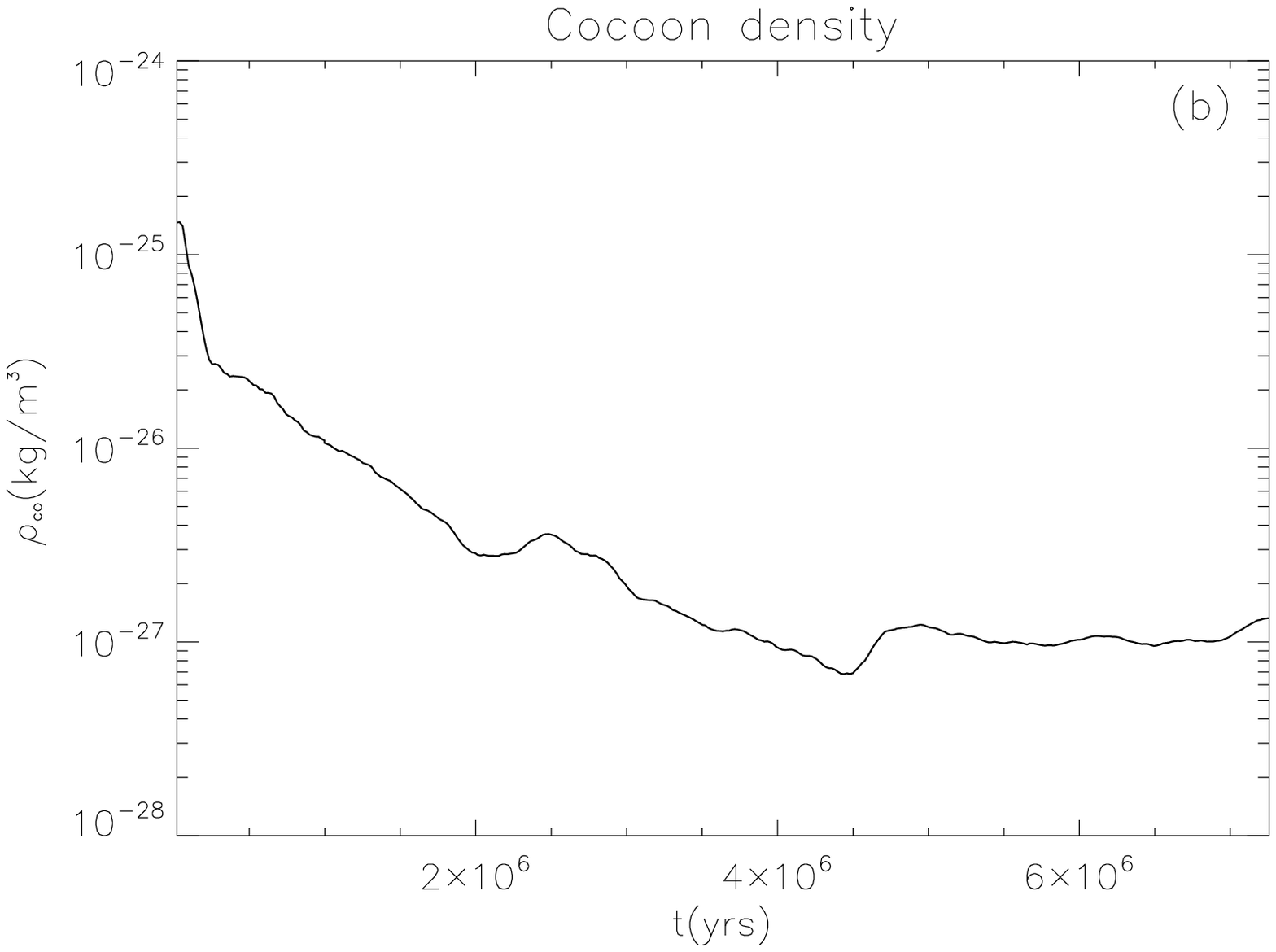}}
\includegraphics[width=0.5\textwidth]{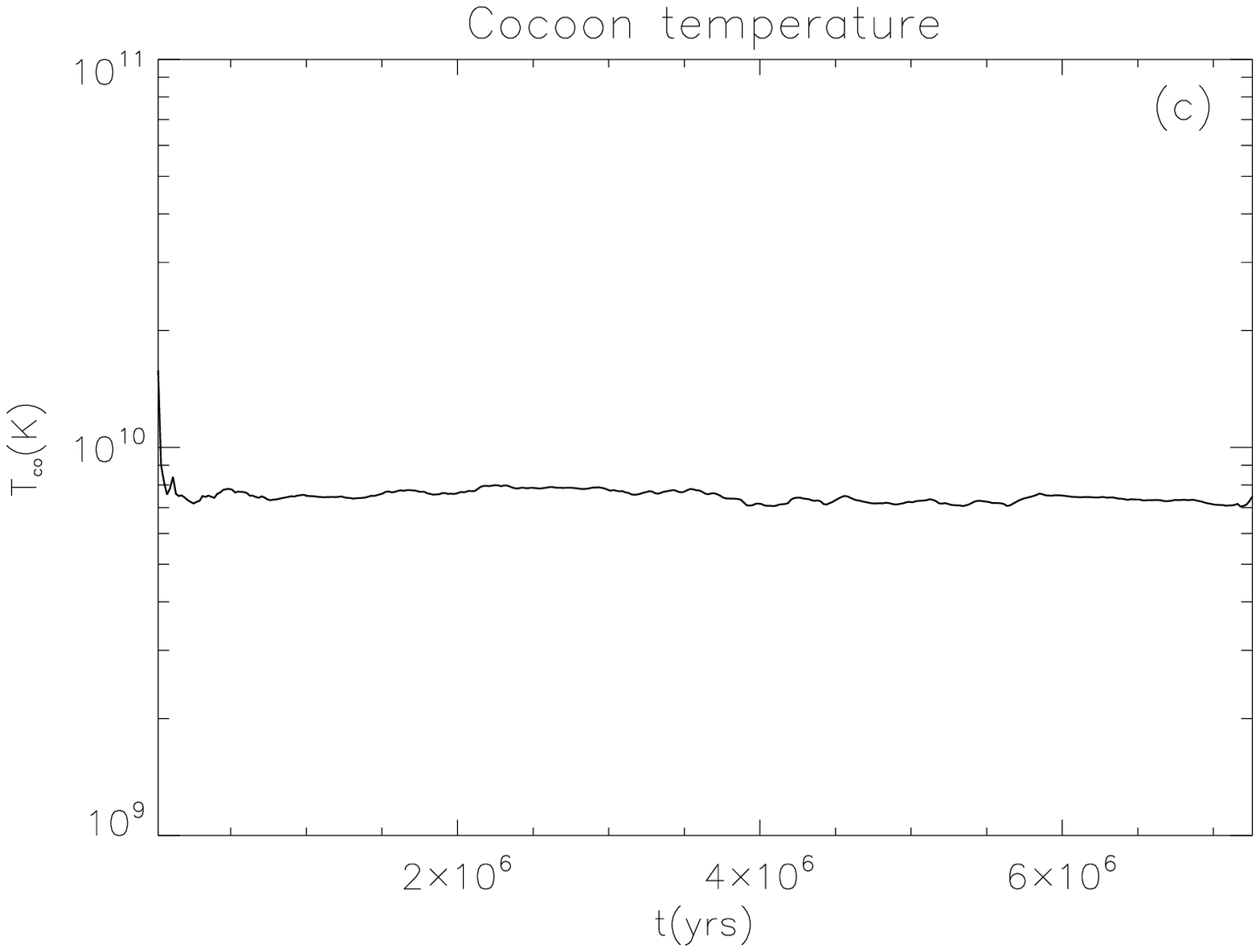}
\caption{Mean pressure (a), density (b) and temperature in the
cocoon as a function of time (c). The cocoon is defined as the
region with tracer values between 0.01 and 0.3.}
\label{fig:evol_cocoon}
\end{figure*}
%
%%%%%%%%%%%%%%%%%%%%%%%%%%%%%%%%%%%%%%%%%%%%%%%%%%%%%%%%%%%%%%%%%%

  Fig.~\ref{fig:evol_cocoon} show the evolution of the mean pressure
(Fig.~\ref{fig:evol_cocoon}a), density
(Fig.~\ref{fig:evol_cocoon}b) and temperature
(Fig.~\ref{fig:evol_cocoon}c) in the cocoon as a function of time.
Remarkably, the cocoon temperature is almost constant in the long
term evolution. Also interesting is the fact that the pressure in
the cocoon follows a similar evolution to that in the whole
shocked region (see Fig.~\ref{fig:evol1}a). This last fact can be
explained taking into account that the sound speed in the
cocoon/shocked-ambient-medium region, is about one or two orders
of magnitude larger than its expansion velocity, hence allowing
for an almost instantaneous (with respect to the dynamical time
scale) adjustment of the pressure. An example of this is seen in
the pressure map at the end of the simulation to be discussed in
the next section.

  The isothermal evolution of the cocoon can be explained by assuming that
the almost self-similar evolution of the shocked region extends also to
its internal structure. As discussed in the previous paragraph,
for the pressure in the cocoon, $P_c$, we have, at any time,

\begin{equation}
  P_c \sim P_s \propto \frac{L_s}{v_{bs} A_s}.
\end{equation}

\noindent The density in the cocoon, $\rho_c$, can be estimated,
in a similar way, as the quotient of total mass injected in the
cocoon, $J_c t$, and its volume. If $A_c$ is the transversal
cross-section of the cocoon, then

\begin{equation}
  \rho_c = \frac{J_c}{v_{bs} A_c}.
\end{equation}

\noindent On the other hand, for the temperature in the cocoon,
$T_c$, we have $T_c \propto P_c/\rho_c$, and, from the two
previous equations,

\begin{equation}
  T_c \propto \frac{L_s}{J_c} \frac{A_c}{A_s}.
\end{equation}

\noindent Now, the independency of $T_c$ follows if the fluxes of
energy and mass through the jet terminal shock are constant and
the evolution of $A_c$ and $A_s$ with time is the same.

\subsection{Fate} \label{sec:fate}

\begin{figure*}
\centerline{
\includegraphics[width=0.9\textwidth]{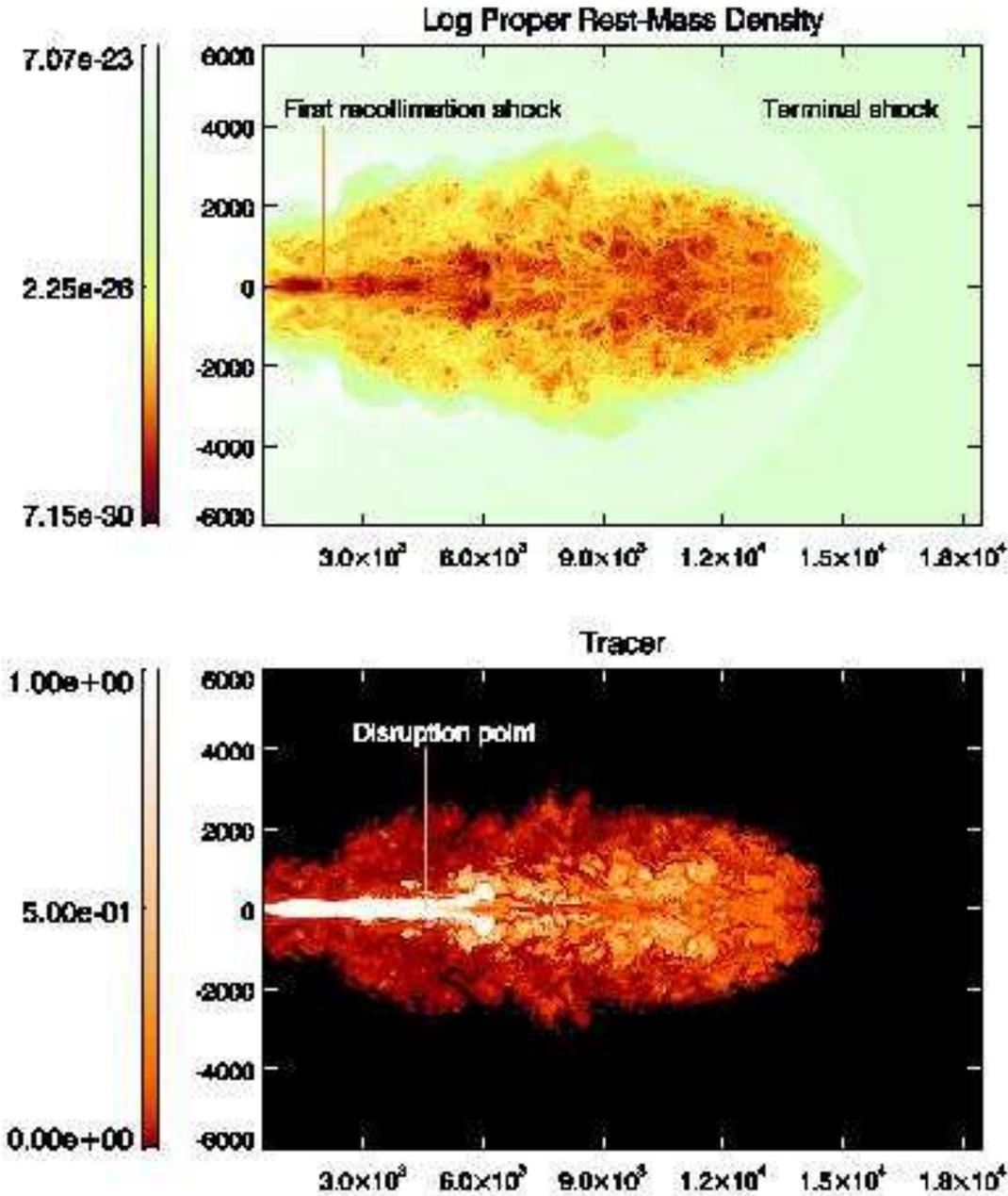}}
\caption{Logarithm of rest mass density (top panel) in kg/m$^3$
and jet mass fraction (used as a tracer for the jet plasma; bottom
panel) at the last frame of the simulation in this work. The
locations of the first recollimation shock and the terminal (or
bow) shock are indicated in the rest mass density map, and the
disruption point is indicated in the jet mass fraction map.
Coordinates are in parsecs.} \label{fig:3c31rhot}
\end{figure*}

\begin{figure*}
\centerline{
\includegraphics[width=0.9\textwidth]{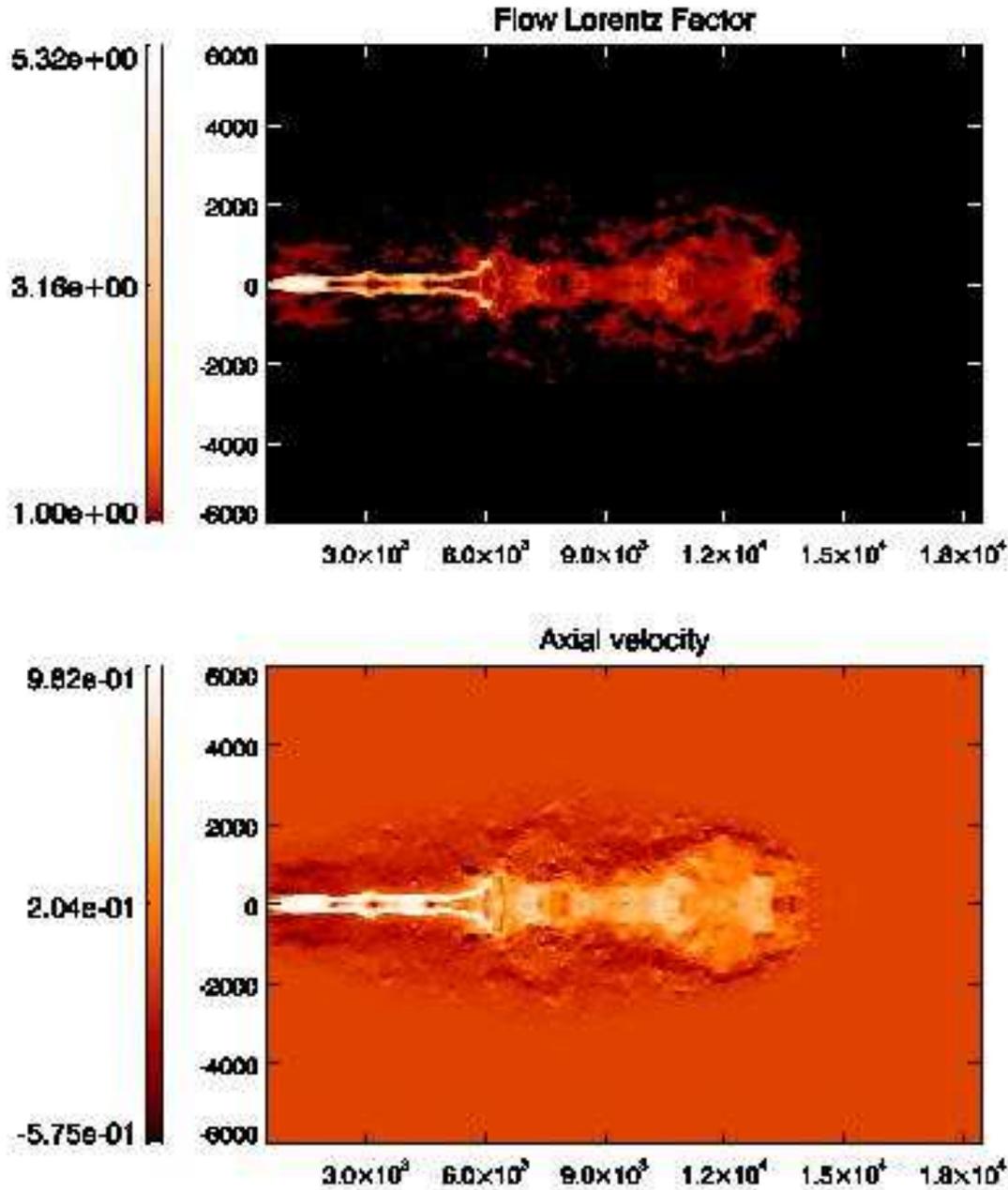}}
\caption{Lorentz factor (top panel) and axial velocity (bottom
panel) in units of $c$ at the last frame of the simulation in this
work. Coordinates are in parsecs.} \label{fig:3c31lofe}
\end{figure*}

\begin{figure*}
\centerline{
\includegraphics[width=0.9\textwidth]{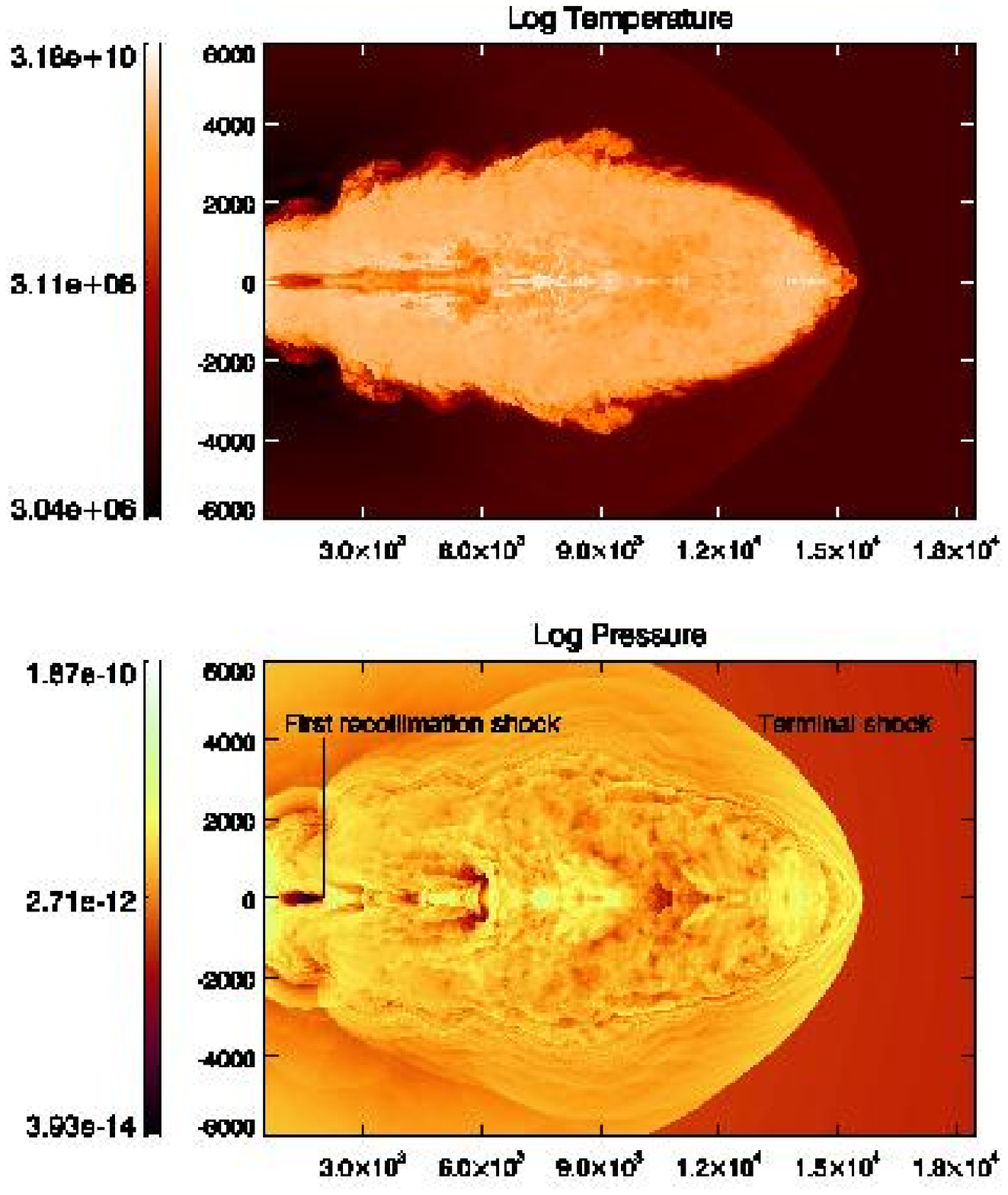}}
\caption{Logarithm of temperature (top panel) in K and logarithm
of pressure (bottom panel) in Pa at the last frame of the
simulation in this work. The position of the first recollimation
shock and the terminal (or bow) shock are indicated in the
pressure map. Coordinates are in parsecs.} \label{fig:3c31temp}
\end{figure*}

  Figs.~\ref{fig:3c31rhot}-\ref{fig:3c31temp} show several maps of
different quantities at the end of the simulation
($t=7.26\,10^6\,\rm{yrs}$). The morphological features observed in
the panels are those of a typical jet \citep[see,
e.g.,][]{mart97}: a bow-shock propagating through the unperturbed
ambient gas (clearly seen in the top panel of
Fig.~\ref{fig:3c31rhot} displaying the logarithm of the rest-mass
density), a cocoon composed of mixed jet and ambient matter, and
the jet itself propagating inside. The cocoon, formed by mixed jet
material, is better observed in the bottom panel of
Fig.~\ref{fig:3c31rhot} displaying the jet mass fraction. Both
maps show how the jet initially expands up to a distance $z \sim
1.5\,\rm{kpc}$ from the source\footnote{Throughout this section
and in the discussion of results, $z$ will refer to distances to
the galactic source, located 500 pc away from the injection in the
grid.} and then recollimates and oscillates until it is disrupted,
i.e., until the ambient medium material reaches the jet axis (at
$z \sim 4.5\,\rm{kpc}$), due to entrainment of the external
medium. Fig.~\ref{fig:3c31lofe} shows the flow Lorentz factor (top
panel) and the axial velocity (bottom panel). A relativistic jet
(with maximum Lorentz factor of 5.3) is seen up to the disruption
point at $z \sim 4.5\,\rm{kpc}$. After this point, strong
deceleration of the flow associated with mass loading of the jet
is observed in both maps. The axial velocity plot reveals a mildly
relativistic backflow in the cocoon. However, these high speeds
can be an artifact of the imposed axisymmetry, as discussed in
\cite{alo99} on the basis of 3D simulations of relativistic jets.
The use of open boundaries on the symmetry plane at the jet base
could also maintain artificially large pressure gradients along
the cocoon leading to high speed backflows (see the discussion in
sect.~\ref{ss:analmod}). Fig.~\ref{fig:3c31temp} shows maps of the
logarithm of temperature (top panel) and pressure (bottom panel).
The bow shock, cocoon and jet structure are also clearly seen in
these maps. As can be seen in the top panel of
Fig.~\ref{fig:3c31temp}, the ambient temperature increases by a
factor of a few behind the bow shock and by two orders of
magnitude in the cocoon due to the mixing with hot plasma injected
through the terminal shock at the head of the jet. Also in this
panel, Kelvin-Helmholtz instabilities arising in the interface
between the outer subsonic backflow in the cocoon and the shocked
ambient medium are observed. Finally, the bottom panel of
Fig.~\ref{fig:3c31temp} shows a shocked region with an homogeneous
pressure distribution and a series of waves emanating from the
outer part of the cocoon.

%%%%%%%%%%%%%%%%%%%%%%%%%%%%%%%%%%%%%%%%%%%%%%%%%%%%%%%%%%%%%%%%%%%
%
%
%%%%%%%%%%%%%%%%%%%%%%%%%%%%%%%%%%%%%%%%%%%%%%%%%%%%%%%%%%%%%%%%%%%

%%%%%%%%%%%%%%%%%%%%%%%%%%%%%%%%%%%%%%%%%%%%%%%%%%%%%%%%%%%%%%%%%%%
%
%
%%%%%%%%%%%%%%%%%%%%%%%%%%%%%%%%%%%%%%%%%%%%%%%%%%%%%%%%%%%%%%%%%%

%%%%%%%%%%%%%%%%%%%%%%%%%%%%%%%%%%%%%%%%%%%%%%%%%%%%%%%%%%%%%%%%%%
%
%
%%%%%%%%%%%%%%%%%%%%%%%%%%%%%%%%%%%%%%%%%%%%%%%%%%%%%%%%%%%%%%%%%%

\begin{figure*}
\centerline{
\includegraphics[width=0.5\textwidth]{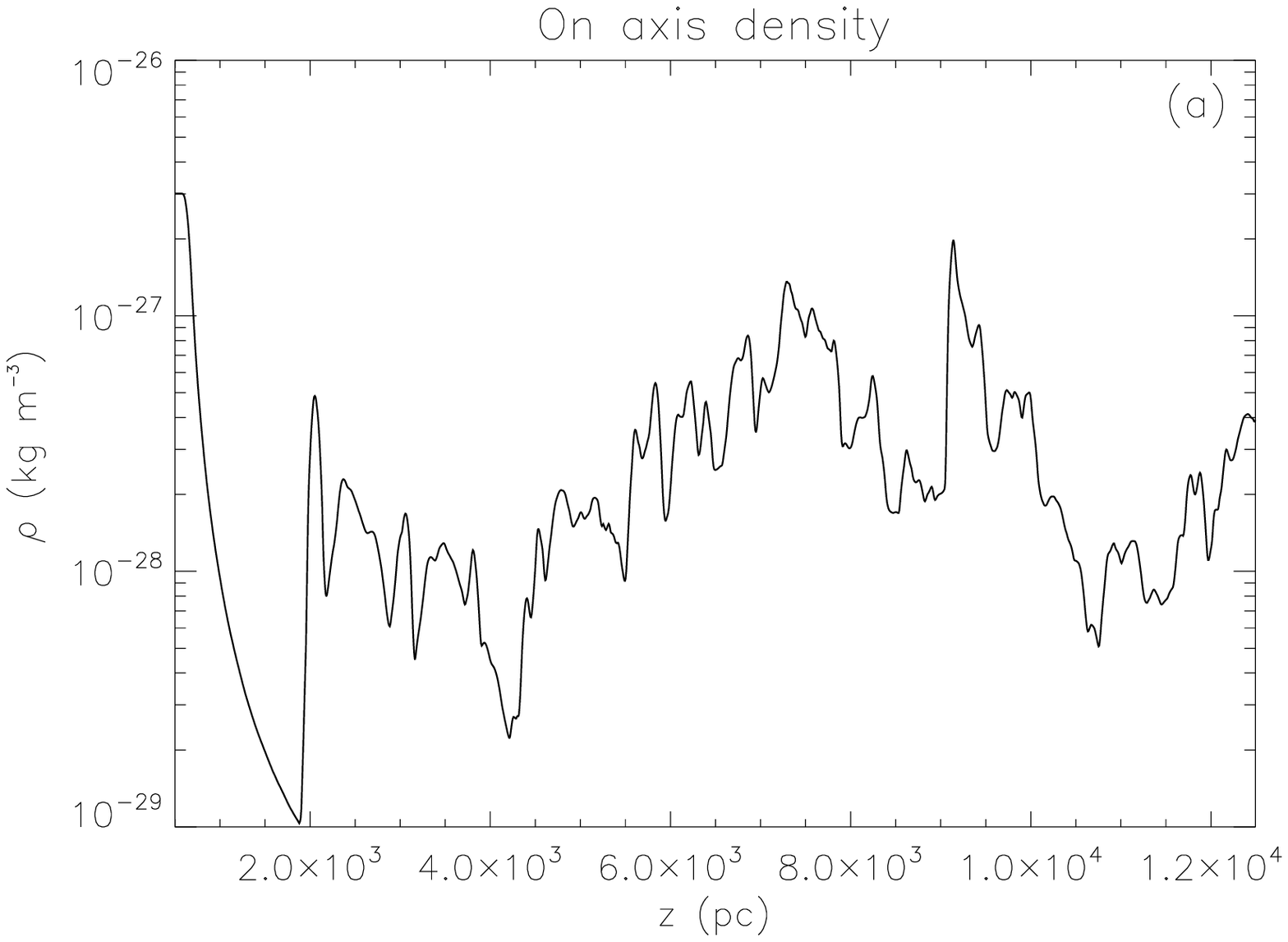}
\includegraphics[width=0.5\textwidth]{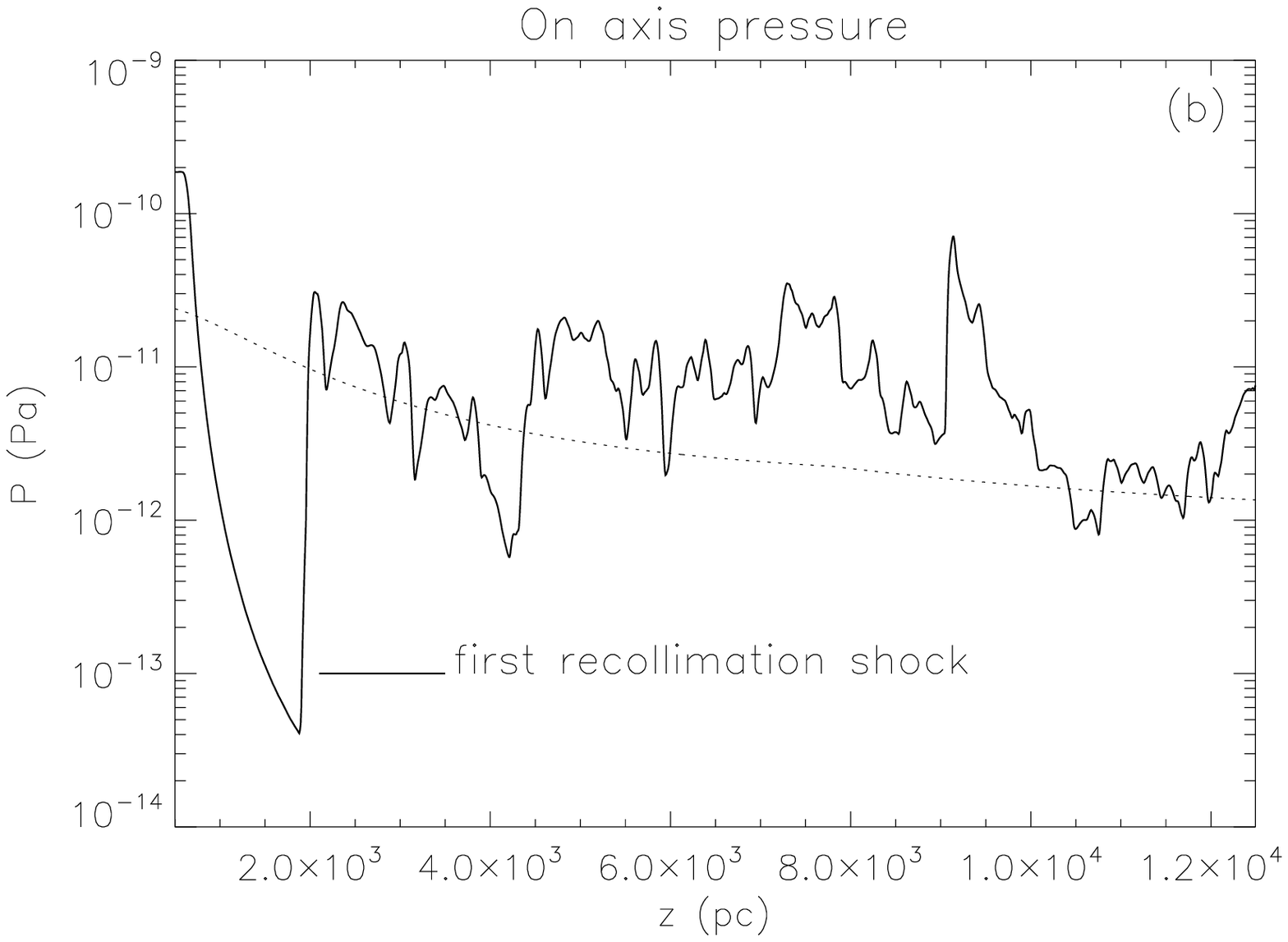}}
\centerline{
\includegraphics[width=0.5\textwidth]{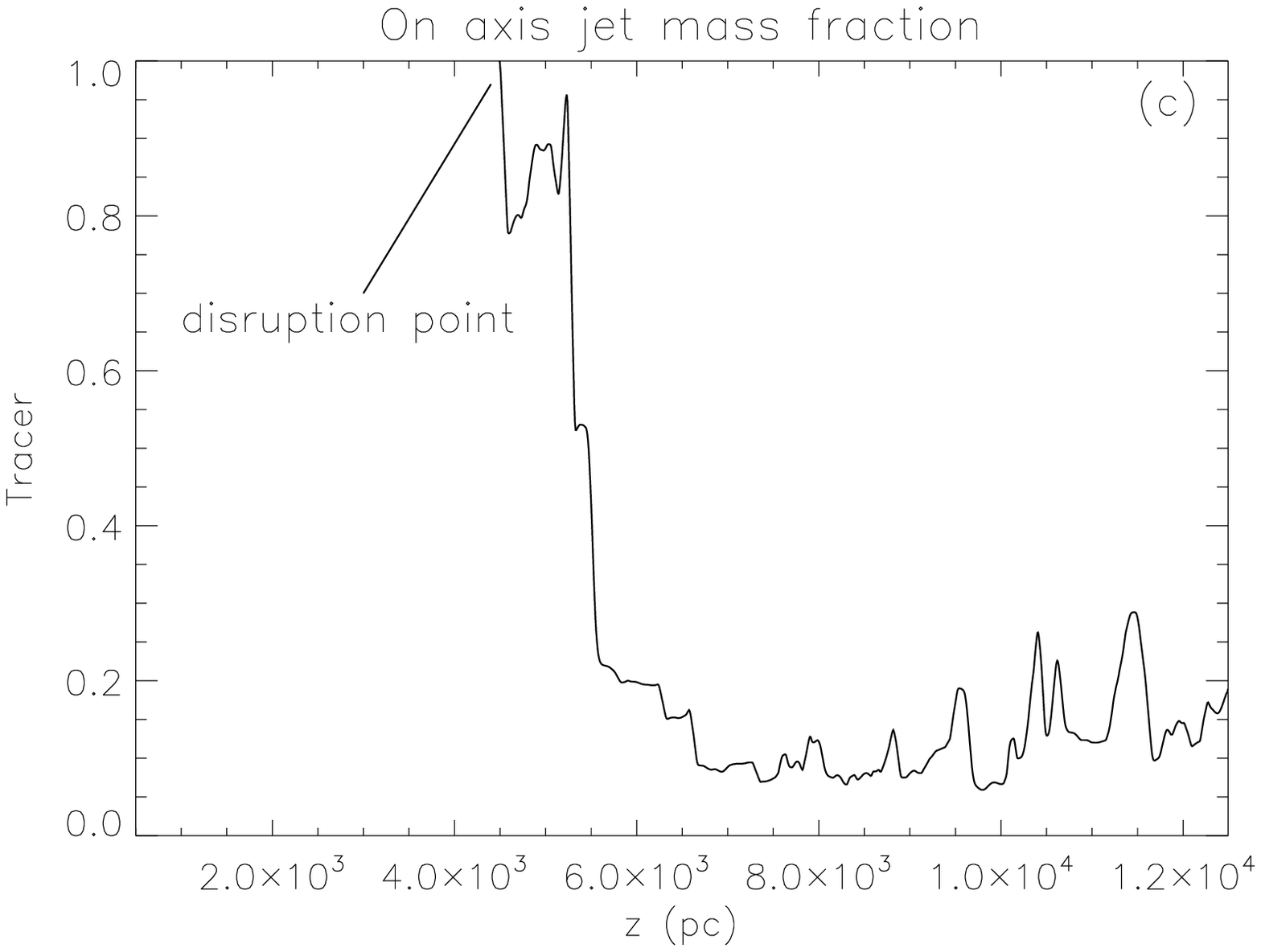}
\includegraphics[width=0.5\textwidth]{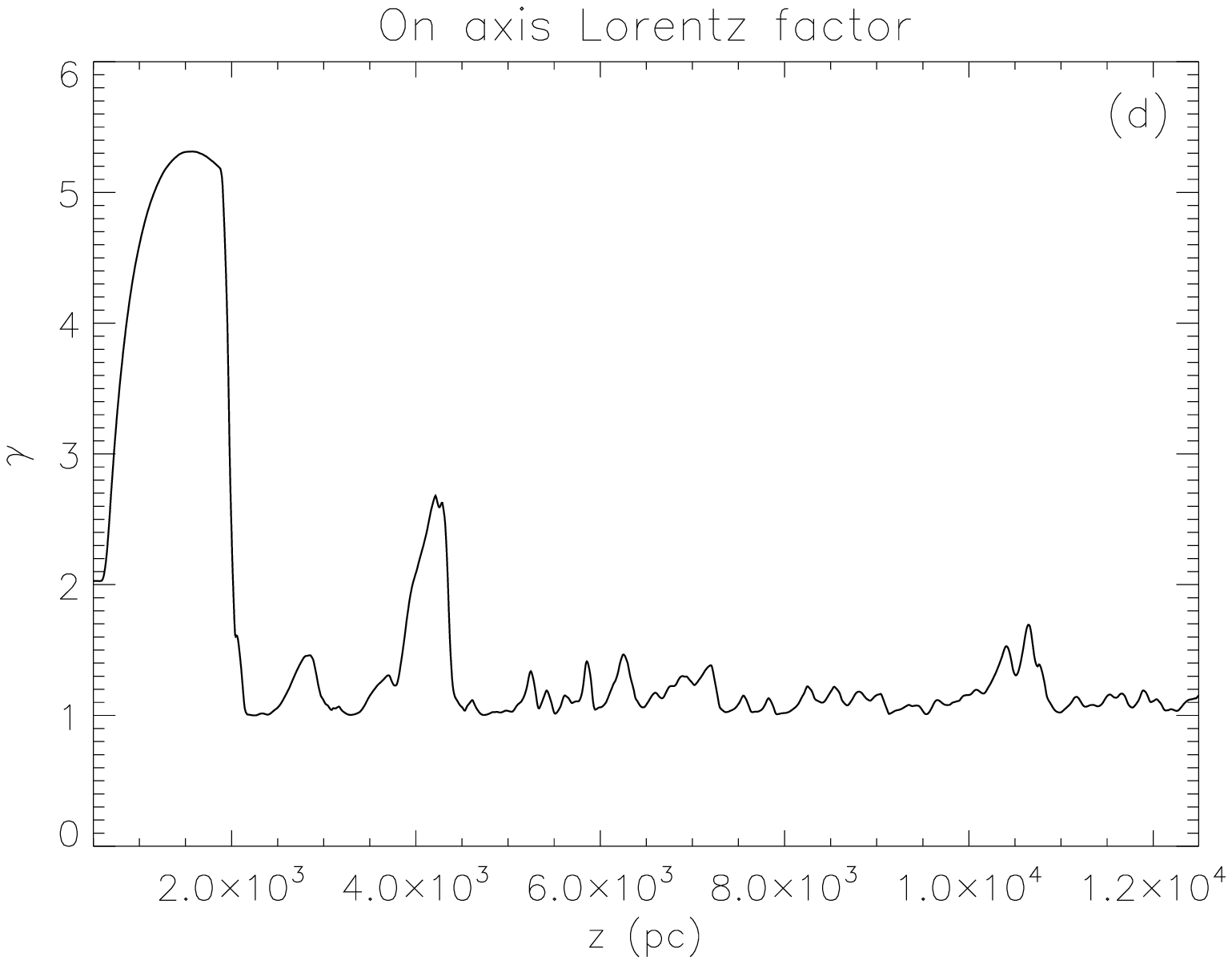}}
\centerline{
\includegraphics[width=0.5\textwidth]{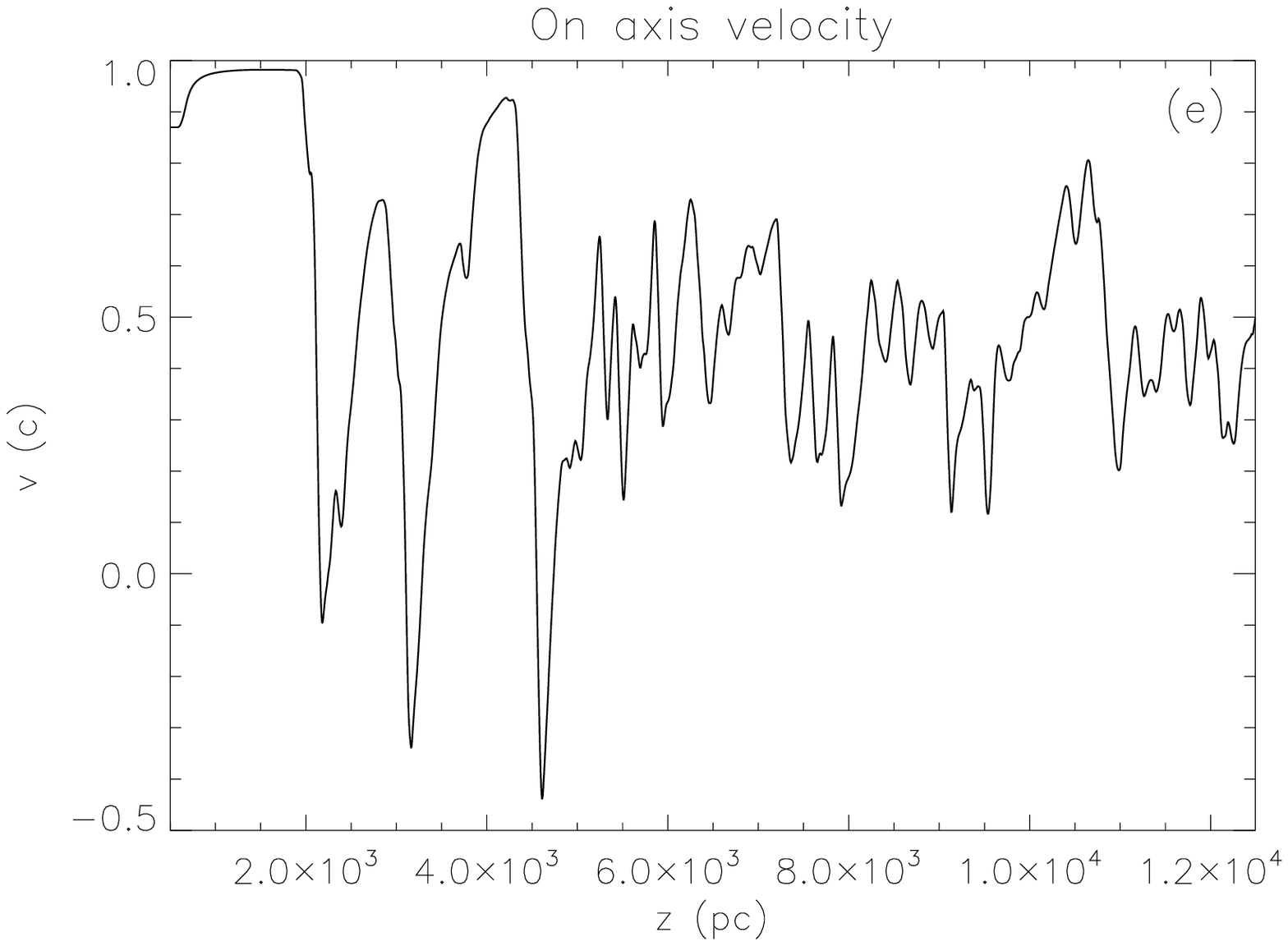}
\includegraphics[width=0.5\textwidth]{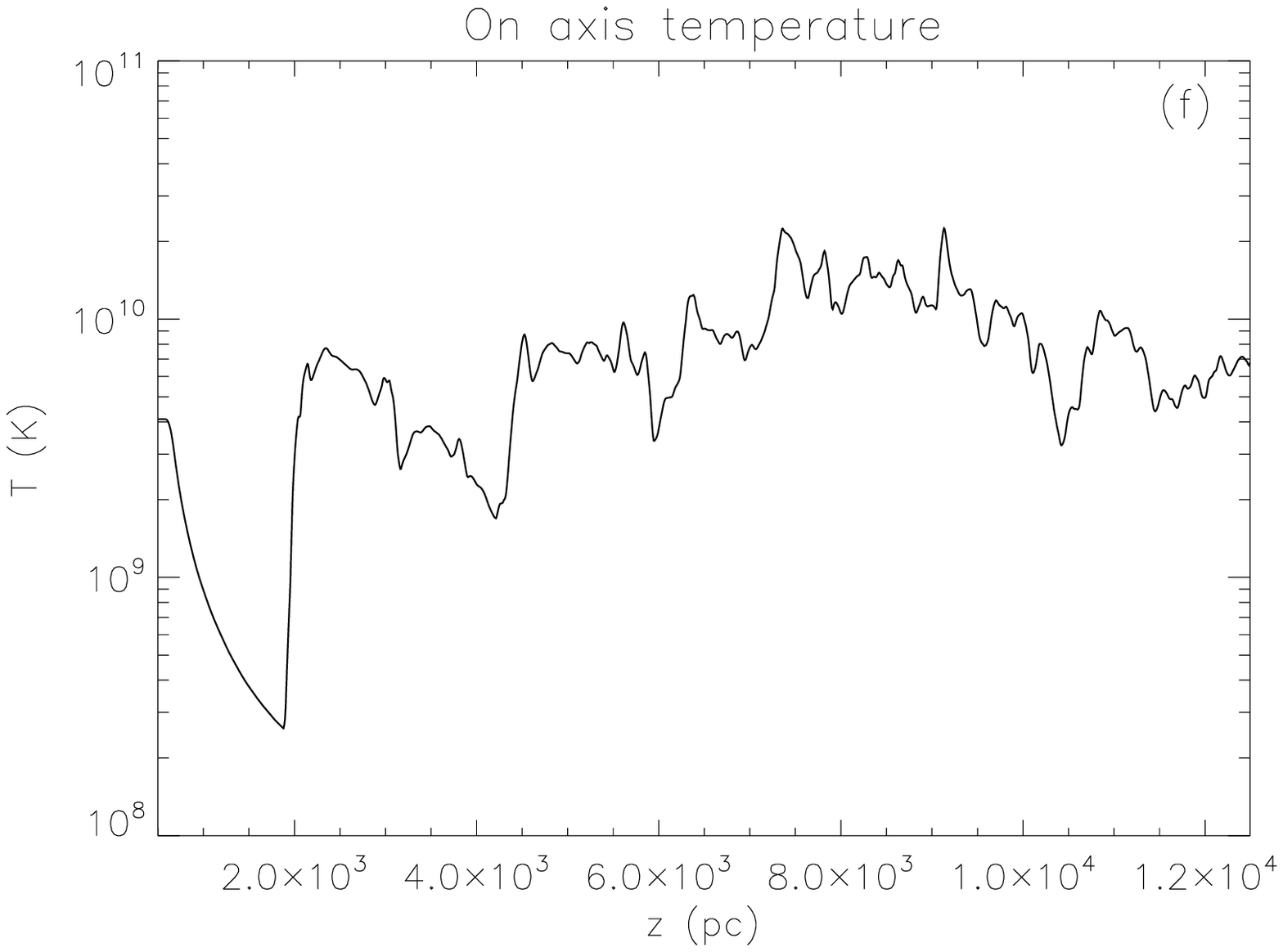}}
\caption{Different profiles of variables on the jet axis at the
end of the simulation. The different panels show the rest mass
density (a), pressure (solid line) and original ambient medium
pressure on the axis (dotted line; b), jet mass fraction (c), flow
Lorentz factor (d), axial flow velocity (e) and temperature (f).
The location of the first recollimation shock is indicated in the
pressure plot (panel b) and the location of the disruption point
is indicated in the jet mass fraction plot (panel c).}
\label{fig:fin0}
\end{figure*}

\begin{figure*}
\centerline{
\includegraphics[width=0.5\textwidth]{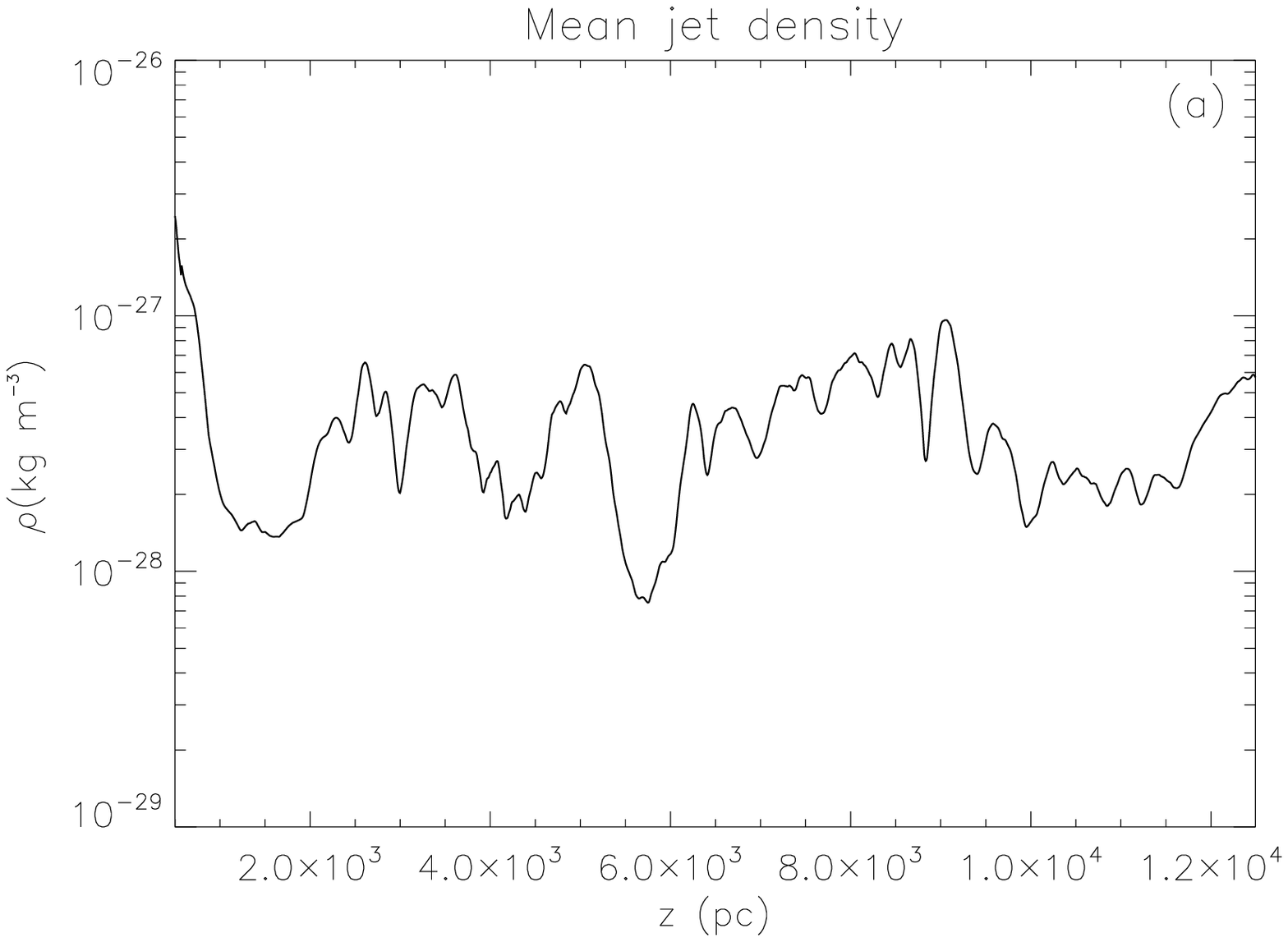}
\includegraphics[width=0.5\textwidth]{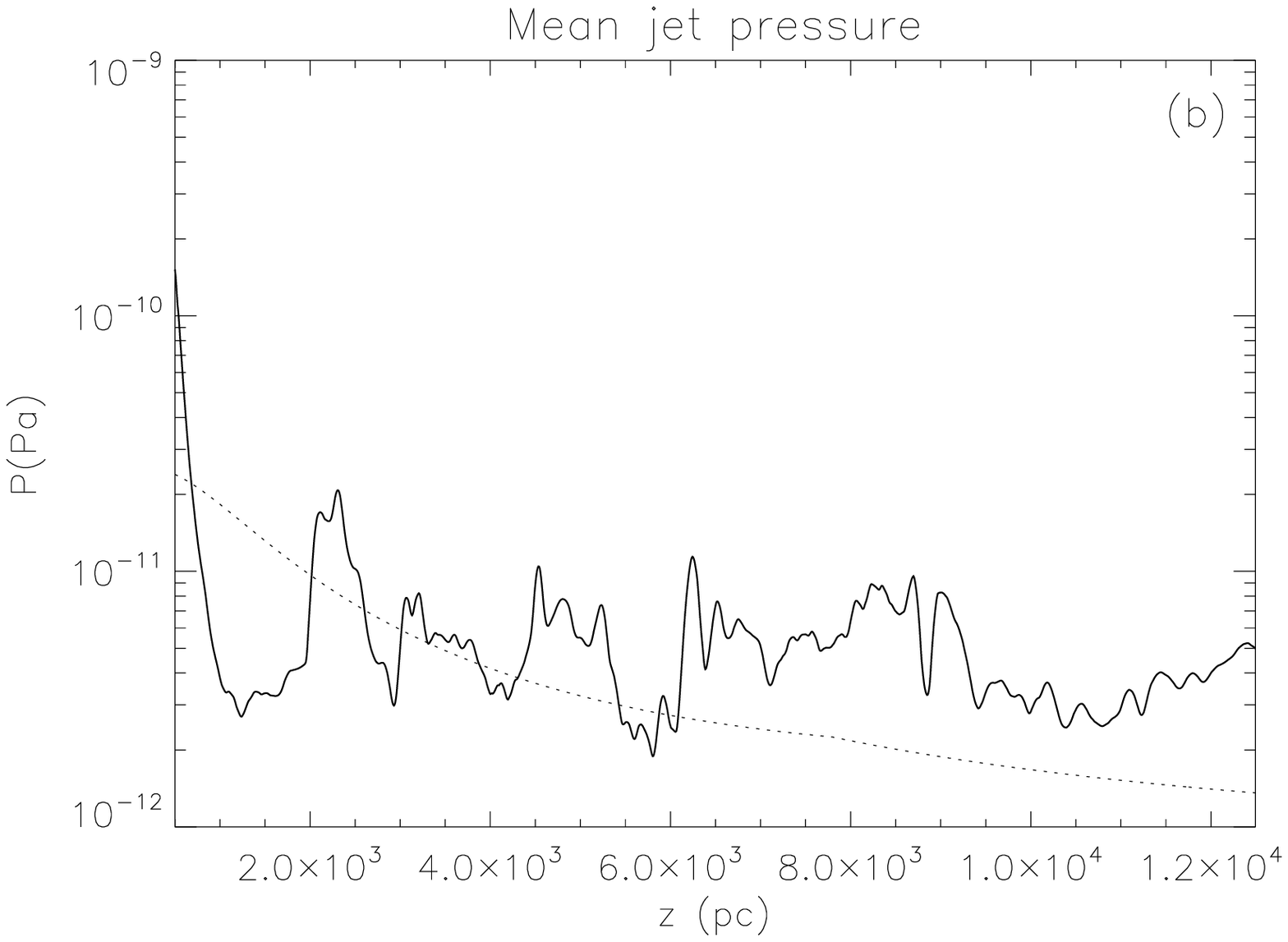}}
\centerline{
\includegraphics[width=0.5\textwidth]{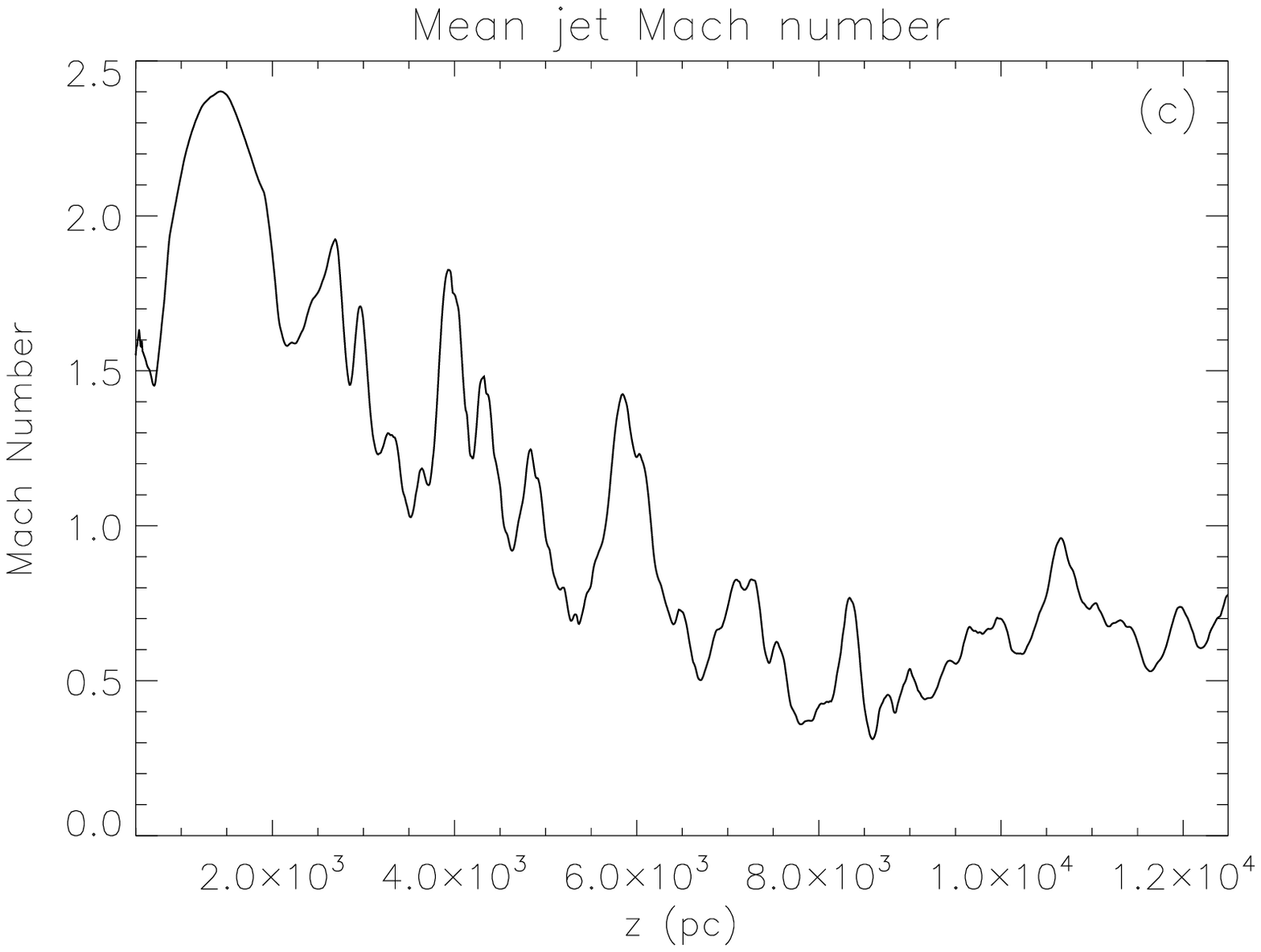}
\includegraphics[width=0.5\textwidth]{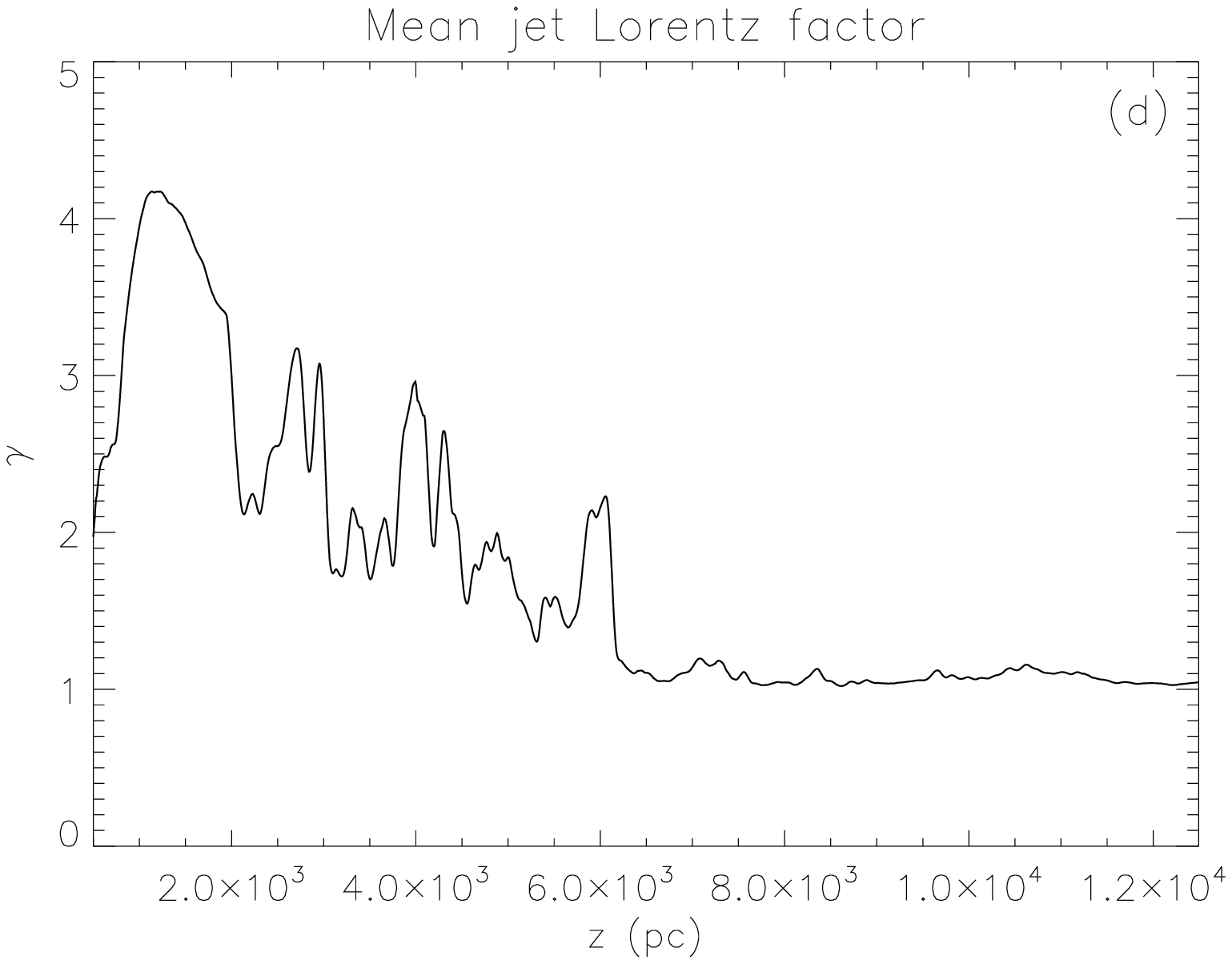}}
\caption{Profiles of radially averaged variables, weighted with
the tracer and counting only those cells where the axial velocity
is significantly larger than zero ($v^z > 10^{-3} \, c$). The
different panels show the rest mass density (a), pressure (solid
line) and original ambient medium pressure on the axis (dotted
line; b), Mach number (c), and Lorentz factor (d).}
\label{fig:fin1}
\end{figure*}

 The profiles of several variables along the jet axis (rest-mass
density, pressure, jet mass fraction, Lorentz factor, axial
velocity and temperature) at the end of the simulation are plotted
in Fig.~\ref{fig:fin0}. Fig.~\ref{fig:fin1} shows averaged values
of rest-mass density, pressure, Mach number and flow Lorentz
factor along the jet at the same time. From these plots and the
maps in Figs.~\ref{fig:3c31rhot}-\ref{fig:3c31temp} we can obtain
a general picture of the evolution of the jet in terms of its
dynamics. Close to the injection point, the flow can be considered
as steady. The jet initially expands in the ambient density
gradient. The expansion accelerates (Fig.~\ref{fig:3c31lofe}),
rarefies (Fig.~\ref{fig:3c31rhot}) and cools
(Fig.~\ref{fig:3c31temp}) the flow. This is seen in more detail in
the cuts along the jet axis shown in Fig.~\ref{fig:fin0}. After a
short distance, in which the variables remain constant, a strong
adiabatic expansion produces a fast decrease in density, pressure
and temperature on the jet axis, up to $z = 1.5-2.0 \, \rm{kpc}$.
At the same time, there is a strong acceleration seen in the plots
of Lorentz factor and axial velocity. The jet overpressures again
in a standing shock at $z \sim 2 \, \rm{kpc}$ from the source. The
shock is seen in the maps and plots of density, pressure and
temperature as a sudden increase and, in those of velocity and
Lorentz factor, as a strong deceleration of the flow. The
overpressure of the jet with respect to the ambient (see the
pressure map in Fig.~\ref{fig:3c31temp}, and the plots in the
Figs.~\ref{fig:fin0}b and \ref{fig:fin1}b) results in a new phase
of expansion. Each expansion is followed by a recollimation shock
that produces significant deceleration of the jet (see
Fig.~\ref{fig:3c31lofe} and Figs.~\ref{fig:fin0}d and
\ref{fig:fin0}e). Up to three recollimation shocks (at $z \sim 2
\, \rm{kpc}$, $z \sim 3 \, \rm{kpc}$ and $z \sim4 .5 \, \rm{kpc}$)
are observed before the jet is disrupted. These shocks can be
clearly identified in the Lorentz factor and velocity plots in
Fig.~\ref{fig:fin0} (panels d and e). The planar/conical shape of
these shocks could be an artefact of axisymmetry, but at the
distances from the central source at which these shocks are formed
(i.e. a few kpc), 3D effects are probably small. After the first
shock, the jet is less overpressured with respect to the ambient
medium and it expands in an atmosphere with a smoother density
gradient. This makes the next standing shocks milder (see the
pressure plots in Fig.~\ref{fig:fin0}b and Fig.~\ref{fig:fin1}b).
At $z > 6 \, \rm{kpc}$ (i.e., behind the terminal shock), the jet
is overpressured with respect to the ambient
(Fig.~\ref{fig:fin1}b) due to the heating of the flow in shocks,
as seen in Fig.~\ref{fig:fin0}f.

 The Lorentz factor is between 2 and 5.3 in the whole
section of the jet up to the first recollimation shock, at $z \sim
2 \, \rm{kpc}$ (see Fig.~\ref{fig:3c31lofe}). After this shock,
the flow is relativistic in filaments far from the axis, as the
portions of the jet closer to the axis are decelerated by the
shock. The averaged Lorentz factor (Fig.~\ref{fig:fin1}d)
decreases from 4.3, after the first adiabatic expansion of the
jet, to a value of 1.15 at the disruption point, which we define
as the point in which ambient material reaches the axis (see the
jet mass fraction map in Fig.~\ref{fig:3c31rhot}, the flow Lorentz
factor map in Fig.~\ref{fig:3c31lofe}, and the corresponding axial
cuts in Figs.~\ref{fig:fin0}c and \ref{fig:fin0}d). The internal
Mach number of the flow (Fig.~\ref{fig:fin1}c) also decreases
along the jet up to the terminal shock where the flow becomes
subsonic. After the terminal shock, the flow is only slightly
relativistic, with velocities around $0.5 \, c$. It is also
remarkable that in the cocoon a backflow with mildly relativistic
speeds is established right down to the central regions of the
parent galaxy.

 The jet remains well collimated and relativistic up to the
disruption point. However, after this point, the subsonic
character of the flow triggers the process of mixing with the
ambient medium, as already pointed out by \cite{bic95}. The
rest-mass density plots (Figs.~\ref{fig:fin0}a and
\ref{fig:fin1}a) show the increase of this variable as the jet
entrains ambient material for $z > 4.5 \, \rm{kpc}$. The position
of the disruption point oscillates throughout the simulation from
$z\sim2.5-3.5\,\rm{kpc}$ to $z\sim6-7\,\rm{kpc}$. This
non-monotonic advance is due to the complex dynamics of the
terminal shock, which changes periodically from planar to conical,
as already noted in previous long-term simulations of jets
\citep[see, e.g.,][]{mart97,sch02}. In the phases in which the
disruption point moves backwards, the further injected plasma
accumulates in the region between injection and disruption, thus
increasing the pressure in the jet, that ultimately \emph{bursts},
expanding the unmixed jet outwards and bringing the disruption
point farther from injection.

  Fig.~\ref{fig:3c31temp} shows that the temperature in the cocoon
is of the same order, but larger than that of the jet ($T \sim
10^9-10^{10}$ K, cf $T_j=4.1\,10^9$ K at injection). In addition,
the gas in this high temperature region can be up to an order of
magnitude denser than the jet (top panel in
Fig.~\ref{fig:3c31rhot}) due to the loading with baryons from the
ambient (jet mass fraction, $f<0.2$). However, despite this large
baryon load, the number of leptons is still between 20 and 200
times larger than the number of baryons in this region. Leptons
are heated in shocks inside the jet (see top panel of
Fig.~\ref{fig:3c31temp}) and at the terminal shock. The high
temperatures achieved in the cocoon are a consequence of the small
amount of baryons. The adiabatic exponent in the cocoon remains
close to the relativistic limit of 4/3.

  In a recent paper, Kino, Kawakatu and Ito (2007) conclude that the
temperature of the cocoon gas for an FRII jet with Lorentz factor
10 could be of about $1.2 \, 10^{10} \, \rm{K}$, a value close to
that obtained in our simulation. In the case of FRII jets, the
particles are heated in the hot spot, the heating depending on the
strength of the shock. In a electron-positron FRI jet like that
simulated here, the particles are heated in the strong
recollimation shocks along the jet. This effect can compensate the
lack of heating at the terminal point of the FRI jet due to its
low power. Thus, the temperature in the cocoon remains high as
long as the pollution by baryons is low.

%%%%%%%%%%%%%%%%%%%%%%%%%%%%%%%%%%%%%%%%%%%%%%%%%%%%%%%%%%%%%%%%%%%%%%%%
%
%
%%%%%%%%%%%%%%%%%%%%%%%%%%%%%%%%%%%%%%%%%%%%%%%%%%%%%%%%%%%%%%%%%%%%%%%%

%%%%%%%%%%%%%%%%%%%%%%%%%%%%%%%%%%%%%%%%%%%%%%%%%%%%%%%%%%%%%%%%%%%%%%%%
%
%
%%%%%%%%%%%%%%%%%%%%%%%%%%%%%%%%%%%%%%%%%%%%%%%%%%%%%%%%%%%%%%%%%%%%%%%%

\subsection{Additional simulations} \label{sec:newsims}

  The modelling of 3C~31 presented in LB02b divides the inner
$12 \, \rm{kpc}$ of the jet into three regions: inner, flaring and
outer region. The authors suggest that the boundary between the
inner and flaring regions (at $1.1 \, \rm{kpc}$ from the source)
consists of a discontinuity in velocity, density and pressure,
which is the cause of the sudden increase of radio emission. In
\cite{can04}, \cite{can05} and \cite{la06}, the inner and flaring
regions of the sources studied are defined in terms of emissivity,
but the authors do not invoke the presence of a shock in the
transition from one region to the next. In this paper we identify
that transition as due to the presence of a recollimation shock.
Within this frame, we compare the properties of the observed
discontinuity with those of its counterpart in the numerical
simulation at $z \sim 2 \, \rm{kpc}$.

  LB02b give a speed of the jet of $0.87\,c$ in the inner region,
that we used as the injection speed of the jet in our simulation.
However, the flow in the simulation expands adiabatically and
accelerates in the decreasing density ambient medium, entering the
first recollimation shock with a larger speed ($\sim 0.98 \, c$).
The technique used in LB02a that allows for a fitting of the
velocity of the jet is not adequately constrained in the inner
region, so the initial velocity used here may be in error. The
authors report that jet/counter-jet ratio is slightly smaller in
the inner region than at the start of the flaring region, which
could be interpreted as due to an acceleration of the flow in this
region, as observed in our simulation. Observations with higher
transverse resolution and sensitivity could potentially constrain
any acceleration in the inner region.

  The position and velocity of the flow upstream/downstream of
the standing shock depend on the acceleration of the jet in the
initial adiabatic expansion phase. Similarly, the properties of
the flow upstream the shock determine the properties downstream
the shock and ultimately the jet disruption. In order to study the
influence of jet pressure and injection velocity on the evolution
of the jet, we performed three additional simulations with the
following properties: a) Simulation 2, $v_j = 0.5 \, c$,
$P_j/P_{a,c} = 3.8$; b) Simulation 3, $v_j = 0.6 \, c$,
$P_j/P_{a,c} = 7.8$, and c) Simulation 4, $v_j = 0.5 \, c$,
$P_j/P_{a,c} = 1$. The change in the flow injection velocity and
the jet pressure in these additional simulations produces a change
in the power of the jet. In the case of simulation 2, the change
in the injection parameters produces a reduction of the jet power
of about an order of magnitude with respect to the original
simulation. In the case of simulations 3 and 4, the reduction in
power with respect to the first simulation is of a factor of 3
and 30, respectively. Finally, in these simulations we have used
approximations to the Bessel functions that are solved in the
equation of state -see the Appendix- \citep{pei05,ser86}, in order
to reduce the computational load.

  In simulation 2, the jet flow is accelerated in the pressure
gradient up to $v_j \sim 0.82\,c$. The first recollimation shock
occurs at $z \sim700 \, \rm{pc}$, closer to the source than given
by LB02b, due to the smaller initial pressure ratio at injection
and the smaller injection speed. After this shock, the jet is
strongly mass loaded and decelerated to speeds $v_j < 0.5 \, c$.
The entrained jet expands and accelerates slightly to
$v_j=0.6\,c$, but the latter expansion ends up in further mass
loading and deceleration of the jet at $z \sim 1.1 \, \rm{kpc}$.
Mass loading and deceleration continue downstream up to $z = 4.8
\, \rm{kpc}$, where the velocity of the jet is already
subrelativistic and subsonic. The simulation was stopped when the
head of the jet had reached this distance, at time $t \sim 6.6 \,
10^6 \, \rm{yrs}$.

  The jet in simulation 3 goes through the first recollimation shock
at $z \sim 800 \, \rm{pc}$. The flow enters the shock with a speed
$v_j \sim 0.94 \, c$. After the shock, the jet decelerates and is
disrupted at $z \sim \, 1.5 \, \rm{kpc}$. As in simulation 2, mass
loading continues downstream. The simulation was stopped at $t
\sim 6 \, 10^6 \, \rm{yrs}$, when the jet head is at $z \sim \, 5
\, \rm{kpc}$ and the bow shock is at $z \sim \, 6 \, \rm{kpc}$.

  Simulation 4 differs from simulation 2 in that the jet is in
pressure equilibrium with the ambient medium at injection. This
difference prevents the formation of strong shocks. The jet
velocity oscillates close to the injection value ($v_j = 0.5 \,
c$). Successive expansions and contractions, i.e. smooth
recollimations, in the pressure gradient cause the pinching of the
jet. The growth in amplitude of this pinching as it couples to a
Kelvin-Helmholtz instability mode causes entrainment and the
disruption of the jet. The jet is more stable in this case simply
because the amplitude of the recollimation shocks is reduced.
Nevertheless, the jet is light and slow, which makes it a good
candidate for disruption due to the growth of instabilities
\citep{pe+04b,pe+05}. At the end of the simulation ($t \sim7 \,
10^6 \, \rm{yrs}$) the jet head is at $z \sim 4.8 \, \rm{kpc}$.

\section{Discussion}\label{sec:discus}

  \cite{sch02} performed a series of simulations of the long
term evolution of jets with different compositions evolving in a
uniform density ambient medium. The initial power of those jets
was typical for FRII sources. Compared to their simulations, the
jet in the simulation presented here is 100 times weaker. The bow
shock propagates at a slightly larger mean velocity in our
simulation. This fact is due to the simulated jet being
overpressured and propagating through a decreasing density
atmosphere. The morphology of our jet is close to that of model LH
(leptonic, hot) in \cite{sch02}, although in our case, the jet is
more pinched and presents entrainment behind the head. It is
remarkable that the jet in our simulation is leptonic and hot and
the structure obtained is solely similar to the same case in
\cite{sch02}.

\subsection{Jet advance speed and source age}

 \cite{par02} studied a sample of FRI radio galaxies and
estimated their spectral ages from two-frequency data. Although
these measurements are subject to uncertainties, as pointed out by
\cite{la06} and \cite{kr97}, mainly due to possible confusion
between jet and lobe emission, we have used them in order to make
a rough comparison of the age of the source in the simulation with
those of FRIs. The ages obtained by \cite{par02}, claimed to be
lower limits, range between $10^7$ and $10^8 \, \rm{yrs}$, in
general larger than those of FRII jets. From these estimates, the
authors compute advance velocities for the lobes in the range
$10^{-3} - 10^{-2} \, c$. VLA observations of 3C~31 (e.g., LB02a)
show jet material at projected distances up to $150 \, \rm{kpc}$.
Considering a viewing angle of $52\degr$ (LB02a), this implies
linear distances of about $200 \, \rm{kpc}$. From the advance
velocities found in the simulation, we can put constrains on the
time that the source has been continuously active. The continuity
is supported by the fact that there are no emission gaps in the
images. At the typical advance speeds found in this simulation,
say $5 \, 10^{-3} \, c$ for further advance of the head as an
upper limit, the age of the source would be $t > 10^8 \,
\rm{yrs}$, in agreement with the spectral ages given by
\cite{par02}. The lower limit takes into account further
deceleration of the head, and the curved trajectory of the plasma
observed in the source at distances larger than $\sim 25 \,
\rm{kpc}$. Moreover, low-resolution images show emission out to
300 kpc. The uncertainty in the viewing angle at large distances
from the core makes difficult the use of this value for an
estimate. Nevertheless, this fact makes of the value given here a
strict lower limit. In the frame of intermittency models for the
activity in AGNs, the lower limit of the age of the source given
here puts constraints on the possible intermittency of 3C~31:
either the periodicity is as long as $0.01-0.1$ times the age of
the Universe, or it has been continuously active since the onset
of its activity.

\subsection{Early evolution and the young counterparts of FRIs}

  We have shown, in Fig.~\ref{fig:compact}, that the evolution of
the source in the compact phase is divided in two stages: the
\emph{CSO-like} and the \emph{weak CSS-like} phase. \cite{dra04}
have proposed that low-power CSS sources can be the young
counterparts of large FRI sources. Based on spectral aging, the
authors give age estimates of $\sim 10^5 \, \rm{yrs}$ for several
of these CSS sources with linear sizes of a few kiloparsecs, and
derive expansion velocities of $0.004 - 0.007\,c$. These
velocities are in agreement with those derived from the simulation
(see Fig.~\ref{fig:evol0}). Furthermore, the age of the simulated
jet in the \emph{weak CSS-like} phase ($t \sim 10^5 - 10^6 \,
\rm{yrs}$) is of the order of that estimated by \cite{dra04} for
the observed low-power CSS sources, and its morphology is
irregular, as those authors show to be the case of the low-power
CSS jets. Fig.~\ref{fig:compact} tells us that FRIs could first go
through a CSO stage, characterized by a regular, expanding jet,
and, after developing a strong shock due to underpressuring with
respect to the ambient medium, at a distance of the order of
$1\,\rm{kpc}$ (depending on the properties of the host galaxy and
the jet, as we have seen in Sect.~\ref{sec:newsims}), develop the
irregular structure observed in the maps for the low power CSS
sources. Therefore, powerful CSO sources could be the young FRII
sources, but, from our results, we predict that low power CSOs
could be the young FRI sources.

\subsection{Cocoon temperature and emission}

  The high temperature of the fluid in the cocoon
(Figs.~\ref{fig:3c31rhot}-\ref{fig:3c31temp}) deserves some
discussion. Heating of the jet plasma occurs at the standing
shocks along the jet, as shown in Fig.~\ref{fig:fin0}f. The
efficiency of the heating depends on the strength of these shocks,
which ultimately depends on the jet power. In the case of the
mildly relativistic, hot jets like those considered in the present
work, this last quantity is dominated by the internal energy
density flux (or pressure). This is why simulations 2 and 4, with
values of initial pressure ratio ($P_j/P_{c,a}$) of 3.8 and 1,
respectively, display weaker recollimation shocks and the
temperature of jet material hardly rises above the injection
values. In any case, the temperatures found in the cocoon ($T \sim
10^{10} \, \rm{K}$), in combination with the resulting cocoon
densities ($n_{e^+,e^-} \sim 10^{-3} \, \rm{cm^{-3}}$), would
produce a flux at 1 MeV of $\nu F_\nu \sim 10^{-19} \, \rm{erg \,
cm^{-2}\, s^{-1}}$ for a source located at the distance of 3C~31,
quite below the detection limits of \emph{INTEGRAL} and
\emph{NeXT} \citep[][and references therein]{kin07}, and the
resulting spectrum would also fall below the detection limits of
the X-ray satellite \emph{XMM-Newton}. Moreover, \cite{kin07}
conclude that the bremsstrahlung luminosity decreases with time as
$t^{-1}$, from the results of a cocoon expansion that follows
basically the same time dependence as that found in this
simulation. Thus, the present bremsstrahlung luminosity should be
10 times smaller than that computed here, on the basis of the
calculated age of the source compared to the simulated time. In
this respect, \cite{kin07} show that bremsstrahlung cooling of the
cocoon is only important in the very first stages of the evolution
($t < 200 \, \rm{yrs}$). This validates the adiabatic treatment of
the problem. We want to point out that the bremsstrahlung emission
calculated for the shocked ambient medium is even smaller.
\cite{za03} performed a series of simulations of supersonic and
underdense jets in a decreasing pressure atmosphere and showed
that jets evolve in two different phases regarding their high
energy emission: a phase in which the shell formed by shocked
material is highly overpressured and radiative, and a later phase
in which the shock is weaker and a deficit of X-ray emission is
expected from the lobes. The jet in our simulation is in the
former of the two phases. However, \cite{za03} point out that jets
with low density ratios, as that simulated here, form wide and not
very dense shells from which no strong emission is expected, in
agreement with the results given above. Finally, the recent
discovery of bow shocks in low power radio jets
\citep{kra03,cro07}, moving at similar Mach numbers as those
obtained here, and showing overpressure by more than an order of
magnitude with respect to the ambient medium, gives support to our
results regarding the dynamics of the bow shock (see next
subsection).

\subsection{Bow shock}

  At the end of the simulation, the head of the bow-shock has
reached a distance of $\sim 15 \, \rm{kpc}$ from the injection
position. It expands self-similarly and at basically constant rate
($\sim 7 \, 10^{-3} \, c$ in the axial direction), with a slight
deceleration with time. The bow shock is still supersonic by the
end of the simulation ($M \sim 2.5$), contrary to what is expected
for an FRI jet in theoretical models, which predict trans-sonic
speeds for shocks at such distances. However, recent X-ray
observations by \cite{kra03} and \cite{cro07} show the presence of
bow shocks with Mach numbers between 3 and 8 in the low power
radiogalaxies Centaurus A and NGC~3801 at distances of a few
kiloparsecs from the source. Taking into account that the jets in
3C~31 are among the most powerful ($10^{44} \, \rm{erg/s}$)in FRI
sources, it is plausible that the bow shock is long lived in this
source. Furthermore, as we have shown in the previous paragraph,
we have only simulated less than 10\% of the real evolutionary
time of the source. Thus, it is possible that, after the simulated
time, the bow shocks may naturally decelerate to trans-sonic
velocities. The properties and morphology of the jet would then be
modified, accommodating the further evolution of the flow to the
observed structure of 3C~31, where no bow-shock is observed, and
the emission appears to fade gradually with distance.

  Table~\ref{tab:comp} shows the values of number density,
temperature and pressure at both sides of the bow shock in our
simulation, compared to those given in \cite{kra03} and
\cite{cro07} for Cen~A and NGC~3801, respectively. The jump in
density at the head of the bow shock, in the axial direction, has
a value of 3.7, close to the Rankine-Hugoniot jump condition limit
for strong shocks, similar to the case of NGC~3801. The jump in
pressure in the simulation is smaller than in the observations of
Centaurus~A and NGC~3801, which can be understood in terms of the
age of the jets: the jet in the simulation is older than those in
Centaurus~A and NGC~3801 and therefore the bow shock is slightly
closer to equilibrium with the external medium. The main
differences between the simulation and the observations arise in
the temperature. In a direction transverse to the jet axis, the
temperature of the shocked ambient medium is of the order of that
in Centaurus~A. However, in the head of the bow shock, the
temperatures of the shocked gas reach values of $10^9\,\rm{K}$,
too high compared with observations. This difference may be due to
3C~31 being more powerful than the observed sources and also due
to the lack of cooling mechanisms in the simulation. Apart from
this, the results obtained from the simulation are in fair
agreement with the X-ray observations of the aforementioned
sources. Therefore, we conclude that 3C~31 possibly went through a
similar stage to that observed for Centaurus~A and NGC~3801, and
only in later stages than those simulated here, the bow shock
decelerated to transonic speeds and disappeared.

%%%%%%%%%%%%%%%%%%%%%%%%%%%%%%%%%%%%%%%%%%%%%%%%%%%%%%%%%%%%%%%%%%%%%%%%
%
\begin{table*}
\begin{center}
\caption{Shell (shocked ambient medium in the simulation) and ISM
(unperturbed ambient medium in the simulation) thermodynamic
values in the simulation, compared to those of Centaurus~A and
NGC~3801. The ranges in the values given for the simulation stand
for the different values of the parameters in the transversal and
in the axial directions: the maxima (minima) in the shell
correspond to the axial (transversal) direction and the maxima
(minima) in the ISM correspond to the transversal (axial)
direction. The ranges in the other columns are taken from the
referenced papers. } \label{tab:comp}
\begin{tabular}{@{}cccc}
\hline
&Simulation& Centaurus A$^1$& NGC~3801$^2$\\
\hline
$n_{shell}$ ($m^{-3}$)&$(3.3-5.2)\,10^{4}$&$2.0\,10^{4}$&$(2.0-3.0)\,10^{4}$\\
$n_{ISM}$ ($m^{-3}$)&$(1.4-1.5)\,10^{4}$&$1.7\,10^{3}$&$4.6\,10^{3}$\\
$P_{shell}$ (Pa)&$(0.4-1.8)\,10^{-11}$&$2.1\,10^{-11}$&$(4.2-8.9)\,10^{-12}$\\
$P_{ISM}$ (Pa)&$1.3\,10^{-12}$&$1.0\,10^{-13}$&$3.8\,10^{-13}$\\
$T_{shell}$ (K)&$3.2\,10^7-1.0\,10^9$&$3.34\,10^7$&$(0.8-1.2)\,10^7$\\
$T_{ISM}$ (K)& $1.7\,10^7$&$3.36\,10^6$&$2.67\,10^6$\\
\hline
\end{tabular}

\medskip
$^1$\cite{kra03}. $^2$\cite{cro07}.
\end{center}
\end{table*}
%
%%%%%%%%%%%%%%%%%%%%%%%%%%%%%%%%%%%%%%%%%%%%%%%%%%%%%%%%%%%%%%%%%%%%%%%%%

\subsection{Jet structure and the Laing \& Bridle (2002a,b) model}

  The final structure of the jet is analyzed in the following
paragraphs. We observe a fast adiabatic expansion when the jet
leaves the galactic core, as the jet propagates through the steep
density gradient of the galaxy. This expansion ends at $z \sim 2
\, \rm{kpc}$ and it is followed by a sudden recollimation. This
recollimation generates a new overpressuring of the jet, and thus,
a second expansion. The smoother pressure and density gradient of
the atmosphere in the region where the second expansion takes
place, causes it to appear smoother than the first expansion. This
process ends, as in the previous case, in a recollimation shock,
at $z\sim 3\,\rm{kpc}$. The third standing shock is at $z\sim
4.5\,\rm{kpc}$, after which the jet, strongly entraining and
decelerated by mass loading, generates a wide shear layer.

  The structure of the jet at the end of our simulation is to be compared
with that given in the dynamical model for the jet of 3C~31 in
LB02b. The main caveat for comparison between the simulation and
the observations and modelling in the latter paper is that the jet
in our simulation has not reached a steady-state, but keeps
evolving. This is clearly observed in the overall structure of the
jet seen in Figs.~\ref{fig:3c31rhot}-\ref{fig:3c31temp} compared
to the images of the jets in 3C\,31. The main difference is the
presence of lobes in the simulation: these are not observed.
\cite{la06} have shown that the kinematics of the jets in 3C296
(which are surrounded by their lobes) may be affected by the
absence of a shear layer with the ambient medium.

  In LB02b, the inner $12\,\rm{kpc}$ of the jet are divided into three
regions: inner, flaring and outer. The inner region (up to
$1.1\,\rm{kpc}$) consists of a fast moving flow $v\sim0.8-0.9\,c$
that enters the flaring region ($1.1-3.5\,\rm{kpc}$) through a
discontinuity that decelerates the flow and increases the
emissivity. In this region, the authors observe a spread of the
isophotes and a subsequent recollimation. The outer region
($3.5-12\,\rm{kpc}$) is characterized by a slow decrease in
velocity, and continuous mass-loading. In our simulation, the
discontinuity between the inner and flaring region is identified
with the recollimation shock at $z \sim 2 \, \rm{kpc}$. LB02a
argue that the velocity of the jet is about $0.8\,c$ up to about
$3\,\rm{kpc}$. In our case, however, after the standing shock, the
velocity already drops to about $0.4\,c$. Moreover, LB02b found
that the entrainment rate required to counterbalance the effects
of adiabatic expansion and keep the velocity fairly constant at
the beginning of the flaring region was consistent with that
expected from mass injection by stellar winds. This points towards
a fundamental difference between our simulation and their
modelling, mainly due to the lack of mass load from stellar winds
and the presence of a standing shock in our case. However, such a
shock could explain discontinuities in emissivity found close to
the start of the flaring regions in the observations and modelling
of the sources studied by LB02a,b, \cite{can04}, \cite{can05} and
\cite{la06}, and the observation of high energy emission from the
flaring regions in 3C\,31 \citep{hard02}, NGC\,315 \citep{wor07}
and 3C\,296 \citep{hard05,la06}. On the other hand, strong
recollimation is not evident from the shapes of the jets in the
sources studied, although it could be occurring in the inner parts
of the jet, whereas the outer layers show constant or increasing
radius, as shown by our simulation, where the radius of the outer
layers shows little signs of recollimation, contrary to the case
of the core of the jet (Fig.~\ref{fig:evol2}b and
Figs.~\ref{fig:3c31rhot}-\ref{fig:3c31temp}).

In the simulation, the transition between the flaring and outer
regions occurs at $z\sim 4\,\rm{kpc}$, after the third
recollimation shock, and not as a slight underpressuring and
recollimation of the jet that generates a smooth continuation
between both regions, as required by the model of LB02b. Although
there is no observational evidence for recollimation shocks in
this region, these outer shocks in the simulation are milder than
the first, so no strong emission from such structures would be
expected. The oscillations of the jet pressure around pressure
equilibrium captured by the simulation in the outer region are
still strong enough to generate shocks, though milder than the
previous ones. This difference between the simulation and the
model reflects the disagreement between the assumption of pressure
equilibrium at long distances to the core in LB02b model and the
numerical results that display an overpressured cocoon by the end
of the simulation.

\begin{figure}
\centerline{
\includegraphics[width=0.5\textwidth]{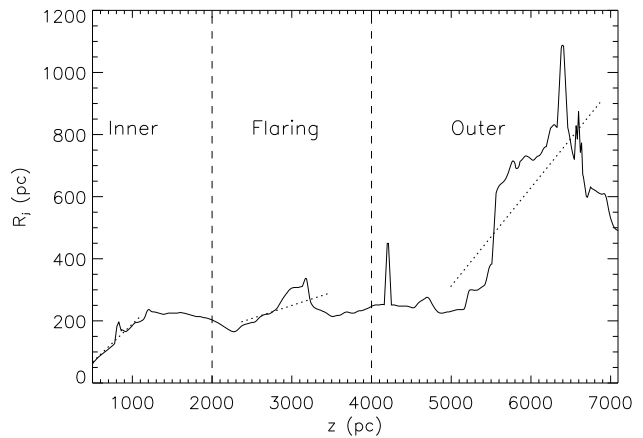}}
\caption{Jet radius versus distance in the last frame. Jet radius
is computed taking the outermost position where the axial velocity
is larger than $0.3\,c$. The boundaries between regions in the
simulation are marked with dashed vertical lines; dotted lines
indicate fitted parts in order to obtain opening angles.}
\label{fig:radj}
\end{figure}

In order to illustrate the discussion in the two previous
paragraphs, Fig.~\ref{fig:radj} shows the jet radius versus
distance. The plot clearly shows three regions for the jet
morphology as in LB02a,b. The transitions between the inner and
flaring regions and between the flaring and outer ones are
indicated in the plot. The initial expansion in the inner region,
expansion -due to the sudden increase in pressure and density
behind the recollimation shock- and contraction in the flaring
region, and further expansion of the jet in the outer region, are
observed here, as pointed out by LB02a (see their Fig.~4). The
oscillations of the jet radius in the outer region are due to the
irregularities of the flow. Dotted lines show linear fits of the
radius of the jet with respect to distance, that allow to give
estimates on local jet opening angles: $13.8\degr$ for the inner
region, $4.8\degr$ for the first half of the flaring region, and
$17.5\degr$ for the outer one. These values differ from those
obtained in LB02a (approximately $8.5\degr$, $18.5\degr$ and
$13.1\degr$ for the inner, flaring and outer regions,
respectively). The main difference appears in the flaring region
and is probably due to the series of recollimation shocks acting
in this region in our case. However it should be also pointed out
that we measure the radius of the jet using velocity as a
reference, whereas LB02a use the emitting material. We have shown
in Sec.~\ref{sec:evol} that the jet is radially stratified in the
simulation (Fig.~\ref{fig:evol2}), so the emitting material could
be surrounded by a non-emitting, slower wind that we consider as
part of the jet when computing its radius (e.g., gas moving with
velocities between $0.3\,c$ and $0.5\,c$).

  The jet in the simulation accelerates to a higher velocity (see
Fig.~\ref{fig:fin0}) than that given in LB02a,b for the jet in the
transition from the inner to the flaring region. In the numerical
simulation, the jet is injected into the grid with the speed given
in LB02b for the jet in the inner region ($0.87\,c$), and
acceleration down the pressure gradient speeds up the jet to
$\sim0.98\,c$. The differences in velocity of the flow and
position of the shocks were studied by means of three additional
simulations with different values for velocity and pressure ratio
with the external medium at injection: Simulation 2, with $v_j =
0.5\,c$ and $P_j/P_{a,c} = 3.8$; Simulation 3, with $v_j = 0.6\,c$
and $P_j/P_{a,c} = 7.8$, and Simulation 4, with $v_j = 0.5\,c$ and
$P_j/P_{a,c} = 1$. In simulations 2 and 3 the first recollimation
shock occurs at $z \sim 700-800\,\rm{pc}$, too close to the source
compared to observations, due to the smaller velocity (simulations
2 and 3) and smaller pressure ratio with respect to the ambient
medium at injection (simulation 2). This result points towards the
values used in simulations 1 (main simulation) and 3 for jet
overpressure ($P_j/P_{a,c} \sim 8$) and speed at injection ($v_j
\sim 0.6-0.87$) at $z \sim 500 \, \rm{pc}$ to be close to those in
the real jet in 3C~31, as we are able to reproduce the transition
between the inner and the flaring regions at about the appropriate
distances, given in LB02b. Fine tuning of the initial jet velocity
and pressure ratio with the external medium can certainly
reproduce the exact observed position of the standing shock in the
jet in 3C~31. As we have already pointed out, the significant
overpressure of the jet at $z\sim500\,\rm{pc}$ is required in
order to produce a discontinuity (recollimation shock) in the
transition from the inner to the flaring region at the observed
position. This is confirmed by results from simulations 2 and 4,
which have lower initial values of jet pressure: In simulation 2
the recollimation shock forms too close to the source compared
with the observations, and in simulation 4 the jet shows neither
significant expansions nor strong recollimation shocks that can
explain the increase in emission at the beginning of the flaring
region. We have not studied in this work the possible influence of
a change in the atmosphere model of \cite{hard02}. However, any
change in the density and pressure gradients would certainly
influence the position of the recollimation shock and the
properties of the jet at this point.

%%%%%%%%%%%%%%%%%%%%%%%%%%%%%%%%%%%%%%%%%%%%%%%%%%%%%%%%%%%%%%%%%%%%%%
%
%
%%%%%%%%%%%%%%%%%%%%%%%%%%%%%%%%%%%%%%%%%%%%%%%%%%%%%%%%%%%%%%%%%%%%%%

\subsection{Mass entrainment}

  From the panel of the tracer distribution at the end of the numerical
simulation (Fig.~\ref{fig:3c31rhot}), strong mass loading of the
flow is observed for distances $z > 4.5 \, \rm{kpc}$. At distances
shorter than $4.5 \rm{kpc}$ ambient material is entrained through
the jet boundary. In the region $z \sim 4.5 - 6 \, \rm{kpc}$ the
ambient material is entrained by the terminal shock. After this
shock, the jet flow is mainly subsonic favouring the entrainment
of ambient material. For comparison, in LB02b model, the strongest
entrainment occurs at $z \sim 3 - 3.5 \, \rm{kpc}$ (see their
Fig.~11) downstream the flaring point. After this local maximum in
the entrainment rate follows a monotonous increase beyond $z = 4
\, \rm{kpc}$. The authors claim that the mass entrainment is due
to stellar mass loss near the flaring point, but due to mixing in
a boundary layer farther away. In any case, the comparison between
the mass entrainment rate as a function of distance to the source
in LB02b model and that derived from our numerical simulation has
to be considered with caution, as the simulation has not reached a
steady state.

  Our simulation allows to conclude that the cause of
the discontinuity between the inner and the flaring region and of
the process of entrainment in the jets in 3C~31, is the generation
of a strong standing shock due to the initial overpressure of the
jet, in agreement with LB02b, and not to the growth of
Kelvin-Helmholtz instabilities to nonlinear amplitudes
\citep[e.g.,][]{roh00,pe+05}.

\section{Conclusions}\label{sec:concl}

  We present here a hydrodynamical relativistic simulation on
the evolution of a FRI jet, using parameters extracted from the
modelling that LB02a,b made for the radio jets of 3C~31. The
simulated jet is purely leptonic and propagates in a decreasing
density and pressure atmosphere with the profiles given in
\cite{hard02} for 3C~31. The simulation was followed up to $t =
7.26 \, 10^6$ yrs (about one tenth of the estimated lifetime of
3C~31).

  The simulation shows that the source can
go through a {\it weak CSS-like} phase in the early evolution. The
expansion speed, ages, sizes, and irregular morphologies are in
agreement with those claimed by \cite{dra04} for low power CSS
sources. The further evolution of the simulated jet into an FRI
morphology supports their claim that weak CSS sources are the
progenitors of large scale FRI jets.  Estimates of the age of the
source derived from the advance velocities measured in the
simulation lead to ages $t > 10^8 \, \rm{yrs}$, in agreement with
recent studies of FRI radio galaxies \citep{par02}. The lack of
emission gaps in the images points towards 3C~31 having been
continuously active since the triggering of its activity phase,
with no shorter-term periodicity.

  At the end of the simulation, the head of the bow shock has reached
a distance of $\sim 15 \, \rm{kpc}$ propagating at an almost
constant speed of $7 \, 10^{-3} c$ up to $t = 4.5 \, 10^6$ yrs,
decelerating afterwards. The region encompassed by the bow shock
can be divided in two parts: the cocoon, fed with jet material,
hot and light; the shocked ambient medium region, cooler and
denser. This structure is reminiscent of the cavity/shell
structure characteristic of the shocked regions surrounding
powerful jets. The pressure in the shocked region decreases with
time with a steeper slope than in the \cite{bc89} model. A simple
generalization of this model that takes into account the evolution
of the shocked regions in decreasing density atmospheres has been
used to explain the fast decrease of pressure as well as the
self-similar expansion of the shocked region. Moreover, assuming
self-similarity also for the cocoon evolution, our model is able
to explain the constant character of temperature in the cocoon
with time.

  At the end of the simulation, the bow shock is still slightly
supersonic (Mach number $\sim2.5$). This result is supported by
recent observations of bow shocks with Mach numbers between 3 and
8 in the low power radio galaxies Centaurus~A \citep{kra03} and
NGC~3801 \citep{cro07}. The fact that the jet power used in our
simulation for the jets in 3C~31 is in the upper part of the range
appropriate for FRI sources could explain the presence of a bow
shock in the simulation at larger distances from the galaxy than
observed in these galaxies. We show that the pressure and number
density jumps across the bow shock in the simulation are
consistent with those given in \cite{kra03} and \cite{cro07} for
Centaurus~A and NGC~3801, respectively. From this fact, we
conclude that the jet in 3C~31 possibly went through the same
stage as that observed for those sources, and that the
deceleration and disappearance of the bow shock should occur at
later times than those simulated here. A possible overestimate of
the bow shock pressure and velocity could be caused by the lack of
radiation cooling in the simulation.

  Focussing on the jet structure and dynamics, the simulation reveals
that the jet expands in the pressure gradient and undergoes
several subsequent recollimation and expansion processes that end
up in the formation of a wide shear layer, mass loading and
complete disruption of the flow. The parameters used in this
simulation succeed in explaining the general structure of the jet,
as modelled by LB02b. However, the exact locations of transition
between model regions in LB02b are not reproduced here. The
internal shocks in the simulated jet are formed at different
positions than given by the model. Also, the model predicts only
one discontinuity in the transition between the inner and flaring
regions, and a smooth transition from the flaring to the outer
region, whereas we find two more discontinuities in the jet, the
latter being the transition from the flaring to the outer region
in the simulation. This difference may be caused by the assumption
of pressure equilibrium at long distances ($z\sim12\,\rm{kpc}$)
made in the modelling of LB02b. It should be also kept in mind
that our simulation has not reached a steady state. This
assumption is validated by the fact that the jets in 3C31 show no
lobes, although these jets could have been overpressured as they
went through the initial stages simulated here, as it happens with
the jets in younger sources like 3C\,296 \citep{la06}. The extra
discontinuities captured by the simulation are a consequence of
the oscillations of the jet around pressure equilibrium with the
external medium, that turn out to be stronger than given by LB02b.
Fine tuning of the injection velocity ($v_j\sim0.6-0.87\,c$) and a
significant overpressure of the jet (as that used in our
simulation, $P_j/P_{a,c}\sim8$) are required at injection in the
grid ($z\sim500\,\rm{pc}$) in order to fit the first recollimation
shock to the position given in LB02b. We conclude that a standing
shock formed due to the initial overpressure of the jet is the
cause for the discontinuity found between the inner and the
flaring regions and for the mass entrainment of the jets in 3C~31.

  The simulation presented here confirms the paradigm for the evolution
of FRI's \citep{bic84,komi90a,komi90b,la93,la96}, in which the
adiabatic expansion of an overpressured jet is followed by
subsequent recollimation in shocks and expansion processes. The
jet is decelerated to transonic and subsonic velocities in these
shocks, and a mixing layer is formed that finally disrupts the
flow. However, recent X-ray observations and this simulation
indicate the existence of bow shocks still in kiloparsec scale FRI
jets. Further two and three-dimensional magnetohydrodynamic
simulations, including radiative cooling and realistic mass
entrainment from stellar mass losses should be performed in order
to disentangle the role of these processes in the structure and
dynamics of the FRI jets. It is important to note that these
simulations can be crucial for a deeper understanding not only of
the evolution of jets in FRI radio galaxies, but also of their
impact on the interstellar and intergalactic media, e.g., through
heating.

\section*{Acknowledgements}

  Calculations for this paper were performed on the SGI Altix 3000
computer \emph{CERCA} at the \emph{Servei d'Inform\`atica de la
Universitat de Val\`encia}. This work was supported in part by the
Spanish \emph{Direcci\'on General de Ense\~nanza Superior} under
grants AYA2001-3490-C02-01 and AYA2004-08067-C03-01 and
\emph{Conselleria d'Empresa, Universitat i Ciencia de la
Generalitat Valenciana} under project GV2005/244. M.P. benefited
from a postdoctoral fellowship in the Max-Planck-Institut f\"ur
Radioastronomie in Bonn and a postdoctoral fellowship of the
\emph{Generalitat Valenciana} (\emph{Beca Postdoctoral
d'Excel$\cdot$l\`encia}). The authors want to thank J.~Ferrando
for useful information on the calculation of the equation of
state, T.~Beckert and the referee of the paper, R.~Laing, for
their constructive criticism and useful comments.

\appendix

\section{The equation of state}

  The equation of state of a relativistic perfect gas can be written
in the form \citep{syn57,fal96}:

\begin{equation}\label{syngeeos1}
  w = \sum_{I=1}^{N} n_I m_I G(\xi_I),
\end{equation}

\begin{equation}\label{syngeeos2}
  p = \sum_{I=1}^{N} n_I m_I \xi_I^{-1},
\end{equation}

\noindent
where, $w =\rho h$, with $\rho$ the proper rest-mass density and $h$ the
specific enthalpy, $p$ is the pressure, $n_I$ is the number density of a
given family of particles with mass $m_I$,

\begin{equation}\label{xi}
  \xi_I = \frac{m_I}{k_B T},
\end{equation}

\noindent
and

\begin{equation}\label{gxi}
  G(\xi) = \frac{K_2(\xi)}{K_3(\xi)} =
  \frac{K_1(\xi)}{K_2(\xi)} + \frac{4}{\xi}.
\end{equation}

\noindent
In the latter equations $k_B$ is the Boltzmann constant, $T$, the
temperature and $K_\nu(\xi)$ are the modified Bessel functions:

\begin{equation}\label{knu}
  K_\nu(\xi) = \int_0^\infty exp(-\xi \cosh \theta) \,
  \cosh \nu \theta \, d\theta.
\end{equation}

  The adiabatic exponent is derived from the definition of sound
speed

\begin{equation}\label{sndsps}
  a^2 = \frac{1}{h} \left(\frac{\partial p}
  {\partial \rho}\right)_{s},
\end{equation}

\noindent
and turns out to be:

\begin{equation}\label{gamma}
  \Gamma = \frac{\sum_{I=1}^{N} n_I G'(\xi_I) \xi_I^2}{\sum_{I=1}^{N}
  n_I (G'(\xi_I) \xi_I^2+1)},
\end{equation}

\noindent
where $G'$ represents the derivative of the function with respect to its argument, $\xi_I$.

  In our case, we deal with two species of particles: leptons
(electrons and positrons) and baryons (protons). From the leptonic
and total proper rest-mass densities ($\rho$ and $\rho_l$,
respectively), charge neutrality allows to obtain the proton
rest-mass density, $\rho_p$, and the corresponding number
densities, $n_l$ and $n_p$ (and mass fractions, $X_l$ and $X_p$).
Then,

\begin{equation}
  w = \rho_l G(\xi_l) + \rho_p G(\xi_p),
\end{equation}

\noindent
and

\begin{equation}\label{syngeeos2}
  p = \rho_l \xi_l^{-1} + \rho_p \xi_p^{-1}.
\end{equation}

  Now, these two last equations together with the equations defining the set of
conserved variables, eqs.~(\ref{eq:D})-(\ref{eq:tau}), form an
implicit system from which the values of $\rho$, $\rho_l$, $p$,
$w$, $T$ and the two components of flow velocity, $v^R$ and $v^z$,
can be derived. With this purpose, an iteration in the temperature
is performed at each numerical cell in every time-step.

  Finally,

\begin{equation}
  \Gamma = \frac{n_l G'(\xi_l) \xi_l^2 + n_p G'(\xi_p) \xi_p^2}
           {n_l (G'(\xi_l) \xi_l^2 + 1) + n_p (G'(\xi_p) \xi_p^2 +
           1)},
\end{equation}
is the adiabatic exponent.

\label{lastpage}

\begin{thebibliography}{99}

\bibitem[Aloy et al.(1999)]{alo99} Aloy, M.A., Ib\'a\~nez, J.M$^{\underline{\rm a}}$,
Mart\'{\i}, J.M$^{\underline{\rm a}}$, G\'omez, J.L., M\"uller, E.
1999, ApJL, 523, 125


\bibitem[Begelman \& Cioffi(1989)]{bc89} Begelman, M.C., Cioffi, D.F. 1989, ApJL, 345,
21

\bibitem[Bicknell(1984)]{bic84} Bicknell, G.V. 1984, ApJ, 286, 68

\bibitem[Bicknell(1994)]{bic94} Bicknell, G.V. 1994, ApJ, 422, 542

\bibitem[Bicknell(1995)]{bic95} Bicknell, G.V. 1995, ApJS, 101, 29


\bibitem[Bowman(1994)]{bow94} Bowman, M. 1994, MNRAS, 269, 137

\bibitem[Bowman et al.(1996)]{bow96} Bowman, M., Leahy, J.P., Komissarov, S.S.
1996, MNRAS, 279, 899

\bibitem[Canvin \& Laing(2004)]{can04} Canvin, J.R., Laing, R.A., 2004, MNRAS, 350,
1342

\bibitem[Canvin et al.(2005)]{can05} Canvin, J.R., Laing, R.A., Bridle, A.H.,
Cotton, W.D. 2005, MNRAS, 363, 1223

\bibitem[Croston et al.(2007)]{cro07} Croston, J.H., Kraft, R.P., Hardcastle, M.J.
2007, astro-ph/0702094

\bibitem[de Berredo-Peixoto et al.(2005)]{pei05} de Berredo-Peixoto, G., Shapiro, I.L.,
Sobreira, F. 2005, Mod. Phys. Lett. A, 35, 2723

\bibitem[De Young(1986)]{you86} De Young, D.S. 1986, ApJ, 307, 62

\bibitem[De Young(1993)]{you93} De Young, D.S. 1993, ApJL, 405, 13

\bibitem[Drake et al.(2004)]{dra04} Drake, C.L., Bicknell, G.V., McGregor, P.J.,
Dopita, M.A. 2004, AJ, 128, 969

\bibitem[Falle \& Komissarov(1996)]{fal96} Falle, S.A.E.G., Komissarov, S.S. 1996, MNRAS 278,
586

\bibitem[Fanaroff \& Riley´s(1974)]{fr74} Fanaroff, B.L., Riley, J.M.  1974, MNRAS, 167, 31

\bibitem[Fanti et al.(1995)]{fan95} Fanti, C., Fanti, R., Dallacasa, D.,
Schilizzi, R.T., Spencer, R.E., Stanghellini, C. 1995, A\&A, 302,
317

\bibitem[Feretti et al.(1999)]{fer99} Feretti, L., Perley, R., Giovannini, G.,
Andernach, H. 1999, Mem. S. A. It., 70,1

\bibitem[Hardcastle et al.(2002)]{hard02} Hardcastle, M.J., Worrall, D.M., Birkinshaw, M., Laing, R.A., Bridle, A.H.
2002, MNRAS, 334, 182

\bibitem[Hardcastle et al.(2005)]{hard05} Hardcastle, M.J., Worrall, D.M., Birkinshaw, M., Laing, R.A., Bridle, A.H.
2005, MNRAS, 358, 843

\bibitem[Katz-Stone \& Rudnick(1997)]{kr97} Katz-Stone, D.M., Rudnick, L. 1997, ApJ, 488,
146

\bibitem[Kino et al.(2007)]{kin07} Kino, M., Kawakatu, N., Ito, H. 2007, MNRAS,
376, 1630

\bibitem[Komissarov(1990a)]{komi90a} Komissarov, S.S. 1990a, Ap\&SS, 165, 313

\bibitem[Komissarov(1990b)]{komi90b} Komissarov, S.S. 1990b, Ap\&SS, 165, 325

\bibitem[Komissarov(1994)]{komi94} Komissarov, S.S. 1994, MNRAS, 269, 394

\bibitem[Komissarov \& Falle(1998)]{kofa98} Komissarov, S.S., Falle, S.A.E.G. 1998, MNRAS, 297,
1087


\bibitem[Komossa \& Boehringer(1999)]{kom99} Komossa, S., B\"ohringer, H. 1999, A\&A, 344, 755

\bibitem[Kraft et al.(2003)]{kra03} Kraft, R.P., V\'azquez, S.E., Forman, W.R.,
Jones, C., Murray, S.S., Hardcastle, M.J., Worrall, D.M.,
Churazov, E. 2003, ApJ, 592, 129

\bibitem[Krause \& Camenzind(2003)]{km03} Krause, M., Camenzind, M. 2003,
astro-ph/0307152

\bibitem[Krause(2005)]{kra05} Krause, M. 2005, A\&A, 431, 45

\bibitem[Kunert-Bajraszewska et al.(2005)]{kun05} Kunert-Bajraszewska,
M., Marecki, A., Thomasson, P., Spencer, R.E. 2005, A\&A, 440, 93

\bibitem[Laing(1993)]{la93} Laing, R.A. 1993, in \emph{Astrophysical Jets}, eds.
Burgarella, Livio, O'Dea, Cambridge Univ. Press, p. 95

\bibitem[Laing(1996)]{la96} Laing, R.A. 1996, in \emph{Proceedings of the 175th IAU Symp.},
eds. Eckers, Fanti, Padrielli, Kluwer, p. 147

\bibitem[Laing \& Bridle(2002a)]{lb02a} Laing, R.A., Bridle, A.H. 2002a, MNRAS, 336, 328

\bibitem[Laing \& Bridle(2002b)]{lb02b} Laing, R.A., Bridle, A.H. 2002b, MNRAS, 336, 1161

\bibitem[Laing \& Bridle(2004)]{lb04} Laing, R.A., Bridle, A.H. 2004, MNRAS, 348,
1459

\bibitem[Laing et al.(2006)]{la06} Laing, R.A., Canvin, J.R., Bridle, A.H., and
Hardcastle, M.J. 2006, MNRAS 372, 510

\bibitem[Mart\'{\i} et al.(1997)]{mart97} Mart\'{\i}, J.M$^{\underline{\rm a}}$, M\"uller, E., Font, J.A.,
Ib\'a\~nez, J.M$^{\underline{\rm a}}$, and Marquina, A. 1997, ApJ,
479, 151

\bibitem[Parma et al.(2002)]{par02} Parma, P., Murgia, M., de Ruiter, H.R.,
Fanti, R. 2002, NewAR, 46, 313

\bibitem[Perucho \& Mart\'{\i}(2002)]{pm02} Perucho, M., Mart\'{\i}, J.~M. 2002,
ApJ, 568, 639

\bibitem[Perucho et al.(2004)]{pe+04b} Perucho, M., Mart\'{\i}, J.~M., Hanasz, M. 2004,
A\&A, 427, 431

\bibitem[Perucho et al.(2005)]{pe+05} Perucho, M., Mart\'{\i}, J.~M., Hanasz, M. 2005,
A\&A, 443, 863

\bibitem[Rosen \& Hardee(2000)]{roh00} Rosen, A., Hardee, P.E. 2000, ApJ, 542, 750

\bibitem[Scheck et al.(2002)]{sch02} Scheck, L., Aloy, M.A., Mart\'{\i}, J.M$^{\underline{\rm a}}$,
G\'omez, J.L., M\"uller, E. 2002, MNRAS, 331, 615

\bibitem[Service(1986)]{ser86} Service, A.T. 1986, ApJ, 307, 60

\bibitem[Synge(1957)]{syn57} Synge, J.L. 1957, \emph{The relativistic gas},
North-Holland, Amsterdam

\bibitem[Worrall et al.(2007)]{wor07} Worrall, D.M., Birkinshaw, M.,
Laing, R.A., Cotton, W.D., Bridle, A.H. 2007, MNRAS, 380, 2

\bibitem[Zanni et al.(2003)]{za03} Zanni, C., Bodo, G., Rossi, P., Massaglia,
S., Durbala, A., Ferrari, A. 2003, A\&A, 402, 949

\end{thebibliography}
\end{document}